\documentclass[review]{elsarticle}

\usepackage{adjustbox}
\usepackage{longtable}

\PassOptionsToPackage{table}{xcolor}
\usepackage[numbers]{natbib}
\usepackage{longtable}
\usepackage{microtype}
\usepackage[utf8]{inputenx}
\usepackage[english]{babel}
\usepackage{floatrow}
\usepackage{subcaption}
\usepackage{url}
\usepackage{colortbl}
\usepackage{amsmath}
\usepackage{mathrsfs}
\usepackage{amssymb}
\usepackage[left]{lineno}
\usepackage[nonumberlist,acronym,sanitize=none]{glossaries}
\glsdisablehyper
\usepackage{comment}
\usepackage[capitalise,nameinlink]{cleveref}
\Crefname{algocf}{Algorithm}{Algorithms}
\Crefname{lemma}{Lemma}{Lemmata}
\crefname{definition}{Def.}{Def.s}
\crefname{example}{Ex.}{Ex.s}
\crefalias{AlgoLine}{line}
\usepackage{tkz-base}
\usetikzlibrary{decorations.pathmorphing,trees,snakes,arrows,shapes,petri,automata}
\usepackage{paralist}
\usepackage{multicol}
\usepackage{booktabs}
\usepackage{enumitem}
\usepackage{multirow}
\usepackage{rotating}
\usepackage{etaremune}
\usepackage[ruled,linesnumbered,algo2e,vlined]{algorithm2e}
\usepackage{mathtools}
\usepackage{ifthen}
\usepackage[normalem]{ulem}
\usepackage{lineno}
\usepackage{soul}
\usepackage{floatrow}
\usepackage{lipsum} 
\usepackage{microtype}
\usepackage{xfrac}
\usepackage{tabto}
\usepackage[scientific-notation=false]{siunitx}
\usepackage{etoolbox}
\robustify\bfseries
\usepackage{wrapfig}
\usepackage{xspace} 

\usepackage{makecell}

\usepackage{adjustbox}

\usepackage{graphicx}

\usepackage{listings}
\crefname{lstlisting}{Listing}{Listings}

\usepackage[textsize=scriptsize,backgroundcolor=yellow!40]{todonotes}
\usepackage{blindtext}
\usepackage{regexpatch}

\makeatletter
\xpatchcmd{\@todo}{\setkeys{todonotes}{#1}}{\setkeys{todonotes}{inline,#1}}{}{}
\makeatother

\usepackage{upgreek}

\newacronym[\glslongpluralkey={Business Processes}]{bp}{BP}{Business Process}
\newacronym{wf}{WF}{workflow}
\newacronym{bpi}{BPI}{Business Process Intelligence}
\newacronym{bpm}{BPM}{Business Process Management}
\newacronym{bpms}{BPMS}{Business Process Management System}
\newacronym{bpmn}{BPMN}{Business Process Model and Notation}
\newacronym{soa}{SOA}{Service-Oriented Architecture}
\newacronym{kpi}{KPI}{Key Performance Indicator}
\newacronym{wfms}{WfMS}{Workflow Management System}
\newacronym{pn}{PN}{Petri net}
\newacronym{cpn}{CPN}{colored Petri net}
\newacronym{xes}{XES}{eXtensible Event Stream}
\newglossaryentry{task}{%
	name={task},description={the non-divisible, elementary activity}}

\newglossaryentry{promod}{%
	name={process model},description={the model of a process}
}
\def\LogAlph {\ensuremath{\Sigma}}

\newglossaryentry{logalph}{
	name={log alphabet},description={the process alphabet, as reflected in a log},%
	symbol={\LogAlph}}
\def\Evt {\ensuremath{e}}
\newglossaryentry{evt}{
	name={event},description={a record of an instantaneous fact during the process enactment},%
	symbol={\Evt}}
\def\EvtTrace {\ensuremath{t}}

\newglossaryentry{evttrace}{
	name={trace},description={a sequence of \glsplural{evt}},%
	symbol={\EvtTrace}}
\def\EvtLog {\ensuremath{L}}

\newglossaryentry{evtlog}{
	name={event log},description={a collection of \glstext{evttrace}s},%
	symbol={\EvtLog}}
\newacronym{po}{PO}{Partial Order}
\newacronym{tl}{TL}{Temporal Logic}  
\newacronym{ltl}{LTL}{Linear Temporal Logic}
\newacronym{ldl}{LDL}{Linear Dynamic Logic}
\newacronym{ldlf}{LDL$_f$}{Linear Dynamic Logic over Finite Traces}
\newacronym{fol}{FOL}{First Order Logic}
\newacronym{ltlf}{LTL$_f$}{Linear Temporal Logic on Finite Traces}
\newacronym{ltlp}{LTLp}{Linear-time Temporal Logic with Past}
\def\ltlpf {\ensuremath{\textrm{LTLp}_f}}
\newacronym{ltlpf}{\ltlpf}{Linear-time Temporal Logic with Past on Finite Traces}
\newacronym{mso}{MSO}{Monadic Second Order Logic}
\newacronym{rex}{RE}{regular expression}
\def\MultiSetFunctor {\ensuremath{\mathbb{M}}}
\newglossaryentry{multiset}{
	name={multi-set},description={a collection possibly containing multiple units of the same element},
	symbol={\MultiSetFunctor}}

\def\PowerSetFunctor {\ensuremath{\mathbb{P}}}
\newglossaryentry{powerset}{
	name={power-set},description={the collection of sets generated by all combinations without repetition of elements in a set},
	symbol={\PowerSetFunctor}}

\def\Autom {\ensuremath{A}}
\newacronym[symbol=\Autom,longplural={finite state automata}]{fsa}{FSA}{finite state automaton}
\newacronym[symbol=\Autom,longplural={deterministic finite-state automata}]{dfa}{DFSA}{deterministic finite-state automaton}
\newacronym[symbol=\Autom,longplural={nondeterministic finite-state automata}]{nfa}{NFSA}{nondeterministic finite-state automaton}
\def\AutomInitState {\ensuremath{s_0}}

\newglossaryentry{fsainit}{name={initial state},description={initial state of the automaton},
	symbol=\AutomInitState}

\newglossaryentry{satis}{name={satisfaction},description={evaluation as true of a formula on a structure}}
\newglossaryentry{verif}{name={verification},description={evaluation of a formula on a structure}}
\def\LLsim {\ensuremath{\mathrm{L2L_{freq}}}}
\def\TTsim {\ensuremath{\mathrm{L2L_{trace}}}}
\def\FEsim {\ensuremath{\mathrm{L2L_{first}}}}
\def\PEsim {\ensuremath{\mathrm{L2L_{2gram}}}}
\def\TEsim {\ensuremath{\mathrm{L2L_{3gram}}}}
\def\Csim {\ensuremath{\mathrm{L2L_{case}}}}
\def\ECSA {\ensuremath{\mathrm{\text{EC-SA}}}\xspace}
\def\ECSAnew {\ensuremath{\mathrm{\text{EC-SA-Data}}}\xspace}
\def\LLsmape {\ensuremath{\mathrm{SMAPE_{ET}}}\xspace}
\def\LLCTsmape {\ensuremath{\mathrm{SMAPE_{CT}}}\xspace}

\def\SEinsdel {\ensuremath{\Delta^{\mathrm{ins}}_{\mathrm{del}}}}

\def\scurr {\ensuremath{S_\mathrm{curr}}}
\def\smax {\ensuremath{S_\mathrm{max}}}
\def\tinit {\ensuremath{\uptau_\mathrm{init}}}
\def\tcurr {\ensuremath{\uptau_\mathrm{curr}}}
\def\fa {\ensuremath{\mathrm{f_{a}}}}
\def\ft {\ensuremath{\mathrm{f_{t}}}}
\def\fr {\ensuremath{\mathrm{f_{r}}}}
\def\fc {\ensuremath{\delta\mathrm{f_{c}}}}
\def\prob {\ensuremath{\mathrm{prob}}}

\newcommand{\abbrev}[1]{\ensuremath{\mathrm{#1}}}
\newcommand{\inst}[1]{\ensuremath{\mathsf{#1}}}
\newcommand{\attr}[1]{\ensuremath{\mathrm{#1}}}
\def\ifcon {\ensuremath{\texttt{IF}}}
\def\thencon {\ensuremath{\texttt{THEN}}}
\def\ifthencon {\ensuremath{\ifcon\textrm{-}\thencon}}

\newcolumntype{d}{>{\columncolor{gray!10}}c}
\newcolumntype{m}{>{\columncolor{gray!10}}l}
\newfloatcommand{capbtabbox}{table}[][\FBwidth]
\newenvironment{iiilist}%
{\begin{inparaenum}[\itshape(i)\upshape]}%
	{\end{inparaenum}}

\RequirePackage{xparse}
\NewDocumentEnvironment{AuthNote}{+o+o}{%
	\IfValueT{#2}{\marginnote{\scriptsize{#2}}}%
	\begin{scriptsize}
		\colorbox{gray}%
		{\color{white} Note\IfValueT{#1}{ (#1)}:}%
		\quad%
		\color{brown}
	}{%
		\normalcolor
	\end{scriptsize}
}

\RequirePackage{siunitx}
\newcolumntype{D}[1]{S[
	table-omit-exponent,
	round-mode=places,
	round-precision={#1}]} 
\newdefinition{definition}{Definition}

\newcommand{\veewedge}{\mathrel{\substack{\vee\\\wedge}}}
\newcommand{\wedgephi}{\widehat{\varphi}}
\begin{document}
\let\WriteBookmarks\relax
\def\floatpagepagefraction{1}
\def\textpagefraction{.001}
\begin{frontmatter}
\title{Event-Case Correlation for Process Mining using Probabilistic Optimization}
%

%
%
%
%

\author[WU,CU]{Dina Bayomie}

\address[WU]{Vienna University of Economics and Business, Austria}
\address[CU]{Cairo University, Egypt}
\ead{dina.sayed.bayomie.sobh@wu.ac.at}

\author[R]{Claudio Di Ciccio}
\address[R]{Sapienza University of Rome, Italy}
\ead{claudio.diciccio@uniroma1.it}
\author[HU]{Jan Mendling}
\address[HU]{Humboldt-Universit{\"a}t zu Berlin, Germany}
\ead{jan.mendling@hu-berlin.de}
\cortext[1]{Corresponding author}

\begin{abstract}
Process mining supports the analysis of the actual behavior and performance of business processes using event logs. 
An essential requirement is that every event in the log must be associated with a unique case identifier (e.g., the order ID of an order-to-cash process). In reality, however, this case identifier may not always be present, especially when logs are acquired from different systems or extracted from non-process-aware information systems. In such settings, the event log needs to be pre-processed by grouping events into cases -- an operation known as event correlation. 
Existing techniques for correlating events have worked with assumptions to make the problem tractable: some assume the generative processes to be acyclic, while others require heuristic information or user input. Moreover, 
they abstract the log to activities and timestamps, and miss the opportunity to use data attributes.
In this paper, we lift these assumptions and propose a new technique called EC-SA-Data based on probabilistic optimization. The technique takes as inputs a sequence of timestamped events (the log without case IDs), a process model describing the underlying business process, and constraints over the event attributes.
Our approach returns an event log in which every event is associated with a case identifier.
The technique allows users to incorporate rules on process knowledge and data constraints flexibly.
The approach minimizes the misalignment between the generated log and the input process model, maximizes the support of the given data constraints over the correlated log, and the variance between activity durations across cases.
Our experiments with various real-life datasets show the advantages of our approach over the state of the art.
\end{abstract}
%
%
%
\end{frontmatter}

\maketitle

\section{Introduction}\label{intro}
Recent years have seen a drastically increasing availability of process execution data from various data sources~\cite{beheshti2016scalable,benatallah2016process,DBLP:journals/is/SofferHKZCKKJSS19}. Process mining offers different analysis techniques that can extract business insights from these data, known as event logs. Each event in a log must have at least three attributes \cite{DBLP:conf/bpm/AalstAM11,PM}: 
\begin{iiilist}
	\item the \emph{event class} referring to a specific activity in the process (e.g., ``Order checked'' or ``Claim assessed''),
	\item the \emph{end timestamp} capturing the occurrence of the event, and
	\item the \emph{case identifier} (e.g., the order number in an order-to-cash process, or the claim ID in a claims handling process).
\end{iiilist} 
Recent process mining algorithms such as 
$\alpha\$$~\cite{guo2015mining},
Inductive Miner~\cite{leemans2014infrequent},
Evolutionary Tree Miner~\cite{BuijsDV12},
Fodina~\cite{vanden2017fodina},
Structured Heuristic Miner~\cite{augusto2017automated},
Split Miner~\cite{augusto2018split}, and the
Hybrid ILP Miner~\cite{van2015ilp}
need all of these three attributes together for discovering a process model. 

\begin{figure}[tb]%
	\includegraphics[width=0.9\textwidth]{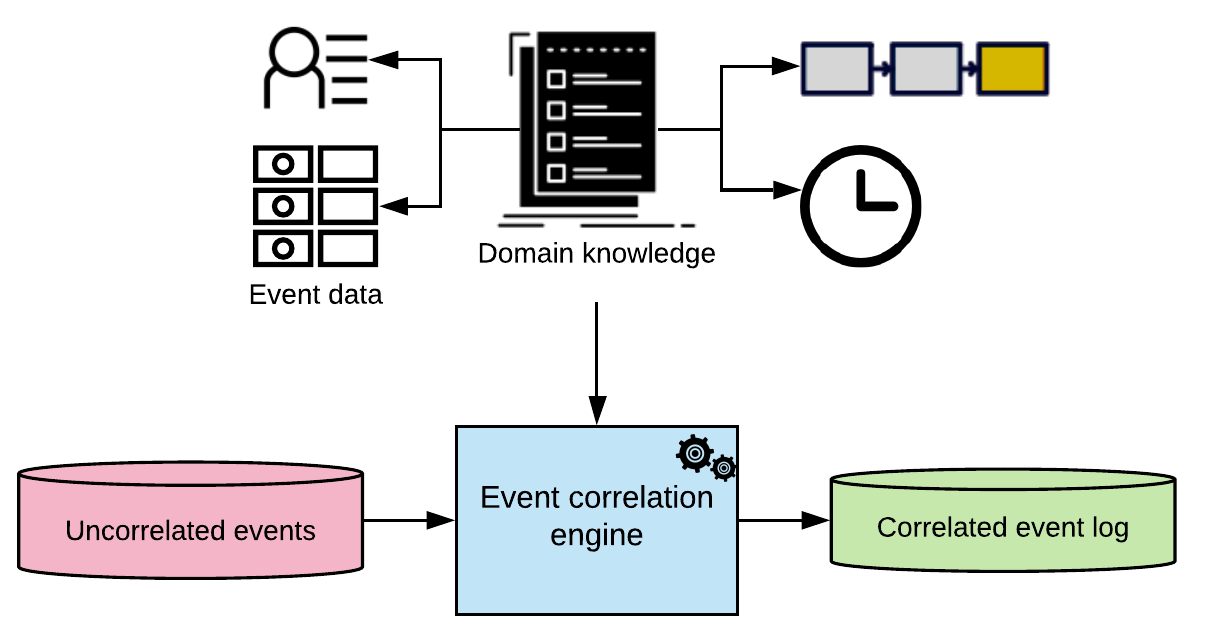}%
	\caption{Overview of the event correlation engine}%
	\label{fig:approach}%
\end{figure}

Various data infrastructures such as data lakes often give more attention to data storage than to structuring them such that process mining can be readily applied~\cite{DBLP:conf/ifip8-1/BalaMSQ18,MeroniCM17}.
Prior research has described the problem of missing case identifiers as a \emph{correlation problem}, because the connections between different events have to be reestablished based on heuristics, domain knowledge or payload data. In essence, the correlation problem is concerned with identifying which events belong to the same case when a unique case identifier is missing. 
This identification can be done by an event correlation engine (see \Cref{fig:approach}).
An event correlation engine constructs a correlated log with case identifiers by using domain knowledge, e.g. about the process control flow, organizational resources, maximal task durations, or other data knowledge over the event attributes.
Existing correlation techniques face the challenge of operating in a large search space. For this reason, previous proposals have introduced assumptions to make the problem tractable. The main assumptions they have in common is abstracting a log on activities and timestamps with ignoring the other attributes. Moreover, some techniques assume the generative processes to be acyclic \cite{cMiner,eMax} while others require heuristic information about the execution behavior of activities in addition to the process model \cite{DCIc}. Beyond that, these approaches suffer from poor efficiency and miss the opportunity to make use of data attributes. 

In previous work \cite{ECSA-ER}, we introduced a probabilistic optimization technique called \emph{\ECSA} (Events Correlation by Simulated Annealing), which is based on a simulated annealing heuristic approach. {\ECSA} addresses the correlation problem as a multi-level optimization problem. 
In this paper, we extend \ECSA to consider a broader spectrum domain knowledge for the correlation process, integrating ideas of mixing process specification paradigms~\cite{van2020conformance}. We call the extension \emph{\ECSAnew}. 
Using the domain knowledge about the event data attributes improves the event correlation process as it decreases the random assigning of events to their corresponding cases.
The technique revolves around three nested objectives.
First, it seeks to minimize the misalignment between the generated log and an input process model.
Second, it aims to maximize the support of the given data constraints over the correlated log.
Third, it targets to minimize the activity execution time variance across cases. The latter objective builds on the assumption that the same activities tend to have similar duration across cases. Our extensive evaluation demonstrates the benefits of our novel technique.

The remainder of this paper is organized as follows.
\Cref{related} discusses prior research on the event correlation problem. 
\cref{preliminaries} defines preliminaries that are essential for our technique.
\Cref{CSA} presents the different phases of our novel \ECSAnew event correlation technique. 
\Cref{evaluation} then discusses the experimental evaluation on real-life logs before \Cref{conclusion} concludes the paper.

\section{Related Work}\label{related}



The correlation problem focuses on identifying two or more variables that are related, even though this relationship is not explicitly define. It received adequate attention in the data science research field for static data \cite{Calude2017TheDO,l2017machine,ASAJournal}. The existing techniques have various objectives to improve the analysis, monitoring and mining of the data. Database correlation techniques work on improving query performance of knowledge discovery. Jermaine \cite{JERMAINE200521} discusses the discovery of correlations between database attributes. He develops measures 
of independence between the attributes to discover the actual correlation, and analyses the computational complexity of correlation techniques. Brown and Haas \cite{BHUNT} present a data-driven technique that discovers the hidden relation between database attributes and integrates them into an optimizer to improve query performance. Though valuable for data, these works cannot be directly applied for process mining event correlation. They lack a notion of temporal order. 

In this paper, we focus on a specific instance of the \emph{event correlation} problem: how to correlate those events that have been generated from the execution of the same process instance? An event correlation technique takes as input at least an uncorrelated log \cite{Diba2019}. Depending on whether or not a technique relies on additional input, we classify existing techniques into four categories: 1) requiring no additional input; 2) requiring a process model; 3) relying on correlation conditions; and 4) relying on event similarity functions.

The first category of techniques uses as input only event names and ignores the other event data attributes. 
Ferreira and Gillbald \cite{eMax} provide an Expectation-Maximization {(E-Max)} approach that builds a Markov model from the uncorrelated log and discovers the possible process behavior in the uncorrelated log. This technique is adversely sensitive to the density of the working cases within the system,i.e., the number of overlapping cases at a given point in time. Also, it does not support concurrency and cyclic process behaviors.
Walicki and Ferreira \cite{seq} provide a {sequence partitioning} approach that searches for the minimal set of patterns that can represent the process behavior in an uncorrelated log. The technique does not support concurrency and cyclic process behaviors.

The second category includes correlation techniques that use timestamps of events and the process model. 
The Correlation Miner {(CMiner)} by Pourmiza et al.~\cite{CMinerConf} 
builds on two matrices, one capturing proceed/succeed relations and another one capturing the time difference between pairs of events. 
An optimal correlation is calculated using integer linear programming. An extension of the CMiner 
\cite{cMiner} 
uses quadratic programming to find the optimal correlation by minimizing the duration between the events. CMiner does not support cyclic processes and relies on quadratic constraints solving, which limits its scalability. Experiments reported in \cite{cMiner} show that the approach can only handle logs with a few dozen cases.
The Deducing Case Ids {(DCI)} approach~\cite{DCI} requires a process model and heuristic information about the activity execution durations. 
The approach utilizes a breadth-first approach to build a case decision tree to explore the solution space. 
In \cite{DCIc}, DCI is extended with a pre-processing step to detect the cyclic behavior and build a relationship matrix for the correlation decision.  
DCI supports cyclic processes. It is sensitive to the quality of the input data, such that if the model and the log have low fitness, then the generated correlated logs will contain noise and missing events. Also, it is computationally inefficient due to the breadth-first search approach. 
In a previous paper~\cite{ECSA-ER}, we propose the Event Correlation by Simulated Annealing {(\ECSA)} approach, which uses the event names and timestamp in addition to the process model. {\ECSA} addresses the correlation problem as a multi-level optimization problem, as it searches for the nearest optimal correlated log considering the fitness with an input process model and the activities' timed behavior within the log. 
The accuracy of the given model affects the quality of the correlated log, and the performance is affected by the number of uncorrelated events.

The third category includes correlation techniques that use event data attributes and apply correlation conditions to correlate the events.
Motahari-Nezhad et al.~\cite{ProcessView} propose a semi-automated correlation approach to correlate the web service messages based on the correlation conditions. The approach derives correlation conditions using the event data from different data layers. Also, it computes the interestingness of the attributes of the events to prune the conditions search space. Thus, it generates several log partitions and possible process views. The approach requires user-defined domain parameters and intermediate domain expert feedback to guide the correlation process.
Engel et ¨\cite{EDImessages} propose the {EDImine} framework, which allows for the usage of process mining over electronic data interchange (EDI) messages. EDIminer resorts to message flow mining and physical activity mining methods to generate the events from EDI messages. Then, EDIminer employs user-defined correlation rules to correlate the events to their cases. This framework is limited to EDI messages and depends on the user-defined parameters for the event generation and the correlation process.
Reguieg et al.~\cite{EC-mapReduce} propose a MapReduce-based approach to derive the correlation conditions from the service interaction logs. It consists of two stages. The first stage defines the simple correlation conditions. The second stage derives more complex correlation conditions and correlates the events to their cases. In~\cite{Reguieg2015-mapReduce}, the authors extend their previous work to improve scalability and efficiency. They introduce two strategies to perform the log partition and explore the complex correlation condition space. The main challenges of the approach are the log partitioning and the vast amount of network traffic communication.
Cheng et al.~\cite{Cheng2017-mapReduce} propose the Rule Filtering and Graph Partitioning {(RF-Grap)} approach, which follows the filtering and verification principle to improve the efficiency of event correlation using distributed platforms. RF-Grap prunes a large number of uninteresting correlation rules in the filtering step. Accordingly, not all the derived rules are investigated, but only the interesting rules that fulfil the criteria. In the verification steps, they use graph partitioning to decompose the correlation possibilities over the clusters.
De Murillas et al.~\cite{redoLog} propose an approach to extract the event log from a database based on the redo logs that contain the events of data manipulation. They use a data model to define the relation between the events. In~\cite{caseNotation}, the authors provide a way to automatically generate different event logs from a database by defining the case notion based on the data relations in the data model. A case notion defines which events should be considered for the correlation based on the selected data objects that represent the investigated cases.
They measure the interestingness of the generated logs and recommend the highest one to the user.

The fourth category includes the correlation techniques that rely on event similarity or the case identifier in the log. 
Djedović et al.~\cite{ECruleBased} propose an algorithm to compute the similarity between pairs of events to correlate events with higher similarity. Event similarity function defined over the equality between the attributes over the events. The optimal correlation represents the highest similarity score between the case's events. The approach expects the existence of main attributes that do not change over the case.
Abbad Andaloussi et al.~\cite{automatedEventLabeling2018} address the correlation problem under the assumption that event data already contains the case identifier. For each event log attribute A, the technique discovers a process model assuming that A is the case identifier. The resulting models are compared based on four quality measures. The attribute that yields the highest-quality model is taken as the case identifier. Bala et al.~\cite{DBLP:conf/ifip8-1/BalaMSQ18} follow a similar direction based on the idea that identifiers are repetitive in the log.
Burattin and Vigo~\cite{DecorativeAtt} propose a framework that generalized from a real business case. They search the activities attributes over the log to define the possible process instance attributes. Then they used the equality relation between these attributes to correlate the events to their cases. This method relies on a-prior knowledge about the application domain and user-defined heuristic parameters, such as the number of events within a case and characteristics of the candidate attributes.



In summary, the approaches in the first category assume that the process is acyclic. Those in the second category expect that a full process model is available. The third category require correlation rules to be provided or discovered heuristically. Approaches in the fourth category assume that there is a case identifier attribute or a similarity function that allows to group events. 
The technique presented in this paper extends \ECSA approach. It integrates the strengths of the first and second category. Unlike the first category, it can handle cyclic behavior. It also allows users to provide additional domain knowledge that will support the correlation process. In this way, it relaxes the dependence on the control-flow knowledge. 

\section{Preliminaries}\label{preliminaries}
%
In this section, we discuss the fundamental notions that our approach builds upon.  \Cref{sec:notation} describes the data structures we handle. \Cref{sec:modeling} outlines our process modeling and execution language and notations. \Cref{sec:sa} illustrates the fundamental mechanisms underpinning simulated annealing, a core technique for our solution .

\begin{figure*}[tbp]%
	\includegraphics[width=0.75\textwidth]{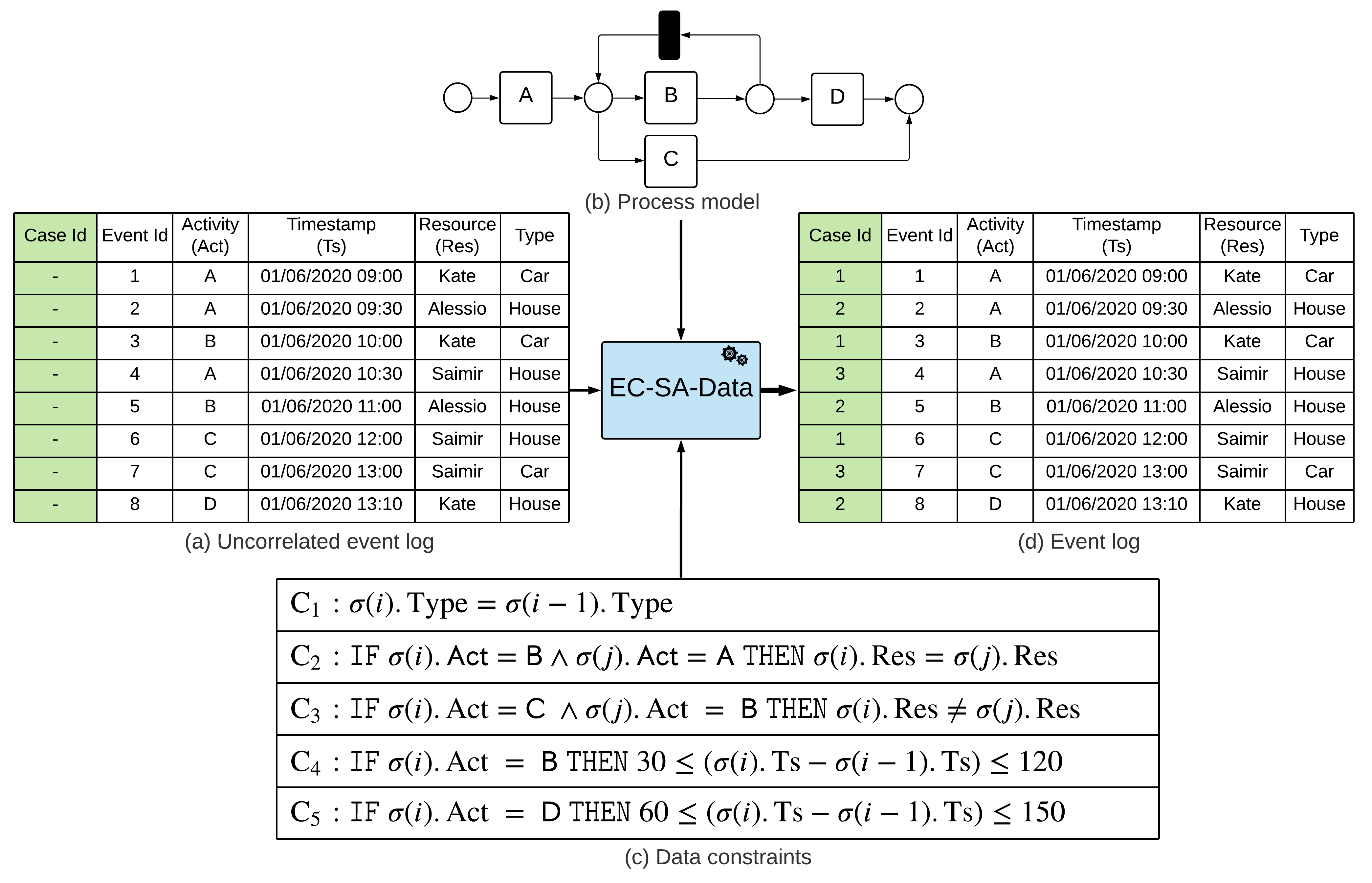}%
	\caption{Running example of a sample loan application check process}%
	\label{fig:example}%
\end{figure*}%

\subsection{Process Event Data Structures}\label{sec:notation}
Starting with the basic notion of event (i.e., the atomic unit of execution), we introduce the uncorrelated event log, case and projection of a case over an event attribute.
Thereupon, we present the definitions of event log and trace.
\begin{definition}[Event]\label{def:event}
	Let $E \ni \Evt$ be a finite non-empty set of symbols. We refer to $e$ as \emph{event} and to $E$ as the \emph{universe of events}.
\end{definition}
\begin{definition}[Attribute]\label{def:attribute}
	Given a non-empty set of domain values $\textrm{Dom} \in \mathfrak{D}$, an \emph{attribute} $ \textrm{Attr} \in \mathfrak{A}$ is a partial function $\textrm{Attr} : E \nrightarrow \textrm{Dom} $ mapping events to domain values. We indicate the value mapped by $\textrm{Attr}$ to an event $e$ by using a dot notation, i.e., $e.\textrm{Attr}$.
\end{definition}
In the following, we assume without loss of generality that attributes $\attr{Act}$ and $\attr{Ts}$ are \emph{total} functions defined over events. The range of the former is a finite subset of strings we interpret as \emph{activity} names and the range of the latter is a finite subset of integers to be interpreted as \emph{timestamps}. 
For example, $\Evt_1$ in \cref{fig:example} is mapped to four different values, one per attribute:
$\Evt_1.\attr{Act} = \inst{A}$ represents the executed activity,
$\Evt_1.\attr{Ts} = $``$\inst{07/06/2020\ 09:00}$'' represents the completion timestamp,
$\Evt_1.\attr{Res} = \inst{R}1$ represents the operating resource,
and
$\Evt_1.\attr{Type} = \inst{Home}$ represents additional data knowledge about the event context. 
\begin{definition}[Uncorrelated log] \label{def:uncorrelatedLog}
	Let $E \ni \Evt$ be the universe of events, $\mathfrak{D} \ni \textrm{Dom}$ the universe of domain values and $\mathfrak{A} \ni \textrm{Attr}$ a set of attributes defined over $E$ as per \cref{def:event,def:attribute}.
	Given a total order defined over $E$, $\preccurlyeq \; \subseteq E \times E$ (henceforth referred to as \emph{event ordering}),
	the \emph{uncorrelated log} is a tuple $\abbrev{UL} = (E, \mathfrak{D},  \mathfrak{A}, \preccurlyeq)$.
\end{definition}
We assume the mapping of $\attr{Ts}$ to be coherent with $\preccurlyeq$, i.e., if $e \preccurlyeq e'$ then $e.\attr{Ts} \leqslant e'.\attr{Ts}$.
Considering the total ordering as a mapping from a convex subset of integers, we can assign to every event a unique integer index (or \emph{event id} for short), induced by $\preccurlyeq$ on the events. We shall denote the index $i \in [1,|E|]$ of an event $e$ as a subscript, $e_i$.
For example, \cref{fig:example}(a) depicts an uncorrelated log and $\Evt_3$ is its third event. 

\begin{definition}[Event log] \label{def:eventLog}
	Let $\abbrev{UL} = (E, \mathfrak{D},  \mathfrak{A}, \preccurlyeq)$ be an uncorrelated log as per \cref{def:uncorrelatedLog} and $I \ni \iota$ be a universe of \emph{case identifier}s (or \emph{case id}'s for short).
	An event log is a triple $\EvtLog = (\abbrev{UL}, I, \ell)$ where $\ell: E \to I$ is a total surjective function.
\end{definition}
\noindent
Notice that $\ell$ induces a partition of $E$ into $|I|$ subsets.
\Cref{fig:example}(d) illustrates an event log. Notice that $I = \{1,2,3\}$ and $\ell$ maps events
$\Evt_1$, $\Evt_3$ and $\Evt_6$ to $1$,
$\Evt_2$, $\Evt_5$ and $\Evt_8$ to $2$, and
$\Evt_4$ and $\Evt_7$ to $3$.
We name the sequences of events that stem from mapping $\ell$ and preserve $\preccurlyeq$ as \emph{cases}.

\begin{definition}[Case] \label{def:case}
	Given a case id $\iota \in I$ and an event log $L$ as per \cref{def:eventLog}, a \emph{case} $ \sigma $ defined by $\iota$ over $ L $ is a finite sequence
	$ \langle e_{\sigma,1}, \ldots, e_{\sigma,n} \rangle \in E^*$ of length $|\sigma| = n \in \mathbb{N}$ of events $e_{\sigma,i} \textrm{ with } 1 \leqslant i \leqslant n$ such that (\emph{i}) $\ell(e_{\sigma,i}) = \iota$ and (\emph{ii}) the sequence is induced by $\preccurlyeq$, i.e., $e_{\sigma,j} \preccurlyeq e_{\sigma,i}$ for every $j$ s.t.\ $j \leqslant i \leqslant n$. 
\end{definition}
\noindent
For example, the event log depicted in \cref{fig:example}(d) is comprised of \num{3} cases. Case $\sigma_1$ defined by case id $1$ is $\langle \Evt_1,\Evt_3,\Evt_6 \rangle$. Notice that it preserves the order of the events within the case.
We name the projection of a case over the activities of its events as \emph{trace}.

\begin{definition}[Trace] \label{def:trace}
	Given a case $\sigma = \langle e_{\sigma,1}, \ldots, e_{\sigma,n} \rangle \in E^*$ and the total attribute function $\attr{Act} : E \to \mathrm{Dom}_\attr{Act}$, a trace $t \in \mathrm{Dom}_\attr{Act}^*$ is the sequence induced by case $\sigma$ through the mapping of $\attr{Act}$, 
	i.e., $t = \langle e_{\sigma,1}.\attr{Act}, \ldots, e_{\sigma,n}.\attr{Act} \rangle$. 
\end{definition}
\noindent
In our example, the trace corresponding to $\sigma_1$ is $\langle \inst{A}, \inst{B}, \inst{C} \rangle$.

\subsubsection*{Short-hand notations}
For the sake of readability, we shall use the following short-hand notations:
\begin{description}
	\item[$L(\iota)$] indicates the case defined by $\iota$ over $ L $; in the example of \cref{fig:example}(d), e.g., $L(2) = \langle \Evt_2,\Evt_5,\Evt_8 \rangle$;
	\item[$S(L)$] denotes the set of all cases defined over event log $ L $, i.e., $S(L) = \{ L(\iota) | \iota \in I \}$; it follows that $|I| = |S(L)|$; in our example, $S(L) = \{ \sigma_1, \sigma_2, \sigma_3 \}$ where
	$\sigma_1 = \langle \Evt_1,\Evt_3,\Evt_6 \rangle $, 
	$\sigma_2 = \langle \Evt_2,\Evt_5,\Evt_8 \rangle $, 
	$\sigma_3 = \langle \Evt_4,\Evt_7 \rangle $;
	\item[$\sigma(i)$] refers to the $i$-th event within a case $\sigma$ (e.g., the first event in $\sigma_1$ is denoted as ${\sigma_1(1)}$, whereby $\sigma_1(1) = \Evt_1$ in our example);
	\item[{$\sigma[i,j]$}] indicates the segment of case $\sigma$ from $i$ to $j$, having $1 \leqslant i \leqslant j \leqslant |\sigma|$ (e.g., $\sigma_1[2,3] = \langle \Evt_3,\Evt_6 \rangle $ );
	\item[$\attr{Act}(\sigma)$] denotes with a slight abuse of notation the trace induced by case $\sigma$ trough the mapping of $\attr{Act}$ (e.g., $\attr{Act}(\sigma_1) = \langle \inst{A}, \inst{B}, \inst{C} \rangle$);
	\item[$\Evt \in \sigma$] indicates that there exists an index $i$, $1 \leqslant i \leqslant n$ such that $\sigma(i) = \Evt$ (e.g., $\Evt_5 \in \sigma_2 $);
	\item[$\langle \Evt, \Evt', \Evt'', \ldots, \Evt^{(m)} \rangle \subseteq \sigma$] indicates  that there exists an index $j$ with $1 \leqslant j \leqslant n - m $ such that $\sigma(j) = \Evt, \sigma(j+1) = \Evt', \sigma(j+2)= \Evt'', \ldots, \sigma(j+m) = \Evt^{(m)}$, where $n$ is the length of $\sigma$ and $m \leq n$ (e.g., $\langle \Evt_2, \Evt_5 \rangle \subseteq \sigma_2$);
	\item[$\sigma = \sigma'$] holds if and only if $\sigma \subseteq \sigma'$ and $\sigma' \subseteq \sigma$.
\end{description}

\subsection{Process Modeling and Execution}\label{sec:modeling}
%
In this section we outline the main notions we shall adopt in the remainder of this paper about process modeling and execution.

\begin{figure*}
		\includegraphics[width=0.75\textwidth]{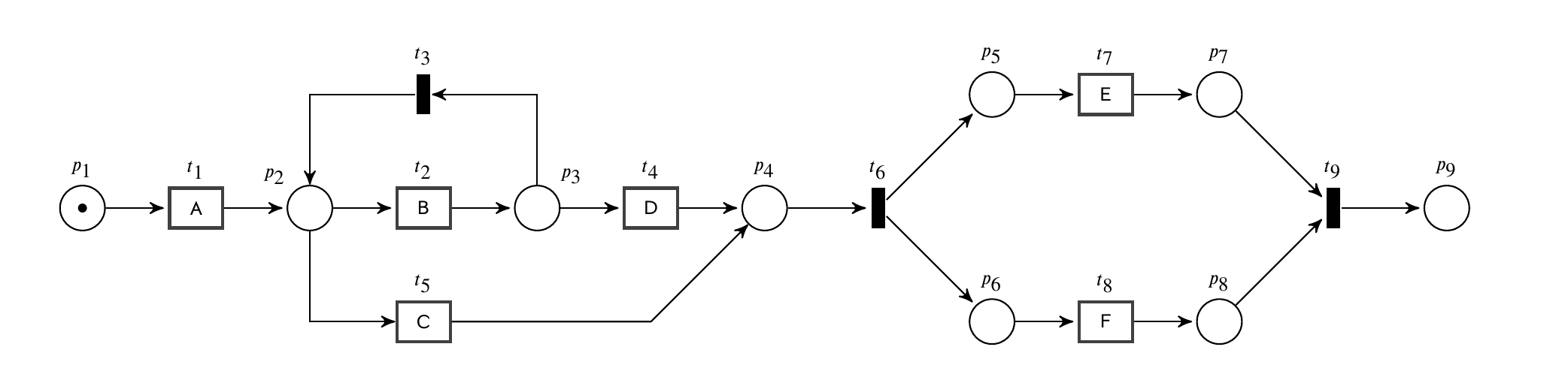}%
	\caption{A workflow net}
	\label{fig:wfnet}
\end{figure*}

Alongside uncorrelated event logs and data constraints, our approach takes as input a behavioral model for processes. We require that the process model has a starting activity and an accepting state. In the context of this paper, we resort to workflow nets~\cite{Aalst/JCSC1998:WorkflowNets} such as the one in \cref{fig:wfnet}. 
Workflow nets are bipartite graphs consisting of
\begin{iiilist}
	\item a finite non-empty set of nodes partitioned into places (graphically depicted as circles -- e.g., $\{p_1, \ldots, p_9\}$ in \cref{fig:wfnet}) and transitions (rectangles -- e.g., $\{t_1, \ldots, t_9\}$), and
	\item a \emph{flow} relation, namely arcs connecting places to transitions and transitions to places (e.g., $p_1 \to t_1$, $t_1 \to p_2$ and $t_5 \to p_2$); the set of places at the tail of flow arcs toward a transition are the \emph{preset} of that transition (e.g., the presets of $t_2$ and $t_9$ are $\{p_2\}$ and $\{p_7, p_8\}$, respectively); places at the head of flow arcs are the \emph{postset} of the transition (e.g., the postsets of $t_2$ and $t_6$ are $\{p_3\}$ and $\{p_5, p_6\}$, respectively).
\end{iiilist}

Transitions can be annotated with an activity name (e.g., $t_1$ is annotated with \inst{A}); otherwise, they are \emph{silent} (in which case, they are graphically depicted as a solid black rectangle as with $t_5$, $t_6$ and $t_9$ in the figure). The set of places contain exactly one \emph{input} and one \emph{output} place, namely a source and a sink node for the graph, respectively -- e.g., $p_1$ and $p_9$. Every node in the workflow net is reachable through a directed walk from the input place.

The execution semantics are specified by the production and consumption of \emph{tokens} (represented as black dots). A transition can be executed (i.e., it is \emph{enabled}) only if a token is in every place of its preset. The execution implies that a token is consumed from every place in its preset and a token is produced in every place in its postset. In the beginning, a token resides in the input place ($p_1$ in the figure) by default -- therefore, it is sometimes omitted from the graphical representation. Transition $t_1$, in the example, is the only one enabled at the beginning. Its execution consumes the token from $p_1$ and produces a token in $p_2$. Then, both $t_2$ and $t_3$ are enabled. Notice that their execution is thus \emph{mutually exclusive} and that $t_3$ enacts a \emph{cyclic} behavior. Executing $t_2$ and then $t_4$ enables $t_6$ which, in turn, can consume a token from $p_4$ and produce two tokens (one in $p_5$ and the other in $p_6$), thereby enabling both $t_7$ and $t_8$. Notice that $t_9$ is enabled only if a token is assigned to $p_7$ and another one to $p_8$. Therefore, $t_7$ and $t_8$ have to be executed, regardless of the order. The execution of $t_9$ consumes a token from $p_7$ and a token from $p_8$ to ultimately produce one in $p_9$, the output place. The sequence of executions of enabled transitions from the initial state (with a token in the input place) to the final one (with a token in the output place) is a \emph{run} of the Workflow net. The physical time it take for a run to complete, from start to end, is called \emph{cycle time}.

As previously discussed, a \emph{trace} is a sequence of activities. The transcription of activities decorating the sequentially executed transitions determines a trace too. Notice that the execution of silent transitions such as $t_5$ in the example does not occur in a trace. Traces that correspond to runs of the Workflow net in \cref{fig:wfnet} are, e.g.,
$\langle \inst{A}, \inst{C}, \inst{E}, \inst{F} \rangle$,
$\langle \inst{A}, \inst{B}, \inst{B}, \inst{D}, \inst{E}, \inst{F} \rangle$ and
$\langle \inst{A}, \inst{B}, \inst{D}, \inst{F}, \inst{E} \rangle$.

Traces can be replayed to check if they correspond to a run of a Workflow net. If an activity in the sequence cannot be bound to the execution of an enabled transition, or requires one or more (non-silent) transitions to be previously executed though they are not recorded in the trace, we say an \emph{asynchronous move} occurs. The computation of those \emph{alignments}~\cite{Adriansyah.etal/ISEM2015:OneAndBestAlignPrecision} is at the basis of a well-known technique for conformance checking~\cite{Carmona.etal/2018:ConformanceChecking}. For example, $\langle \inst{A}, \inst{B}, \inst{B}, \inst{B}, \inst{D}, \inst{F}, \inst{E} \rangle$ conforms with the process model in \cref{fig:wfnet}, thus the alignment consists of sole synchronous moves. Instead, $\langle \inst{A}, \inst{B}, \inst{C}, \inst{E}, \inst{F} \rangle$ does not conform with it as $\inst{B}$ is a move in the log that cannot correspond to a move in the model. Similarly, $\langle \inst{A}, \inst{E}, \inst{F} \rangle$ requires the execution of transition $\inst{C}$ before the occurrence of $\inst{E}$ although it is not recorded in the trace. Intuitively, the fewer asynchronous moves occur, the higher the fitness of model and log is.
The computation of alignments and quality measures for process mining are beyond the scope of this paper. The interested reader can find a detailed examination of these topics in~\cite{Carmona.etal/2018:ConformanceChecking}.

\subsection{Simulated Annealing}\label{sec:sa}
%

We address the correlation problem as an optimization problem and, to solve it, we resort to simulated annealing. 
\emph{Simulated Annealing} (SA) is a metaheuristic algorithm that explores the optimization problem's search space to find the nearest approximate global solution by simulating the cooling process of metals through the annealing process~\cite{SA,EA}. SA applies the stochastic perturbation theory \cite{Stewart2006} to search for an approximate global solution by randomly changing the next individuals in order to skip the iterations' local optimal solution\cite{Henderson2006}.

A population-based SA \cite{PSA} allows for the use of multiple individuals in the same iteration of the annealing process.
SA explores the search space through the following steps. 
It starts by creating the initial \emph{population}. A population ($\mathrm{pop}$) is a non-empty set of individuals ($|\mathrm{pop}|\geq1$).
The population is formed by generating random \emph{individual}s. Then the SA algorithm initializes the current step $\scurr$ as $\scurr=1$ and the current temperature with a given initial temperature, $\tcurr = \tinit$.
The annealing process begins with the generation of a neighbor solution $x'$ for the current individual $x$. Therefore, SA considers as a memory-less algorithm because it disregards the historical individuals, and only focuses on $x$ and $x '$.  
Next, SA computes the \emph{energy cost function} based on $x$ and $x'$, namely $\fc(x,x')$.

The \emph{acceptance probability} of the new neighbor solution $\prob(x')$ is computed using $\fc(x,x')$ and $\tcurr$.
In particular, $\prob(x')$ determines whether the new neighbor, $x'$, can be used as the next individual.
Notice that $\prob(x')$ may select $x'$ even though it performs worse than $x$ 
in order to increase the chances to skip the local optimum and let the algorithm explore the search space further, especially with high temperatures. At each iteration, SA compares the global optimal solution $x_\mathrm{G}$, i.e., the best solution over the iterations $[0,\scurr[$, with the local optimal solution $x_\mathrm{L}$ in $\mathrm{pop}$ at $\scurr$ based on $\fc(x_\mathrm{G},x_\mathrm{L})$. Therefore, SA can return the best solution over all the iterations. 
Finally, SA uses a cooling schedule \cite{Nourani1998} that defines the rate at which the temperature ($\tcurr$) cools down, and increments $\scurr$ by \num{1}.
SA repeats the annealing and cooling process until $\scurr$ reaches the maximum number of 
iterations ($\smax$).

To sum up, SA has a set of parameters that influence the annealing process:
\begin{iiilist}
	\item the initial temperature ($\tinit$),
	\item the maximum number of steps ($\smax$), and
	\item the population size ($|\mathrm{pop}|$).
\end{iiilist}
In addition to these parameters, SA requires the following main functions to be defined:
\begin{iiilist}
	\item the cooling schedule,
	\item the creation of a new neighbor ($x'$),
	\item the energy cost function ($\fc(x,x')$), and
	\item the acceptance probability ($\prob(x')$).
\end{iiilist}

\section{The \ECSAnew Solution}\label{CSA}
Equipped with the definitions and main notions defined in the above section, we define the input (I1, I2, I3), preconditions (P1, P2), output (O1) and effects (E1) of \ECSAnew.
\begin{description}
	\item[I1.] An uncorrelated log $\abbrev{UL}$, as per \cref{def:uncorrelatedLog} (depicted, e.g., in \cref{fig:example}(a)).
	\item[I2.] A process model (e.g., the Workflow net depicted in \cref{fig:example}(b)).
	\item[P1.] The process model is required to have exactly one start activity (e.g., \inst{A} in \cref{fig:example}(b)), which is enabled only at the beginning of the run, and thus cannot be part of any cycle.
	\item[P2.] The process model is required to have a final state 
	(such a state, e.g., is reached after \inst{C} or \inst{D} are executed with the model in \cref{fig:example}(b)).
	\item[I3.] A set of domain knowledge rules, i.e., data constraints on process data $\mathfrak{C}= \{C_1,\dots,C_m\}$. In the following, we will interchangeably use the terms ``rule'' and ``data constraint''. Constraints are propositions~\cite{Pace/2012:MathematicsofDiscreteStructuresforCS} exerted on resources, time or additional event attributes.
	Syntax and semantics of the constraint expression language will be described in \cref{sec:dataconstraints}.
	\Cref{fig:example}(c), e.g., illustrates four such constraints. 
	\item[O1.] \ECSAnew generates an event log $L$ as per \cref{def:eventLog}.
	\item[E1.] The event log partitions $\abbrev{UL}$ into a set of cases, i.e., for every event $e \in \abbrev{UL}$ there exists one and only one case $\sigma \in L$ s.t.\ $e \in \sigma$ (see, e.g., \cref{fig:example}(d)).
\end{description}
%

%

\medskip

The key idea of the \ECSAnew technique is to treat the event correlation problem as a multi-level optimization problem. \ECSAnew 
has three nested objectives: 
\begin{iiilist}
	\item minimizing the misalignment between the generated 
	event log and the input process model,  
	\item minimizing the constraints violations over the cases, and
	\item minimizing the activity execution time variance across the cases.
\end{iiilist}

\begin{figure*}[tbp]
	\centering
	\includegraphics[width=0.75\textwidth]{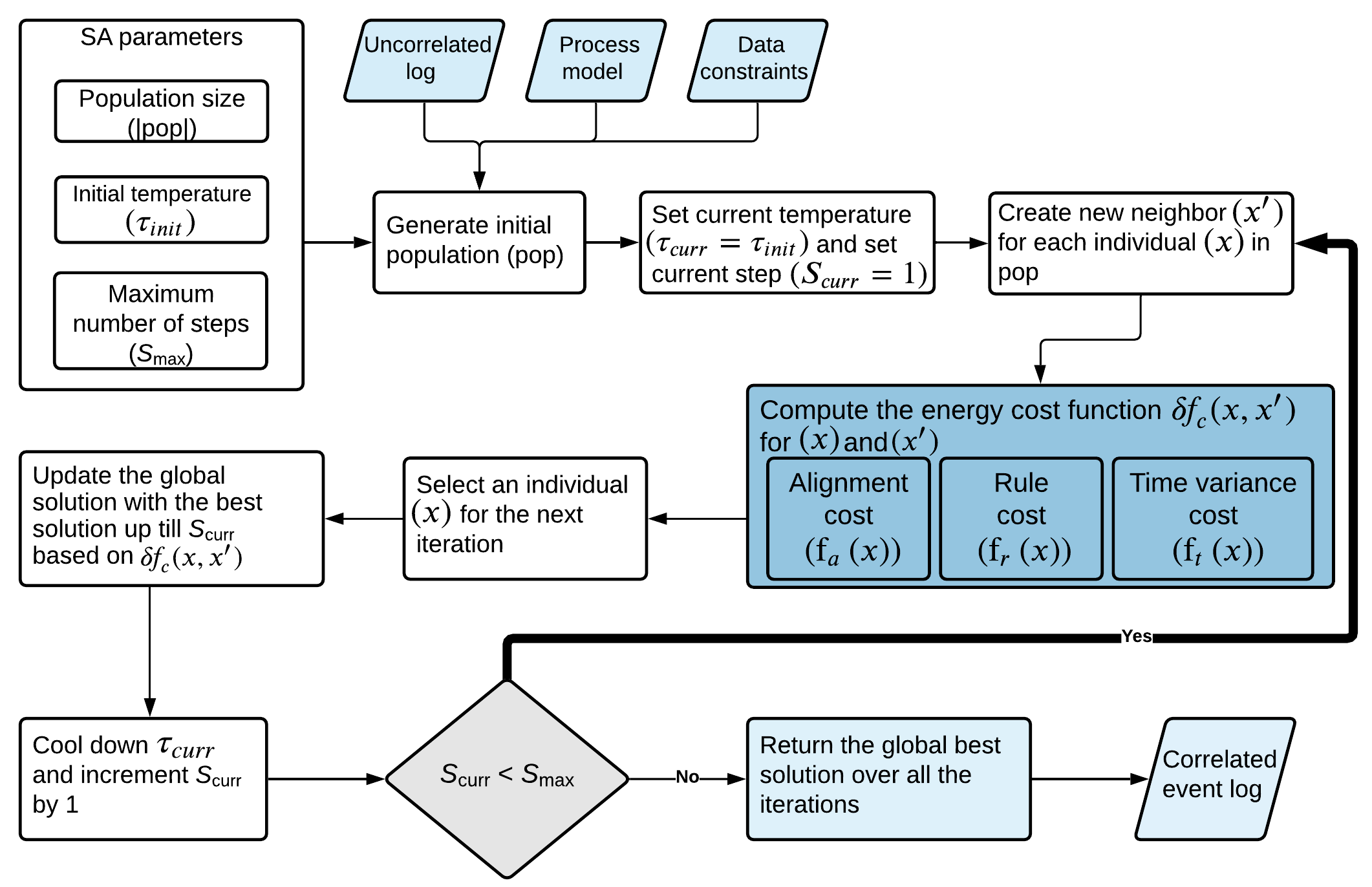}
	\caption{The \ECSAnew technique overview}
	\label{fig:alg-steps}
\end{figure*}

\ECSAnew uses multi-level simulated annealing (SA) as an optimization technique~\cite{PSA}. Using SA to solve the event correlation problem helps to find an \textit{approximate} global optimal correlated log in a reasonable time. To apply SA to solve the event correlation problem, we define the functions required by the SA:
\begin{enumerate}[label=(\roman*)]
	\item The cooling schedule $\tcurr$;
	\item The creation of a new neighbor ($x'$);
	\item The energy cost function ($\fc(x,x')$);
	\item The acceptance probability ($\prob(x')$).
\end{enumerate}
\Cref{fig:alg-steps} shows the steps of \ECSAnew. We describe them in detail in the following subsections.


\subsection{Initial population}
\begin{figure}[tb]
	\centering
	\includegraphics[width=\textwidth]{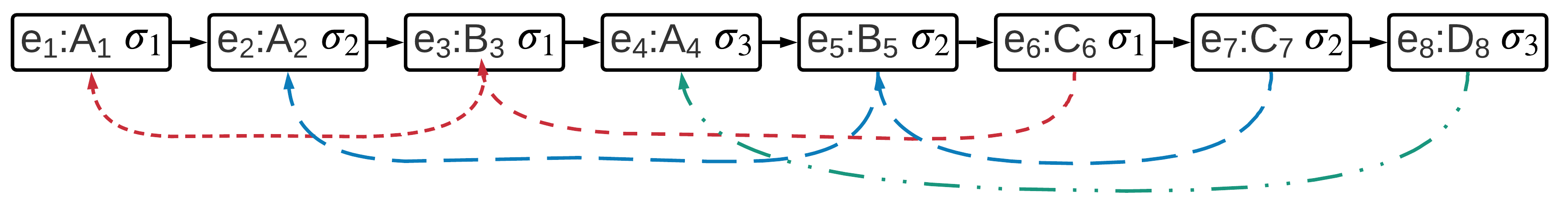}
	\caption{Log graph $LG_1$ of individual $x$}
	\label{fig:lg1:h}
\end{figure}

As shown in \cref{fig:alg-steps}, the first step is the generation of the initial population, $\mathrm{pop}$, of size $|\mathrm{pop}| \geq 1$.
An individual $x \in \mathrm{pop}$ is a candidate event log as defined as per \cref{def:eventLog}.
For the sake of readability, we graphically depict the individuals as graphs such as that of \cref{fig:lg1:h}.
In the following, we shall name these graphs as \emph{log graphs}, or LG's for short.

\begin{figure*}[tbp]
	\centering
	\includegraphics[width=.75\textwidth]{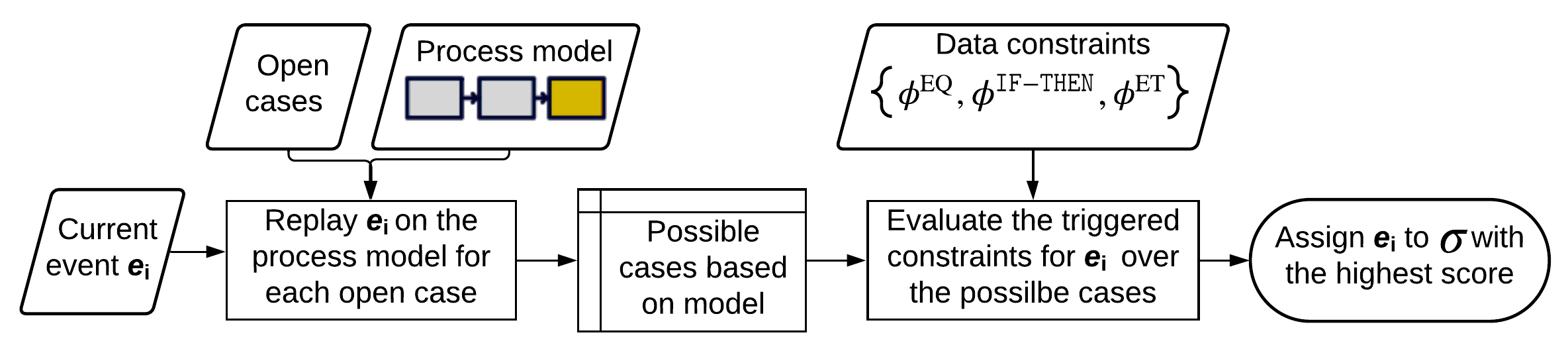}
	\caption{Event correlation decision process for event $\Evt_i$}
	\label{fig:alg-steps:corrDecision}
\end{figure*}

As shown in \cref{fig:lg1:h}, 
every node in the log graph represents an event within the uncorrelated log, its assigned case and the related activity.  For example, the third node represents $\Evt_3$, assigned with case $\sigma_1$ and annotated with $\inst{B}_3$ as its activity $\Evt_3$.\attr{Act} is \inst{B}. We keep the event index as a subscript to distinguish the position in which activities recur in the log (e.g., $\inst{B}_3$ and $\inst{B}_5$ denote the occurrence of activity $\inst{B}$ with events $\Evt_3$ and $\Evt_5$). 
The log graph depicts two different flow connections.
The first flow is represented via forward solid arcs connecting every event to its direct successor according to the temporal order in the uncorrelated log.
The second flow connection (depicted with backward dashed arcs) 
connects every event to its direct predecessor in the same case.
For example, backward dashed arcs connect $\Evt_7$ to $\Evt_5$ and $\Evt_5$ to $\Evt_2$ in the figure as these events belong to case $\sigma_2 = \langle \Evt_2, \Evt_5, \Evt_7 \rangle$. 

The data structure underlying the log graph is generated by replaying the uncorrelated event log on the process model and verifying the data constraints over the possible case assignments. This step is repeated based on the population size. In our example, we assume $|\mathrm{pop}|=1$ for readability purposes.

\Cref{fig:alg-steps:corrDecision} shows the steps taken to correlate an event $e_i$.
The first step filters out the possible candidate cases for $\Evt_i$ based on the process model replay.
Then, we rank the candidate cases based on the number of data constraints thereby satisfied by $\Evt_i$ in those cases. These steps are repeated for every event in \abbrev{UL}.

\subsection{Process model based correlation}
Every run of the process model from the initial activity to the termination conditions corresponds to a case.
We name the cases corresponding to non-terminated runs as \emph{open cases}.
We figure the following three scenarios when replaying an event $\Evt$ over the input process model.
\begin{compactenum}
	\item Event $\Evt$ corresponds to the execution of the start activity of the process model (we name it \emph{start event}). Then, a new run starts and a new case is open, accordingly. 
	\item Event $\Evt$ corresponds to the execution of an enabled (non-starting) activity for one or more cases (\emph{enabled event}). If only one run enables $\Evt$, it is assigned to the case of that run. Otherwise, $\Evt$ is assigned to the case satisfying the highest number of data constraints. 
	\item Event $\Evt$ does not correspond to any enabled activity (\emph{non-enabled event}). Then, $\Evt$ is assigned to the case satisfying the highest number of data constraints. 
	This way, we guarantee that all events are correlated, even if the log deviates from the model.
\end{compactenum} 

\begin{figure}[tbp]
	\centering
	\begin{subfigure}{.4\textwidth}
		
		\includegraphics[width=\textwidth]{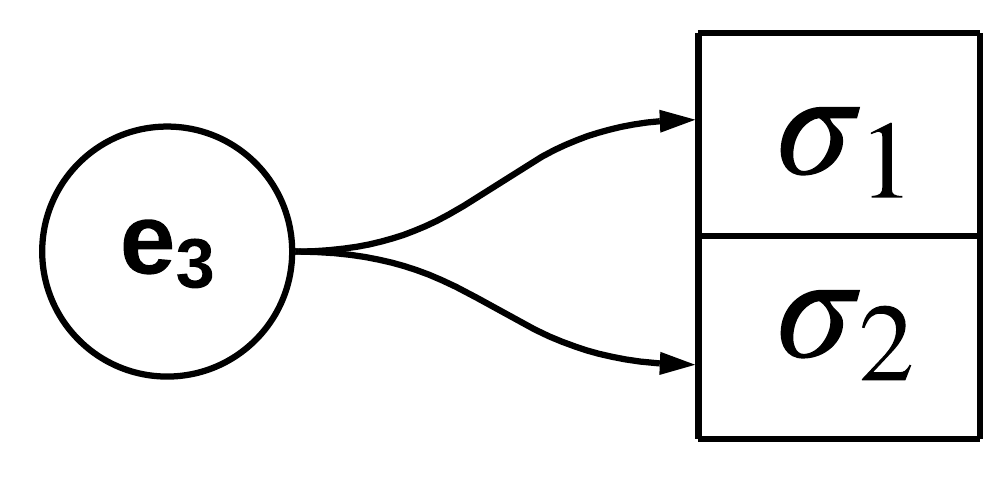}
		
		\caption{Possible cases for $\Evt_3$ }
		\label{fig:b3PM}
		
	\end{subfigure}
	\qquad
	\begin{subfigure}{.48\textwidth}
		
		\includegraphics[width=0.75\textwidth]{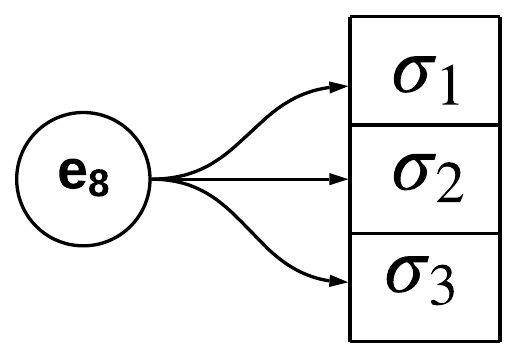}
		\caption{Possible cases for $\Evt_8$}
		\label{fig:d8PM}
		
	\end{subfigure}
	\caption{Process model based assignment of cases for events $\Evt_3$ and $\Evt_8$}
	\label{fig:correlation:PM}
\end{figure}

\noindent
For example, considering \abbrev{UL} in \cref{fig:example}(a) and the process model in \cref{fig:example}(b), $\Evt_1$ is a start event as it executes the start activity (\inst{A}). Thus, it opens a new case, $\sigma_1$. The same goes for $\Evt_2$ and $\Evt_4$, which start $\sigma_2$ and $\sigma_3$, respectively.
There are two open cases within the log before $\Evt_3$: both $\sigma_1$ and $\sigma_2$ expect the execution of activity {\inst{B}} or activity {\inst{C}}.
Therefore, $\Evt_3$ is an enabled event
and both cases are considered as possible candidate assignments for the data-constraint based correlation step, as shown in \cref{fig:b3PM}.
On the other hand, based on the assigned cases in \cref{fig:lg1:h}, none of the cases $\sigma_1$, $\sigma_2$ and $\sigma_3$ expect the execution of activity {\inst{D}}. Thus, $\Evt_8$ is a non-enabled event and all the three cases are considered for data constraints based correlation step in order to reduce the randomization of the correlation decision, as shown in \cref{fig:d8PM}.
Following \cref{fig:alg-steps:corrDecision}, we identify the possible case-assignments of an event based on the model that will be used in the next step.

\subsection{Data constraint based correlation}
\label{sec:dataconstraints}
Knowledge about data attributes provides an additional source of information for correlation. Intuitively, we use data constraints to filter and rank the cases to which an event can be assigned. 
In the context of this paper, we focus on the following types of data constraints.
To define them, we take inspiration from the work of Nehzad et al.~\cite{ProcessView,cMiner} and extend it with {\ifthencon} rules. Though not intended to constitute an exhaustive set, we focus on these templates as we experimentally found them to be a good trade-off between expressiveness and tractability. Further studies on the suitability and effectiveness of the data constraints in use are beyond the scope of this work and subject to future research. 

\begin{definition}[Equality constraint] \label {def:constraints:eq}
	Let $\sigma(i)$ and $\sigma(i-1)$ be two consecutive events in a case $\sigma$ as per \cref{def:case}. 
	A data-attribute equality constraint (henceforth \emph{equality constraint} for short) $\phi^{\textrm{EQ}}$ is a predicate over variables $\sigma(i)$ and $\sigma(i-1)$ formulated as follows: 
	\begin{flalign*}
	\phi^{\textrm{EQ}} & \coloneqq \sigma(i).\attr{Attr}\: = \:\sigma(i-1).\attr{Attr}
	\end{flalign*}
\end{definition}
\noindent
For example, ${C_1}$ in \cref{fig:example} (i.e., $\sigma(i).\attr{Type}~=~ \sigma(i-1).\attr{Type}$) enforces the equality between the $\attr{Type}$ attribute values. As depicted in \cref{fig:b3PM}, $e_3$ reports the execution of activity $\inst{B}$. One of the possible cases, based on the process model, is $\sigma_1$. The constraint is evaluated over $\sigma(i) = e_3$ and $\sigma(i-1)= e_1$.

\begin{definition} [{\ifthencon} constraint]\label{def:constraints:ifthen}
	Let $\sigma(i)$ be an event in a case $\sigma$ as per \cref{def:case} and $j$ an index $1 \leq j < i$; let $E$ be a universe of events and $\textrm{Attr}$ an attribute $\textrm{Attr} : E \nrightarrow \textrm{Dom} $ as per \cref{def:attribute}.
	An {\ifthencon} constraint $\phi^{\ifthencon}$ is a predicate over pairs of events consisting of two propositional clauses: the \emph{antecedent} (the {\ifcon} clause) and \emph{consequent} (the {\thencon} close).
	We define the syntax of well-formed {\ifthencon} constraints as follows.
	\\
	\begin{flalign*}
	%
	%
	\phi^{\ifthencon} & \coloneqq \mathtt{\ifcon} \: \varphi_{\ifcon}^{i} \: {\thencon}\: \varphi_{\thencon}^{i,j} \wedge j = i-1 \\
	~& \: \:\:\:\:\:\:|\: {\ifcon} \: \varphi_{\ifcon}^{i,j} \: {\thencon} \: \varphi_{\thencon}^{i,j}  \wedge j < i \\
	\varphi_{\ifcon}^{i} & \coloneqq \sigma(i)\mathtt{.}\textrm{Attr} \lessgtr a \: | \:
	\varphi_{\ifcon}^{i} \: \wedge \: \sigma(i)\mathtt{.}\textrm{Attr} \lessgtr a 
	\\
	\varphi_{\ifcon}^{i,j} & \coloneqq  \varphi_{\ifcon}^{i} \: \wedge \: \sigma(j)\mathtt{.}\textrm{Attr} \lessgtr a \: \\
	~& \: \:\:\:\:\:\:| \: \varphi_{\ifcon}^{i,j} \: \wedge \: \sigma(j)\mathtt{.}\textrm{Attr} \lessgtr a \\
	\varphi_{\thencon}^{i,j} & \coloneqq \wedgephi_{\thencon}^{i,j} \:|\: \varphi_{\thencon}^{i,j} \veewedge \wedgephi_{\thencon}^{i,j} \\
	\wedgephi_{\thencon}^{i,j} & \coloneqq \sigma(j)\mathtt{.}\textrm{Attr} \lessgtr a \:|\:
	\sigma(i)\mathtt{.}\textrm{Attr} \lessgtr \sigma(j)\mathtt{.}\textrm{Attr}^\prime \\
	\lessgtr & \coloneqq <\:|\: >\:|\:\geq\:|\:\le\:|\:=\:|\:\neq\\
	\veewedge & \coloneqq \vee\:|\: \wedge
	\end{flalign*}	
\end{definition}

\noindent
In the following, we may collectively refer to $\varphi_{\ifcon}^{i}$ and $\varphi_{\ifcon}^{i,j}$ as $\varphi_{\ifcon}$, and to
$\varphi_{\thencon}^{i}$ and $\varphi_{\thencon}^{i,j}$ as $\varphi_{\thencon}$ for the sake of readability.

\medskip

$\phi^{\ifthencon}$ is an association rule in which $\varphi_{\ifcon}$ and $\varphi_{\thencon}$ act as selection and verification criteria, respectively.
In other words, $\varphi_{\ifcon}$ seeks a pair of events, i.e., an event $\sigma(i)$ and a preceding event $\sigma(j)$, that satisfy the {\ifcon} clause. \ECSAnew gives priority to the closest $j$ to $i$ in the selection process. 
Notice that if $\varphi_{\ifcon}$ is in the $\varphi_{\ifcon}^{i}$ form, we impose $j=i-1$ by default (see the grammar rule above).
Subsequently, $\varphi_{\thencon}$ is evaluated on the selected events to check whether $\sigma(i)$ and $\sigma(j)$ satisfy $\phi^{\ifthencon}$.
For instance, the evaluation of $C_2$ in \cref{fig:example}
(i.e., {\ifcon} $\sigma(i).{\inst{Act}} = \inst{B}$ and $\sigma(j).{\inst{Act}} = {\inst{A}}$ {\thencon} $\sigma(i).{\attr{Res}} = \sigma(j).\attr{Res}$) selects two events based on their activities ($\sigma(i).{\inst{Act}} = \inst{B}$ and $\sigma(j).{\inst{Act}} = {\inst{A}}$) and then checks whether the resources are equal for those events ($\sigma(i).{\attr{Res}} = \sigma(j).\attr{Res}$). In the example above, let us take $e_3$. We notice that $e_3.\attr{Act}$ is $\inst{B}$ and one of the assignable cases (based on the process model) is $\sigma_1$ as illustrated in \cref{fig:b3PM}. Considering the log graph illustrated in \cref{fig:lg1:h}, we pick $\sigma(2) = e_3$ and select $\sigma(1) = e_1$ as $e_1.\attr{Act} = \inst{A}$. Then, the $\thencon$ clause is evaluated over the two events: $(e_3.\attr{Res} = e_1.\attr{Res}) \equiv \mathrm{True}$.
We conclude that  $C_2$ is satisfied by $e_3$ and $e_1$ in case $\sigma_1$.

%
\begin{definition}[Event-time constraint] \label{def:constraints:ET}
	Let $\sigma(i)$ and $\sigma(i-1)$ be two consecutive events in a case $\sigma$ as per \cref{def:case}. An event-time constraint $\phi^{\textrm{ET}} $ is a predicate over $\sigma(i)$ and $\sigma(i-1)$ consisting of two propositional clauses: an antecedent ({\ifcon} clause) that selects $\sigma(i)$ and a consequent ({\thencon} clause) that bounds the event execution duration $( \sigma(i).\attr{Ts} - \sigma(i-1).\attr{Ts} )$ to a given minimum duration (\textrm{dur}) and a given maximum duration (\textrm{dur}') as follows.  %
	%
	\begin{flalign*}
	\phi^{\textrm{ET}} & \coloneqq \ifcon \: \varphi_{\ifcon}^{i} \: \thencon \: \varphi_{\thencon}^{\textrm{ET}} \\
	\varphi_{\thencon}^{\textrm{ET}} & \coloneqq \textrm{dur} \leq \left( \sigma(i).\attr{Ts} - \sigma(i-1).\attr{Ts} \right) \leq \textrm{dur}'  
	\end{flalign*}

\end{definition}

\noindent $\phi^{\textrm{ET}}$ is based upon the grammar of {\ifthencon} constraints to express rules on the activity execution duration. In particular, the {\ifcon} clause selects $\sigma(i)$ based on propositional formulas, while $\sigma(j)$ is the directly preceding event $\sigma(i-1)$, taken in order to compute the event duration $( \sigma(i).\attr{Ts} - \sigma(i-1).\attr{Ts} )$.

For example, $C_4$ in \cref{fig:example} (i.e., ${\ifcon~} \sigma(i).\attr{Act} = \inst{B} {~\thencon~}$ $30 \leq (\sigma(i).\attr{Ts} - \sigma(i-1).\attr{Ts}) \leq 120 $) requires that the duration of activity \inst{B} is between \SI{30}{\minute} and \SI{120}{\minute}.
$e_3$ reports the execution of activity $\inst{B}$. One of the assignable cases as per the process model is $\sigma_1$, as illustrated in \cref{fig:b3PM}. Considering the log graph illustrated in \cref{fig:lg1:h}, we select $\sigma(i) = e_3$ and compute the duration based on $\sigma(i-1)= e_1$, which amounts to \SI{60}{\min}. Thus, the $\thencon$ clause holds true as $30 \leq 60 \leq 120$.

\bigskip

We collectively refer to equality, {\ifthencon} and event-time constraints as data constraints.
\begin{definition}[Data constraint] \label{def:constraints} 
	Let $\EvtLog$ be an event log  defined over events in $E \ni e$ and let $\sigma \in S(L)$ be a case as per \cref{def:eventLog,def:case}.
	A data constraint $C$ is a predicate $\phi$ over events that belongs to either of the following three types:
	data-attribute equality constraint (as per \cref{def:constraints:eq}, hitherto indicated with the expression ``$C \textrm{ is } \phi^{\textrm{EQ}}$''),
	{\ifthencon} constraint (as per \cref{def:constraints:ifthen}, ``$C \textrm{ is }  \phi^{\ifthencon}$''), or
	event-time constraint (\cref{def:constraints:ET}, ``$C \textrm{ is }  \phi^{\textrm{ET}}$'').
	The set of all possible constraints over $E$ is the \emph{universe of data constraints} $\mathfrak{C} \ni C$.
\end{definition}

\ECSAnew employs data constraints 
to rank possible case assignments based on the number of satisfied ones. 
To pursue our objective, we first need to restrict the evaluation of constraints to a current event  under analysis. Therefore, we introduce the notion of $i$-preassignment: rather than seeking pairs of events at position $i$ and $j < i$ in a case $\sigma$ that satisfy a constraint, we fix $i$ to an event $e \in \sigma$ so as to be able to check whether any $j$ exists such that the constraint is satisfied by $\sigma(i)$ and $\sigma(j)$ in $\sigma$.

\begin{definition}[$i$-Preassignment]\label{def:ipreasgn}
	Let $L$ be an event log over the universe of events $E \ni e$ and $S(L) \ni \sigma$ be its case set as per \cref{def:event,def:eventLog,def:case}.
	Let $C \in \mathfrak{C}$ be a data constraint expressed by formula $\phi$ as per \cref{def:constraints}.
	Denoting with $\overrightarrow{\imath}(\sigma, e)$ the index of $e$ in case $\sigma$, the \emph{$i$-preassignment} of $\phi$ with $e$ is the assignment of $\sigma(i)$ with $e$ in its formula, i.e., $\phi[i/\,\overrightarrow{\imath}(\sigma, e)]$.
\end{definition}
\noindent The $i$-Preassignment predetermines the $i$-th event to be considered for the evaluation of the constraint. In the case of {\ifthencon} and event-time rules, this entails that only $j$, with $1 \leqslant j < i \leqslant |\sigma|$, is sought for in order to have $C$ satisfied by $\sigma(i)$ and $\sigma(j)$.
For example, consider case $\sigma_1$ as illustrated in \cref{fig:example}(d) and constraint $C_2 = {\ifcon}~\sigma(i).{\attr{Act}} = {\inst{B}} \wedge \sigma(j).{\attr{Act}} = {\inst{A}}~{\thencon}~\sigma(i).\attr{Res} = \sigma(j).{\attr{Res}}$ (\cref{fig:example}(c)).
With a slight abuse of notation, we extend the notion of $i$-preassignment to clauses within the formulation of a constraint (e.g., $\varphi_{\ifcon}[i/\,\overrightarrow{\imath}(\sigma, e)]$).
For example, although the {\ifcon} clause of $C_2$ would be satisfied in $\sigma_1$ by setting $i=2$ and $j=1$ (i.e., considering $e_3$ and $e_1$), but it would not be satisfied if $i$-preassigned with $e_6$ as $e_6.\attr{Act} = \inst{C}$.

Equipped with this notion, we define how to check if an $i$-preassigned constraint is satisfied in a case. Notice we consider {\ifthencon} and event-time constraints as satisfied if and only if both the {\ifcon} and {\thencon} clauses are satisfied, unlike the common interpretation of an ``if-then'' implication may suggest. This design choice is motivated by the goal to avoid that the \emph{ex falso quod libet} statement applies (i.e., that in case the antecedent evaluates to false, the constraint holds true regardless of the consequent), as it reportedly leads to an overestimation of the support of a given rule, as explained in detail in the context of declarative process mining~\cite{DBLP:conf/caise/MaggiBA12,DBLP:journals/is/CiccioMMM18,DBLP:conf/bpm/CecconiCGM18}.
Considering the example above, $C_2$ $i$-preassigned with $e_6$ is not satisfied because its {\ifcon} clause is not satisfied either.

The score function, formalized in the following, counts the number of satisfied constraints that are $i$-preassigned with an event.
\begin{definition}[Score function]
	Let $E$ be the universe of events as per \cref{def:event}, $S(L)$ represent the cases of log $L$ as per \cref{def:case}  and $ \mathfrak{C}$ be the universe of constraints as per \cref{def:constraints}.
	Considering the $i$-preassignment with $e$ $\phi[i/\,\overrightarrow{\imath}(\sigma, e)]$ as per \cref{def:ipreasgn},
	%
	let $\mathrm{eSat} : \mathfrak{C} \times E \times S(L) \to \{0, 1\}$ be a function that indicates whether an $i$-preassigned constraint is satisfied given $e \in E$ in case $\sigma \in S(L)$ as follows:
	%
	%
	%
	\begin{equation}\label{eq:e:eval}
	\resizebox{.99\columnwidth}{!}{$%
		\mathrm{eSat}(C,e,\sigma) = \begin{cases}
		1 & \textrm{ if } C \textrm{ is } \phi^{\textrm{EQ}} \: \textrm{and} \:  \phi^{\textrm{EQ}}[i/\,\overrightarrow{\imath}(\sigma, e)] \equiv  \mathrm{true} \\
		1 & \textrm{ if } C \textrm{ is } \phi^{\ifthencon} \: \textrm{and} \: \varphi_{\ifcon}[i/\,\overrightarrow{\imath}(\sigma, e)] \equiv  \varphi_{\thencon}[i/\,\overrightarrow{\imath}(\sigma, e)] \equiv \mathrm{true} \\
		1 & \textrm{ if } C \textrm{ is } \phi^{\textrm{ET}} \: \textrm{and} \: \varphi_{\ifcon}[i/\,\overrightarrow{\imath}(\sigma, e)] \equiv  \varphi_{\thencon}^{\textrm{ET}}[i/\,\overrightarrow{\imath}(\sigma, e)]  \equiv \mathrm{true} \\
		0 & \textrm{ otherwise}
		\end{cases}     $}
	\end{equation}
	Let $\mathscr{C} \subseteq \mathfrak{C}$ be a set of constraints.
	%
	The \emph{score function} $\mathrm{score} : 2^\mathfrak{C} \times E \times S(L) \to \mathbb{N}$ counts the satisfied constraints within a set $\mathscr{C} \in \mathfrak{C}$ given event $e \in E$ in case $\sigma \in S(L)$ as follows:
	\begin{equation}\label{eq:score}
	\mathrm{score}(\mathscr{C},e,\sigma) =\sum_{C \in \mathscr{C}}  {\mathrm{eSat}(C,e,\sigma)}  
	\end{equation}
\end{definition}

We use the data constraints to guide the selection of the most proper case for an event $e$. We identify the best candidate as the first one ranked by the score. 
If multiple cases share the highest score, we randomly select one of them. Randomness is legitimate in this context as it helps to escape the local optimal solution over subsequent SA-iterations.

\begin{figure}
	\centering
	\begin{subfigure}{.85\textwidth}
		
		\includegraphics[width=\textwidth]{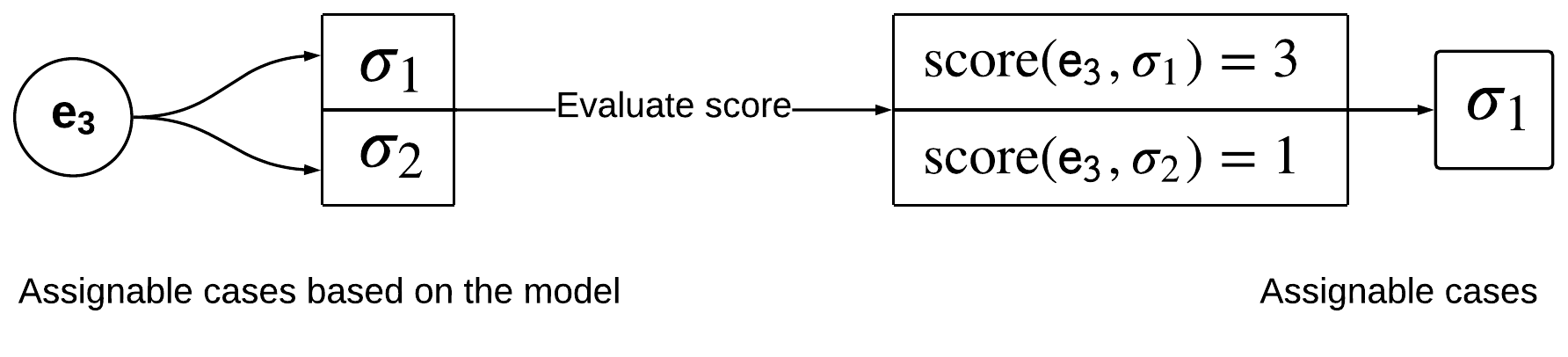}
		\caption{Correlation decision for $e_3$}
		\label{fig:b3}
		
	\end{subfigure}
	\quad
	\begin{subfigure}{.85\textwidth}
		
		\includegraphics[width=\textwidth]{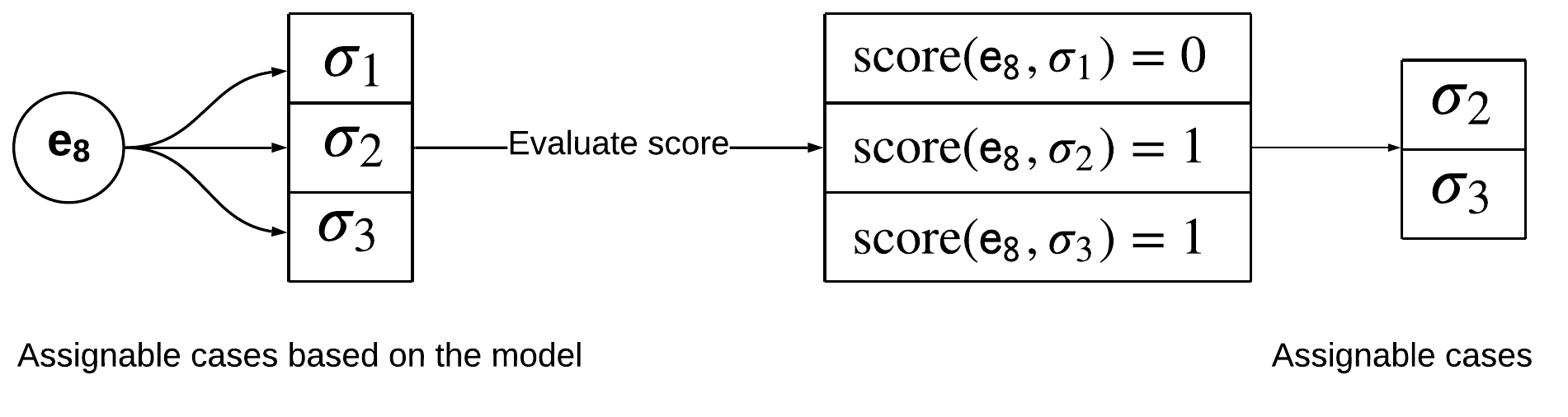}
		\caption{Correlation decision for $e_8$}
		\label{fig:d8}
		
	\end{subfigure}
	\caption{Event correlation decision for events $\Evt_3$ and $\Evt_8$}
	\label{fig:correlation}
	
\end{figure}

For example, considering the process model in \cref{fig:example}(b) and the data constraints in \cref{fig:example}(c), {\ECSAnew} generates the individual represented as log graph $\abbrev{LG}_1$ in \cref{fig:loggraph1} from the uncorrelated log in \cref{fig:example}(a).
Scanning the events, we initially correlate $\Evt_1$ with $\sigma_1$, and upon $\Evt_2$, $\sigma_2$ starts.
At this stage, there are two open cases before $\Evt_3$: both $\sigma_1$ and $\sigma_2$ expect the occurrence of activity {\inst{B}} or activity {\inst{C}} because both are enabled as per the process model.
Therefore, $\Evt_3$ can belong to each of those cases, as shown in \cref{fig:b3}.
Considering the data constraints in \cref{fig:example}(c), let us focus on $C_1$, $C_2$ and $C_4$ in particular. They pertain to $\Evt_3$ as $C_1$ verifies the equality of the \attr{Type} attribute of an event with that of the preceding one (for all events), while $C_2$ and $C_4$ check that $ \sigma(i).\attr{Act} = \inst{B} $ in their {\ifcon} clause (and $e_3.\attr{Act} = \inst{B}$). 
As for $C_1$ (i.e., $\sigma(i).\attr{Type} = \sigma(i-1).\attr{Type}$) and its $i$-preassignment with $e_3$, we observe that it is satisfied in
$\sigma_1$ as $\Evt_1.\attr{Type}$ is \inst{Home} like $\Evt_3.\attr{Type}$.
Instead, the $i$-preassiged constraint is violated in $\sigma_2$ as $\Evt_2.\attr{Type}$ is \inst{Car}, unlike $\Evt_3.\attr{Type}$.
Constraint $C_2$
(i.e., ${\ifcon}~\sigma(i).{\attr{Act}} = {\inst{B}} \wedge \sigma(j).{\attr{Act}} = {\inst{A}}~{\thencon}~\sigma(i).\attr{Res} = \sigma(j).{\attr{Res}}$), once $i$-preassigned with $e_3$, is satisfied in
$\sigma_1$ as the {\ifcon} and {\thencon} clauses
evaluate to $\mathrm{true}$ and $\Evt_1.{\attr{Res}} = \Evt_3.{\attr{Res}} = {\inst{Kate}}$.
Again, it is violated in $\sigma_2$ because $\Evt_1.{\attr{Res}} \neq \Evt_3.{\attr{Res}}$.
As far as $C_4$ (${\ifcon~} \sigma(i).\attr{Act} = {\inst{B}} {~\thencon~} 30 \leq (\sigma(i).\attr{Ts} - \sigma(i-1).\attr{Ts}) \leq 120 $) and its $i$-preassignment with $e_3$ are concerned, the execution duration of activity \inst{B} in $\sigma_1$ is $\Evt_3.\attr{Ts} - \Evt_1.\attr{Ts} = \SI{60}{\minute}$. 
Also, in $\sigma_2$ we have 
$\Evt_3.\attr{Ts} - \Evt_2.\attr{Ts} = \SI{30}{\minute}$.
Therefore, the $i$-preassigned $C_4$ is satisfied both in $\sigma_1$ and $\sigma_2$.

Then, we compute the score as in \cref{eq:score} 
to rank the possible cases. 
As shown in \cref{fig:b3}, $\sigma_1$ gets the highest score as it supports the three constraints unlike $\sigma_2$.
Therefore, $e_3$ is assigned with $\sigma_1$ as graphically depicted in \cref{fig:lg1:h}.
Following this procedure, we observe that also $\Evt_6$ is assigned with the same case later on.

To guarantee that even all events are associated to a case, we consider all cases as assignable also to non-enabled events. 
Again, we rank the cases based on their score to decide the assignment.
For instance, $\Evt_8$ is a non-enabled event, because neither of the three running cases $\sigma_1$, $\sigma_2$ and $\sigma_3$ expects the execution of activity {\inst{D}}. Therefore, $\Evt_8$ is assigned based on the data constraints.
Considering the constraints in \cref{fig:example}(c), we focus on the $i$-preassignment of $C_1$ and $C_5$ with $\Evt_8$ 
and evaluate them over $\sigma_1$, $\sigma_2$ and $\sigma_3$. As shown in \cref{fig:d8}, $\sigma_1$ gets the lowest score, as both $C_1[i/\,\overrightarrow{\imath}(\sigma_1, e_8)]$ and $C_5[i/\,\overrightarrow{\imath}(\sigma_1, e_8)]$ are violated. 
In $\sigma_2$ and $\sigma_3$, instead, the $i$-preassigned $C_1$ is satisfied. Still, both $C_5[i/\,\overrightarrow{\imath}(\sigma_2, e_8)]$ and $C_5[i/\,\overrightarrow{\imath}(\sigma_3, e_8)]$ are violated as in both cases activity {\inst{D}} exceeds the allowed maximum duration. Therefore, $\Evt_8$ is randomly assigned with $\sigma_3$ as one of the cases with the highest score, as illustrated in \cref{fig:lg1:h}.


\subsection{Neighbor solution}

As illustrated in \cref{fig:alg-steps}, simulated annealing explores the search space by creating a new neighbor individual ($x'$) based on the current individual ($x$) at the beginning of each iteration.
We generate a neighbor ($x'$) by altering the current individual ($x$).
In particular, we modify the assignments of the events in the totally ordered set from a given point (henceforth named as \emph{changing point}) on.
In order to form a different solution, we reassign the events from the changing point till the end of the events in \abbrev{UL} by following the event correlation decision steps in \cref{fig:alg-steps:corrDecision}. 

The current step ({\scurr}) determines the extent wo which the new neighbor should differ. Thus, a changing point is selected based on the current step {\scurr}. For instance, when $\scurr = 1 $, the changing point is randomly selected among the first few events, to widely explore the space. Instead, when $\scurr = \smax-1$, the changing point is randomly picked among the last few events as we seek to converge toward a solution. The increment of {\scurr} reduces the number of events to be re-evaluated at each iteration; this is in line with the cooling down mechanism of the annealing process.

\begin{figure}[ptb!]
	\centering
	\includegraphics[width=\textwidth]{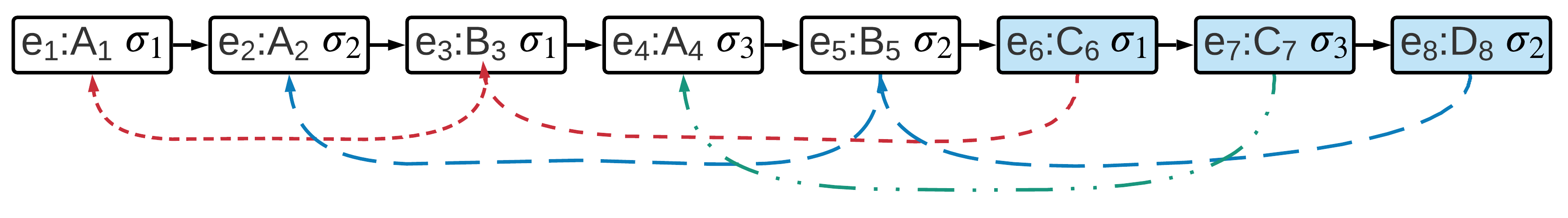}
	\caption{New individual $x'$ based on $\abbrev{LG_1}$}
	\label{fig:lg2:h}
\end{figure}

For example, let $\smax=2$ and $\tinit = 100$. After generating the initial individual in \cref{fig:lg1:h}, we have $\scurr = 1$. Notice that, at this iteration, $\scurr = \smax -1 $, so the changing point is randomly selected from the second half of the events in {\abbrev{UL}}, i.e., from $\Evt_5$ to $\Evt_8$. In this example, the changing point corresponds to $\Evt_6$. 
As shown in \cref{fig:lg2:h}, a new individual $x'$ is thus created and we explore the possible assignments for events $\Evt_6$, $\Evt_7$ and $ \Evt_8$, while the other events are assigned as in $x$.


\subsection{Energy cost function}\label{sec:costFun}

\begin{figure}[ptb!]
	\centering
	\includegraphics[width=0.9\textwidth]{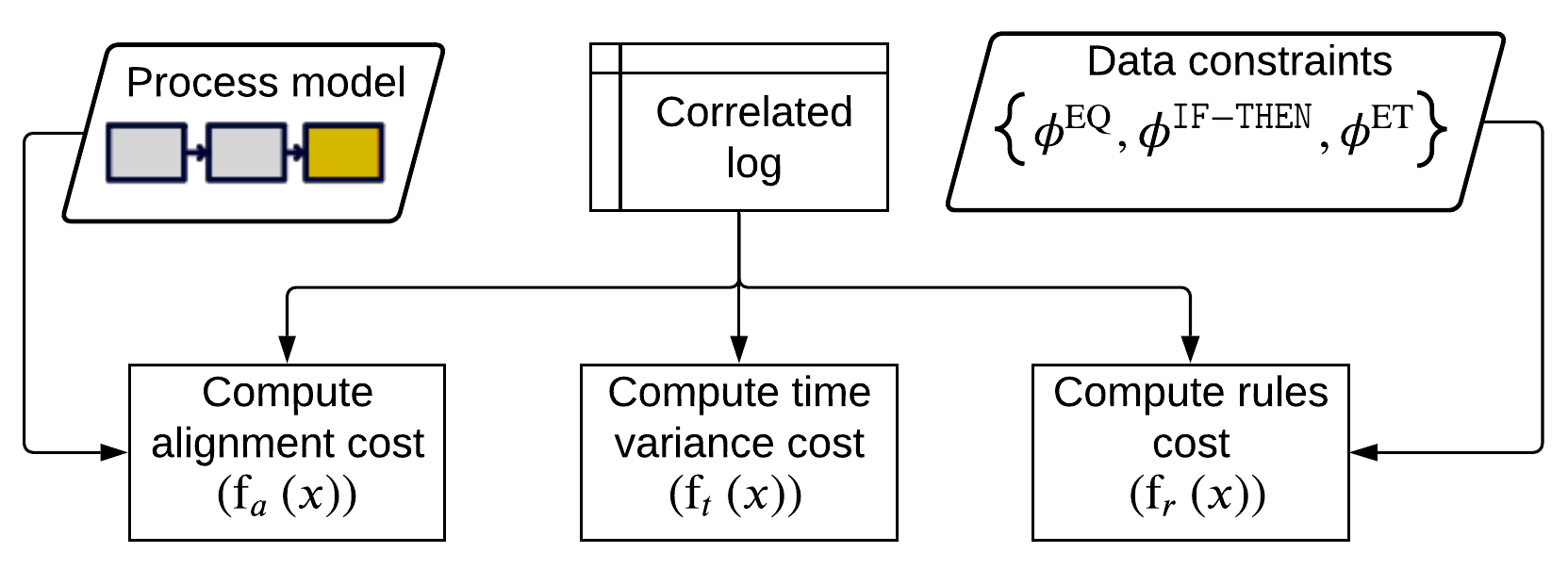}
	\caption{Energy cost functions}
	\label{fig:Energy}
\end{figure}

Simulated annealing uses the energy function to model the optimization problem. We model the event correlation problem as multi-level objectives optimization problem that has three level of objectives: \begin{iiilist}
	\item minimizing the misalignment between the generated	event log and the input process model,  
	\item minimizing the constraints violations over the cases, and
	\item minimizing the activity execution time variance across the cases.
\end{iiilist} 

We define an energy function for each of these objectives, as shown in \cref{fig:Energy}. The first energy function ($\fa(x)$) computes the cost of aligning $x$ and the model. The second energy function ($\fr(x)$) computes the data rules violations cost within $x$. The third energy function ($\ft(x)$) computes the activity execution time variance within $x$. 
They are used to compute the energy cost function between $x$ and $x'$, $\fc(x,x')$, as shown in \cref{fig:alg-steps}.
In the following, we elaborate on the individual energy functions and the computation of $\fc(x,x')$.

\subsubsection{Alignment cost} 

To measure the model-log misalignment we use the well-established \emph{alignment cost} function proposed by Adriansyah et al.~\cite{AdriansyahDA11}. The technique penalizes every asynchronous move between the log and the model, that is to say, it associates a cost to every event that occurs in the trace although the model would not allow for it, or every missing event that the model would require to continue the run though the trace does not contain it.

\begin{figure}[h !]
	\begin{subfigure}{.3\textwidth}
		\includegraphics[width=\textwidth]{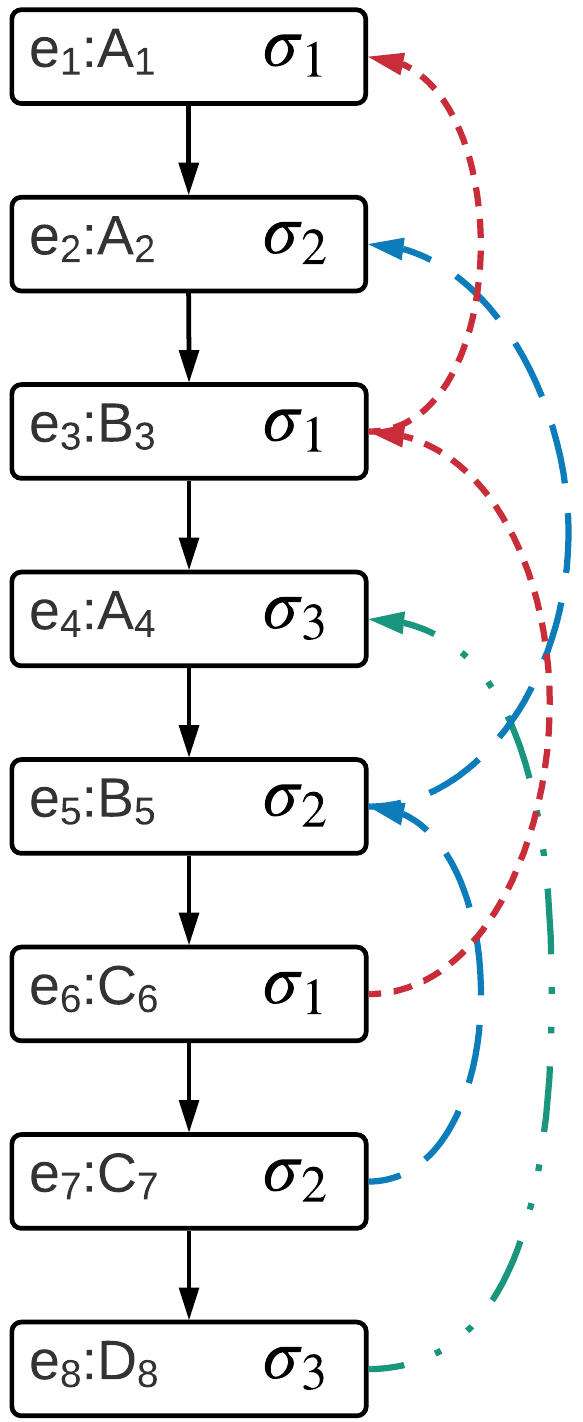}
		\caption{Log graph of $x$}
		\label{fig:loggraph1}
	\end{subfigure}
	\quad
	\begin{subfigure}{.3\textwidth}\centering
		\includegraphics[width=\textwidth]{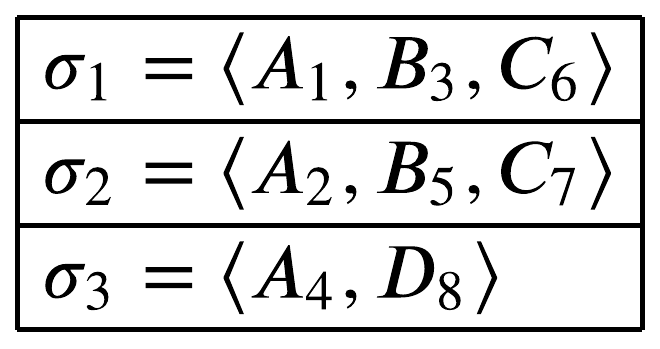}
		\caption{Cases in $x$}
		\label{fig:cases:LG}
		\includegraphics[width=0.9\textwidth]{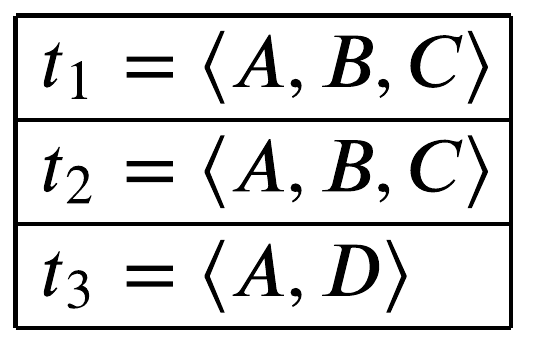}
		\caption{Traces in $x$}
		\label{fig:traces:LG}
		\includegraphics[width=0.9\textwidth]{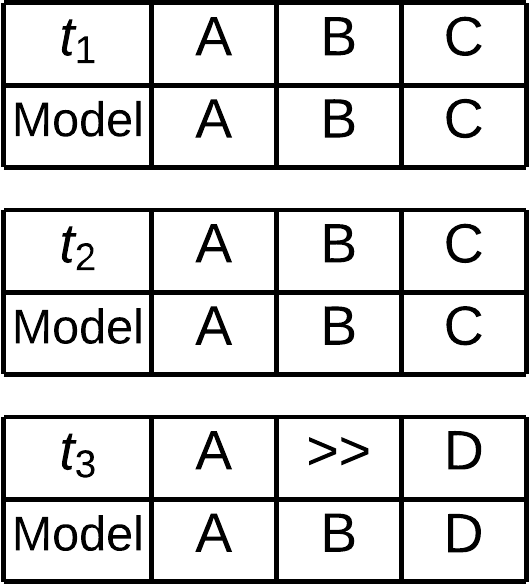}
		\caption{Alignments of traces in $x$}
		\label{fig:alignment:traces}
	\end{subfigure}
	\caption{Computation of the alignments for individual $x$}
	\label{fig:align:ex}
\end{figure}

\Cref{fig:align:ex} shows an example of computing the log alignment cost  $\fa(x)$ over individual $x$ (depicted in \cref{fig:loggraph1}).
The first step is to extract the cases from individual $x$ as shown in \cref{fig:cases:LG}. Then, \ECSAnew deduces the traces from the cases by projecting them over the cases' activities as shown in \cref{fig:traces:LG}. For each trace $\sigma$ in the log, it computes the alignment cost of the trace ($\Delta_\textrm{algn}(\attr{Act}(\sigma))$) with respect to the process model.

%
\footnote{We remark that $\Delta_\textrm{algn}$ requires a trace \emph{and} a process model to be computed. As the process model is given as background knowledge in this context, we omit its explicit mention as an input parameter for the sake of readability.}
The log alignment cost is finally computed as the summation of the trace alignment costs, as shown in \cref{eq:fa}. 
\begin{equation}\label{eq:fa}
\fa(x) = \sum_{\sigma \in S(x)}{\Delta_\textrm{algn}(\attr{Act(\sigma)})}
\end{equation}

For example, \cref{fig:alignment:traces} shows that the execution of activity \inst{D} in $t_3$ is considered as an asynchronous move as it is not enabled by the model (indicated with a guillemet in the figure, $\gg$). Thus, $\Delta_\textrm{algn}(t_3) = 1 $. On the other hand, $\Delta_\textrm{algn}(t_1) =\Delta_\textrm{algn}(t_2)=0$ as $t_1$ and $t_2$ are in complete alignment with the model. The log alignment cost of  $x$ (depicted in \cref{fig:loggraph1}) is $\fa(x) = \Delta_\textrm{algn}(t_1)+ \Delta_\textrm{algn}(t_1)+\Delta_\textrm{algn}(t_3)= 0+0+1 = 1$.
%



\subsubsection{Rule cost}

\ECSAnew measures the cost of data constraint violations in a log (henceforth, rule cost, \fr($x$)) by evaluating the constraints over the log's cases. The first step determines the triggered constraints for each case. A constraint is triggered by a case if 
\begin{iiilist}
	\item it is an equality constraint (hence, always triggered), or
	\item it is an {\ifthencon} constraint or an event-time constraint and the {\ifcon} part is satisfied by at least an event in the case. 
\end{iiilist}

The second step evaluates the triggered constraints over the case. For every violated constraint, a penalty is added for the case. A constraint $C$ is violated by a case $\sigma$ if any of the following holds:
\begin{iiilist}[label=(\roman*)]
	\item $C$ is an equality constraint 
	and at least an event violates it; 
	\item $C$ is an {\ifthencon} or an event-time constraint 
	and the {\ifcon} clause is satisfied while the {\thencon} clause is violated by at least a pair of events in the case. 
\end{iiilist}
We formalize these notions as follows.




\begin{definition}[Rule cost]
	Let $S(L)$ represent the cases of log $L$ as per \cref{def:case} and $\mathfrak{C} \supseteq \mathscr{C} \ni C$ be a universe of constraints as per \cref{def:constraints}. 
	Let $\mathrm{trigger} : \mathfrak{C} \times S(L) \to \{\mathrm{true}, \mathrm{false}\}$ be a function that determines whether a constraint $C \in \mathfrak{C}$ is triggered by a case, as follows:
	\begin{equation}\label{eq:trig:sigma}
	\mathrm{trigger}(C,\sigma) = \begin{cases}
	\mathrm{true}  & \textrm{ if } C \textrm{ is } \phi^{\textrm{EQ}}\\
	\mathrm{true}  & \textrm{ if } C \textrm{ is } \phi^{\ifthencon} \: \: \mathrm{and} \: \:  \varphi_{\ifcon} \equiv \mathrm{true}   \\
	\mathrm{true}  & \textrm{ if } C \textrm{ is } \phi^{\textrm{ET}} \: \: \mathrm{and} \: \:  \varphi_{\ifcon} \equiv \mathrm{true}   \\
	\mathrm{false} & \textrm{ otherwise}
	\end{cases}    
	\end{equation}
	%
	Let function $\mathrm{trig_C} : 2^\mathfrak{C} \times S(L) \to 2^\mathfrak{C}$ return a subset of constraints in $\mathscr{C} \in 2^\mathfrak{C}$ that are triggered by $\sigma \in S(L)$ as follows:
	\begin{equation}\label{eq:trig:constraints:sigma}
	\mathrm{trig_C}(\mathscr{C}, \sigma) = \{ C \in \mathscr{C} : \mathrm{trigger}(C,\sigma) = \mathrm{true}\}
	\end{equation}
	Let $\mathrm{eVio} : \mathfrak{C} \times E \times S(L) \to \{\mathrm{true}, \mathrm{false}\}$ be a function that determines whether an $i$-preassigned constraint $C$ (see \cref{def:ipreasgn}) is violated given $e \in E$ in case $\sigma \in S(L)$ as follows:
	\begin{equation}\label{eq:e:vio}
	\resizebox{8cm}{!}{$%
		\mathrm{eVio}(C,e,\sigma) = \begin{cases}%
		\mathrm{true}  & \textrm{ if } C \textrm{ is } \phi^{\textrm{EQ}} \: \textrm{and} \:  \phi^{\textrm{EQ}}[i/\,\overrightarrow{\imath}(\sigma, e)] \equiv  \mathrm{false} \\
		\mathrm{true}  & \begin{split}{\textrm{ if } C \textrm{ is } \phi^{\ifthencon} \: & \textrm{and} \: \varphi_{\ifcon}[i/\,\overrightarrow{\imath}(\sigma, e)] \equiv \mathrm{true} \\ 
			& \textrm{and} \: \varphi_{\thencon}[i/\,\overrightarrow{\imath}(\sigma, e)] \equiv \mathrm{false}}\end{split} \\
		\mathrm{true}  & \begin{split}{\textrm{ if } C \textrm{ is } \phi^{\textrm{ET}} \: & \textrm{and} \: \varphi_{\ifcon}[i/\,\overrightarrow{\imath}(\sigma, e)] \equiv \mathrm{true} \\ 
			& \textrm{and} \: \varphi_{\thencon}^{\textrm{ET}}[i/\,\overrightarrow{\imath}(\sigma, e)] \equiv \mathrm{false}}\end{split} \\
		\mathrm{false} & \textrm{ otherwise}
		\end{cases}%
		$}
	\end{equation}
	Let $\mathrm{vio} : \mathfrak{C} \times S(L) \to \{0, 1\}$ be an indicator function that returns $1$ if there exists at least an event $\Evt \in \sigma \in S(L)$ such that $\mathrm{eVio}(C,\Evt,\sigma) = \mathrm{true}$ or $0$ otherwise.
	%
	%
	The \emph{rule cost}, $\fr(x)$, is a function computed as the sum of the ratios of triggered constraints that are violated by every case, divided by the number of cases in individual (log) $x$:%
	\footnote{We remark that $\fc(x)$ requires an event log \emph{and} a set of data constraints to be computed. As the set of constraints is given as background knowledge in this context, we keep it as an implicit parameter for the sake of readability.}
	\begin{equation}\label{eq:RLC}
	\fr(x) = \frac{1}{|I|}\sum_{\sigma \in S(x)}{\frac{\sum\limits_{C \in {\mathrm{trig_C}(\mathscr{C}, \sigma_i)}}{\mathrm{vio(C,\sigma)}}}
		{\left|\mathrm{trig_C}(\mathscr{C}, \sigma)\right|}
	}
	\end{equation}
	
\end{definition}

\begin{table}[tbp]
	\resizebox{\textwidth}{!}{%
		\begin{tabular}{|l|l|l|}
			\hline
			\multicolumn{1}{!{\color[rgb]{0.051,0.051,0.051}\vrule}l|}{Cases} & $\mathrm{trig_C(\mathscr{C}, \sigma)}$                                                                                       & $\mathrm{vio}(C,\sigma)$                                                                                                     \\ \hline
			\multirow{8}{*}{$\sigma_1=\langle A_1,B_3,C_6 \rangle$}          & \multirow{2}{*}{$C_1 \textrm{ as it is } \phi^{\textrm{EQ}}$}                                                                & $\mathsf{A_1}.\mathrm{Type} = \mathsf{B_3}.\mathrm{Type} \wedge \mathsf{B_3}.\mathrm{Type} = \mathsf{C_6}.\mathrm{Type}$ \\
			&                                                                                                                              & $ \implies \mathrm{vio}(C_1,\sigma_1)=0$                                                                                             \\ \cline{2-3}
			& \multirow{2}{*}{$C_2 \textrm{ as } \phi^{\textrm{IF}} \textrm{ is satisfied by } \mathsf{B_3} \textrm{ and }\mathsf{A_1}$} & 
			$\textrm{Evaluate }\phi^\mathrm{THEN}~:~\mathsf{B_3}.\mathrm{Res} = \mathsf{A_1}.\mathrm{Res}$                            \\
			&                                                                                                                              & $ \implies \mathrm{vio}(C_2,\sigma_1)=0$                                                                                            \\ \cline{2-3}
			& \multirow{2}{*}{$C_3 \textrm{ as } \phi^{\textrm{IF}} \textrm{ is satisfied by } \mathsf{C_6} \textrm{ and }\mathsf{B_3}$}   & 
			$\textrm{Evaluate }\phi^\mathrm{THEN}~:~\mathsf{C_6}.\mathrm{Res} \neq \mathsf{B_3}.\mathrm{Res}$                         \\
			&                                                                                                                              & $\implies \mathrm{vio}(C_3,\sigma_1)=0$                                                                                             \\ \cline{2-3}
			& \multirow{2}{*}{$C_4 \textrm{ as } \phi^{\textrm{IF}}\textrm {is satisfied by } \mathsf{B_3}$}                               & 
			$\textrm{Evaluate }\phi^\mathrm{THEN}~:30 \le \mathsf{B_3}.\mathrm{Ts}-\mathsf{A_1}.\mathrm{Ts} \le 120$                  \\
			&                                                                                                                              & $\implies \mathrm{vio}(C_4,\sigma_1)=0$                                                                                             \\ \hline
			\multirow{8}{*}{$\sigma_2=\langle A_2,B_5,C_7 \rangle$}          & \multirow{2}{*}{$C_1 \textrm{ as it is } \phi^{\textrm{EQ}}$}                                                                & $\mathsf{A_2}.\mathrm{Type} = \mathsf{B_5}.\mathrm{Type} $ and $~~~\mathsf{B_5}.\mathrm{Type} = \mathsf{C_7}.\mathrm{Type}$                                                                 \\
			&                                                                                                                              & $\implies \mathrm{vio}(C_1,\sigma_2)=0$                                                                                             \\ \cline{2-3}
			& \multirow{2}{*}{$C_2 \textrm{ as } \phi^{\textrm{IF}} \textrm{ is satisfied by }\mathsf{B_5} \textrm{ and } \mathsf{A_2}$}   & 
			$\textrm{Evaluate }\phi^\mathrm{THEN}~:~\mathsf{B_5}.\mathrm{Res} = \mathsf{A_2}.\mathrm{Res}$                            \\
			&                                                                                                                              & $\implies \mathrm{vio}(C_2,\sigma_2)=0$                                                                                             \\ \cline{2-3}
			& \multirow{2}{*}{$C_3 \textrm{ as } \phi^{\textrm{IF}} \textrm{ is satisfied by } \mathsf{C_7} \textrm{ and } \mathsf{B_5}$}  & 
			$\textrm{Evaluate }\phi^\mathrm{THEN}~:~\mathsf{C_7}.\mathrm{Res} \neq \mathsf{B_5}.\mathrm{Res}$                         \\
			&                                                                                                                              & $\implies \mathrm{vio}(C_3,\sigma_2)=0$                                                                                             \\ \cline{2-3}
			& \multirow{2}{*}{$C_4 \textrm{ as } \phi^{\textrm{IF}}\textrm {is satisfied by } \mathsf{B_5}$}                               & 
			$\textrm{Evaluate }\phi^\mathrm{THEN}~:30 \le \mathsf{B_5}.\mathrm{Ts}-\mathsf{A_2}.\mathrm{Ts} \le 120$                  \\
			&                                                                                                                              & $\implies \mathrm{vio}(C_4,\sigma_2)=0$                                                                                             \\ \hline
			\multirow{4}{*}{$\sigma_3=\langle A_4,D_8 \rangle$}               & \multirow{2}{*}{$C_1 \mathrm{as~it~ is~} \phi^{\textrm{EQ}}$}                                                                & $\mathsf{A_4}.\mathrm{Type} = \mathsf{D_8}.\mathrm{Type}$                                                                \\
			&                                                                                                                              & $\implies \mathrm{vio}(C_1,\sigma_3)=0$                                                                                             \\ \cline{2-3}
			& \multirow{2}{*}{$C_5 \textrm{ as } \phi^{\textrm{IF}}\textrm {is satisfied by } \mathsf{D_8}$}                               & 
			$\textrm{Evaluate }\phi^\mathrm{THEN}~:120 \le \mathsf{D_8}.\mathrm{Ts}-\mathsf{A_4}.\mathrm{Ts} \le 240$                 \\
			&                                                                                                                              & $\implies \mathrm{vio}(C_5,\sigma_3)=1  $                                                                                           \\ \hline
		\end{tabular}
		
	}
	\caption{Rule cost computation for individual $x$, considering the data constraints in $\mathscr{C}=\{C_1, \ldots, C_5\}$ as per \cref{fig:example}(c)}
	\label{tbl:RLCexample}
\end{table}

To compute the rule cost over individual $x$, we first identify the constraints in set $\mathscr{C}$ that are triggered by $\sigma$ with $\mathrm{trig_C(\mathscr{C}, \sigma_i)}$. Then, we verify each constraint $C$ in $\mathrm{trig_C}$ over $\sigma$. Finally, we take the average of violations over the log: $\mathrm{RC} = 0$ if no violation occurs.  

\Cref{tbl:RLCexample} shows how we evaluate constraints in our running example, having $\mathscr{C} = \{C_1, \ldots, C_5\}$ (shown in \cref{fig:example}(c)) over $x$ (shown in \cref{fig:loggraph1}). Cases $\sigma_1$ and $\sigma_2$ trigger $C_1$, $C_2$, $C_3$ and $C_4$, while $\sigma_3$ triggers $C_1$ and $C_5$.
We observe that $\sigma_1$ and $\sigma_2$ satisfy their triggered constraints, while  $\sigma_3$ violates $C_5$, as $\mathsf{D_8}.\mathrm{Ts}-\mathsf{A_4}.\mathrm{Ts}~=~\SI{180}{\minute}$ thus exceeding the maximum admissible duration of activity {\inst{D}} ($\SI{150}{\minute}$). Therefore, $\fr(x) = \frac{1}{3}\left(\frac{0}{4}+\frac{0}{4}+\frac{1}{2}\right)= 0.167$.


\subsubsection{Execution time variation cost} 
The third objective is to minimize the activities' execution time variance over the correlated events.  {\ECSAnew} employs the  Mean Square Error (MSE) \cite{Gentle2009} to measure the execution variance over the log for the \ft($x$) energy function. MSE measures the deviation between expected values and actual values. We assume that the activities tend to be carried out similarly across cases. Therefore, we use the activities' average execution time over the log to represent the expected duration. 
We use the events' elapsed time to represent the actual one.
We formalize these concepts as follows.

\begin{definition}[Time variance]
	Let $S(L)$ represent the cases of log $L$ as per \cref{def:case}.
	%
	%
	The event time function $\mathrm{ET}: E^* \times E \to \mathbb{R}^+$ computes the elapsed time 
	of an event $\sigma(i) \in E$ based on the preceding event in the same case $\sigma(i-1) \in E$ as follows:
	
	\begin{equation}\label{eq:elapsedTime}
	\mathrm{ET}(\sigma,\sigma(i)) =
	\begin{cases}
	\sigma(i).\attr{Ts} - \sigma(i-1).\attr{Ts} & \textrm{ if } 1 < i \leqslant n \\
	0 & \textrm{ otherwise}
	\end{cases}
	\end{equation}
	Let $\mathrm{NSE}(L,a)$ be the set of non-starting events in the cases of $S(L)$ that report the execution of $a \in \mathrm{Dom}_\attr{Act}$.\\
	\resizebox{.8\columnwidth}{!}{$%
		\mathrm{NSE}(L,a) = \bigcup\limits_{\sigma \in S(L)}{\left\{\sigma(i) \in \sigma: 1 < i \leqslant |\sigma| \textrm{ and } \sigma(i).\attr{Act} = a\right\}}.
		$}  
	
	Let $\mathrm{T_{avg}}(L,a)$ be the average activity execution duration of activity $a \in \mathrm{Dom}_\attr{Act}$ as per log $L$:
	\begin{equation}\label{eq:tavg}
	\mathrm{T_{avg}}(L,a) = \frac{\sum\limits_{e \in \mathrm{NSE}(L,a)}{\mathrm{ET}(\ell(e),e)}}{\left| \mathrm{NSE}(L,a) \right|}
	\end{equation}
	\noindent
	Given an individual (log) $x$ time variance is the mean square error of the activity execution duration over the events in $x$:
	\begin{equation}\label{eq:MSA}
	\ft(x)  = \frac
	{\sum\limits_{\sigma \in S(x)} \left(
		{\sum\limits_{i = 2}^{|\sigma|}}{\big(
			\mathrm{T_{avg}}(\sigma(i).\mathrm{Act})-\mathrm{ET}(\sigma,\sigma(i))
			\big)^2}
		\right)}
	{|E|-|I|}
	\end{equation}
	
\end{definition}
\noindent
\ECSAnew computes the time variance by measuring the mean square error (MSE) using $\mathrm{T_{avg}}(a)$ as its expected values, and the events' elapsed time as its actual values. MSE is computed for all the events \emph{except} the start events of the cases in log $L$. Therefore, the denominator of $\ft(x)$ is $(|E|-|I|)$  in order not to count the $|I|$ start events.
For example, let us see how \ECSAnew computes $\ft(x)$ for the $x$ individual depicted in \cref{fig:loggraph1}. The first step is computing the average execution time of activities in $x$ to represent the expected values. The average execution times (in minutes) are 
$\mathrm{T_{avg}}(\mathsf{B}) = 75$,
$\mathrm{T_{avg}}(\mathsf{C}) = 120$, and
$\mathrm{T_{avg}}(\mathsf{D}) = 180$.
Then, we compute the elapsed time of the events to represent the actual time and, thereupon, the time variance given the number of non-start events ($ |E|-|I| = 8 - 3 = 5 $). As a result, $\ft(x)=\SI{90}{\minute}$.


\subsubsection{Cost function computation}

Simulated annealing evaluates the energy cost of changing from individual $x$ to a new individual $x'$ in order to decide which one to keep in the next iteration.
\ECSAnew computes the \emph{energy cost function}, $\fc(x,x')$, based on the objective functions $\fa$ (alignment cost), $\fr$ (rule cost) and $\ft$ (time variance), as shown in \cref{eq:cost}. We use these three energy functions to apply the multiple-level optimization as follows: 
\begin{iiilist}
	\item {\fc} is computed based on {\fa} if the alignment cost of $x$ is lower than that of $x'$; else, we compute 
	\item {\fc} is based on {\fr} if $x'$ violates more data rules than $x$ and $x'$ is better aligned with the model; otherwise, 
	\item {\fc} is computed based on {\ft}.
\end{iiilist}
\begin{equation} \label{eq:cost}
\resizebox{.99\columnwidth}{!}{$%
	\fc(x,x')=\left\{
	\begin{array}{lll}
	\fa(x') - \fa(x) & \textrm{ if } \fa(x') \textgreater \fa(x)\\
	\fr(x') - \fr(x) & \textrm{ if } \fa(x') \le \fa(x) \textrm{ and } \fr(x') \textgreater \fr(x)\\
	\ft(x') - \ft(x) & \textrm{ otherwise}.\\
	\end{array}
	\right.
	$}
\end{equation}%

The energy cost function computes the cost of choosing the new neighbor individual $x'$ over $x$. Therefore, $\fc(x,x')$ is computed using an energy function where $x'$ performs worse than $x$ as per \cref{eq:cost}.
For example, \cref{fig:step1} depicts the initial individual, $x$, and the values of the energy functions applied to $x$.  \Cref{fig:step2} shows the new neighbor individual, $x'$. 
The energy cost function $(\fc(x,x'))$ is computed based on time variance $\ft$, as the new neighbor $x'$ has a better alignment cost than that of $x$ ($\fa(x') \le \fa(x)$) and both the individuals have the same rule cost ($\fr(x') = \fr(x)$). Thus, $\fc(x,x')=\ft(x') - \ft(x) = 15.3 - 7.3= 8$. 

\subsection{Selection of the next individual}\label{sec:selection}
\begin{algorithm2e}[bht]
	\SetKwInOut{Input}{input}
	\SetKwInOut{Output}{output}
	\Input{Current individual $x$; new neighbor $x'$}
	\Output{Selected individual}
	\BlankLine
	\lIf{$\fa(x') < \fa(x)$}{
		\Return $x'$
	}
	\ElseIf{$\fa(x') = \fa(x)$}{
		\lIf{$ \fr(x') < \fr(x)$}{
			\Return $x'$
		}
		\ElseIf{$ \fr(x') = \fr(x)$}{
			\lIf{$ \ft(x') < \ft(x) \text{ or } \prob(x') \geq \mathrm{random}(0,1) $}{
				\Return $x'$
			}
		}
		\ElseIf{$ \fr(x') > \fr(x) \text{ and } \prob(x') \geq \mathrm{random}(0,1)$}{
			\Return $x'$
		}
		
	}
	\lElseIf{$\fa(x') > \fa(x) \text{ and } \prob(x') \geq \mathrm{random}(0,1)$}{
		\Return $x'$
	}
	
	\Return $x$
	
	\caption{Selection of the solution for the next iteration}
	\label{alg:determineFun}
\end{algorithm2e}
\Cref{alg:determineFun} shows the full selection procedure of the individual for the next iteration.
Its decision between $x$ and $x'$ is based on the objective functions $\fa$ (first-level), $\fr$ (second-level) and $\ft$ (third-level), together with the acceptance probability, $\prob(x')$.
The latter is computed using $\fc(x,x')$ and the current temperature ($\tcurr$) as shown in \cref{eq:ET:acceptanceP}: 
\begin{equation}\label{eq:ET:acceptanceP}
\prob(x')=\exp^{\frac{-\fc(x,x')}{\tcurr}}
\end{equation}
\ECSAnew compares the value of $\prob(x')$ with a random value in a $[0,1]$ interval to accept or reject the new neighbor, depending on whether $\prob(x')$ is higher or lower than the random value, respectively. In this way, we simulate the annealing process, enforced by the fact that the decrease of temperature $\tcurr$ also reduces the randomness of the choice. Furthermore, notice that the memory-less stochastic perturbation
makes it possible to skip the local optimal.

If the new neighbor ($x'$) has a lower alignment cost, then it is selected. If the new neighbor ($x'$) and the current individual ($x$) have the same alignment cost, then
we check the rule cost energy function. If the new neighbor ($x'$) has a lower rule cost, then it is selected. If the new neighbor ($x'$) and current individual ($x$) have the same rule cost, then we check the time variance energy function. The acceptance probability $\prob(x')$ is computed using $\fc(x,x')$ based on $\ft(x)$ and $\ft(x')$. Then, $x'$ is selected either if $x'$ has a lower time variance or if it is randomly selected based on $\prob(x')$. On the other hand, if the new neighbor has a higher rule cost than the current individual, then we calculate $\fc(x,x')$ based on $\fr(x)$ and $\fr(x')$. The same holds if the new neighbor has a higher alignment cost than the current individual. In that case, we calculate $\fc(x,x')$ based on $\fa(x)$ and $\fa(x')$. We take the final decision based on a random selection weighed by $\prob(x')$ which, in turn, is calculated on the basis of the current temperature, $\tcurr$, and $\fc(x,x')$. This process is repeated for each individual within the population.

For example, \cref{fig:iterations} shows the results through the \ECSAnew iterations. We assume that $\smax =2$ and $\tinit = 100$. \Cref{fig:step1} depicts the initial individual, $x$, and its energy cost functions.  \Cref{fig:step2} shows the new individual, $x'$, generated on the basis of $x$.
$\fc(x,x')$ is computed considering $\ft$: $\fc(x,x')= 8$. According to \cref{alg:determineFun}, $x'$ is selected and replaces $x$ in the population as $\fa(x') \le \fa(x)$. 

\subsection{Global solution update, cooling down and new iteration}
As a final step, \ECSAnew returns the global optimal solution $x_\mathrm{G}$ at $\smax$, namely the solution that has the best $\fa$, $\fr$ and $\ft$ over all iterations, as shown in \cref{fig:example}.
%
%
The cooling schedule simulates the cooling-down technique of the annealing process by controlling the computation of the current temperature, $\tcurr$. 
We use the logarithmic function schedule~\cite{logSchedule} as per \cref{eq:t}. The number of iterations that
the logarithmic schedule goes through to cool down helps to skip the local optimum and explore a wider correlation search space especially in the early phases of the run:
\begin{equation}\label{eq:t}
\tcurr=\frac{\tinit}{\ln\left(1+\scurr\right)}
\end{equation}

\begin{figure}[h!]
	\begin{subfigure}{.48\columnwidth}
		\centering
		\includegraphics[width=.75\textwidth]{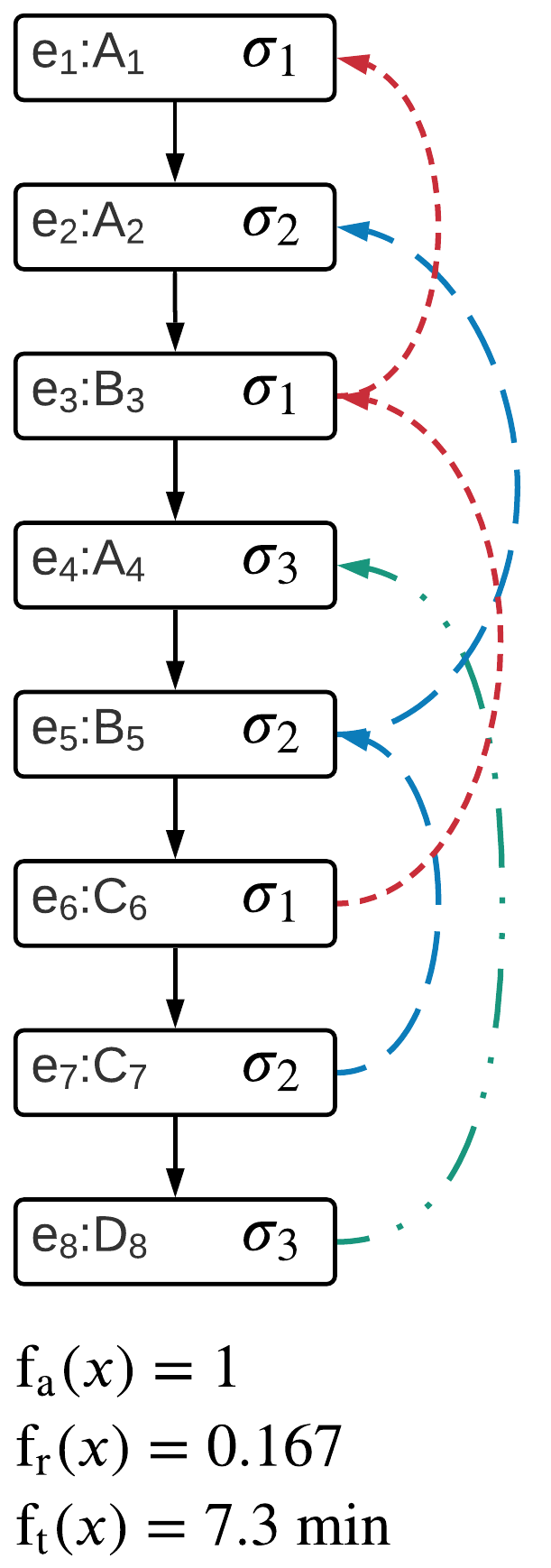}
		\caption{Iteration 1 ($\scurr=1$, $\tcurr =100$, $\prob(x')= 0.99$) }
		\label{fig:step1}
	\end{subfigure}
	\hfill
	\begin{subfigure}{.48\columnwidth}
		\centering
		\includegraphics[width=.71\textwidth]{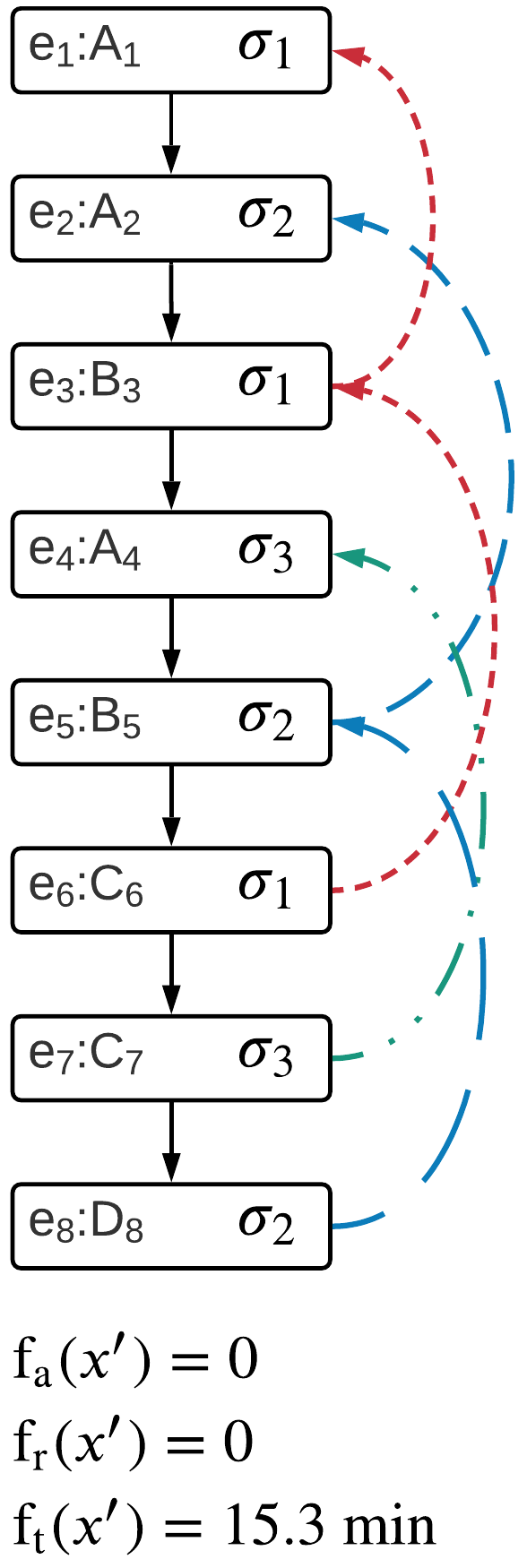}
		\caption{Iteration 2 ($\scurr=2$, $\tcurr =144.9$, $\prob(x')$ is irrelevant)}
		\label{fig:step2}
	\end{subfigure}
	
	\caption{\ECSAnew iterations, with $\smax = \num{2}$ and $\tinit=\num{100}$}
	\label{fig:iterations}
\end{figure}

Following the \ECSAnew steps in \cref{fig:alg-steps}, the algorithm proceeds until $\scurr = \smax$ as shown in \cref{fig:iterations}. At each iteration, it reassigns the events from different changing points in the log to explore the search space. We recall that accepting a worse solution than the current one in some iterations helps to skip the optimal local solution and reach an approximate optimal global solution.

\medskip

In the following section, we discuss the measures we introduce to evaluate the accuracy of our technique based on a pair of logs. We will see next in \cref{evaluation} that the two logs correspond to a golden standard and the one originated by \ECSAnew.
\section{Quality measures}\label{qualitymeasures}

To assess the quality of our technique, we defined measures that quantify the output accuracy.
They are based on criteria that compare pairs of logs. In particular, we can group our measures in two categories. The first category is the \emph{log-to-log similarity}, which takes into account trace-based and case-based distances. The second category is \emph{log-to-log time deviation}, determining the temporal distance of events' elapsed times and cases' cycle times.
The definition of a complete set of measures to compare logs goes beyond the scope of this paper.
The measures we propose here are inspired by related work in the literature~\cite{Navarro,AdriansyahDA11,DezaDeza/2006:DictionaryOfDistances,Devore2012,CMinerConf,ECSA-ER}
and provide a good trade-off between run-time computability and different level of details used in the comparison, experimental results evidenced. A refinement and enrichment of this set can be part of future investigations.

\begin{figure}
	\centering
	\begin{subfigure}{.4\textwidth}
		
		\includegraphics[width=\textwidth]{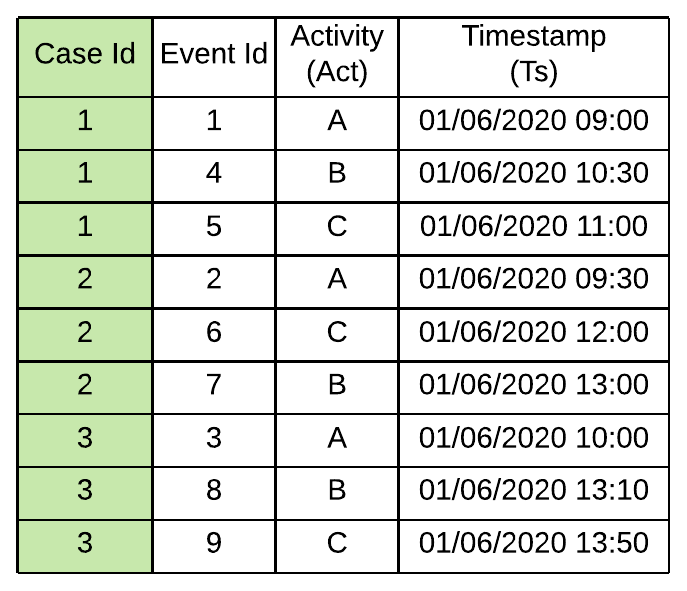}
		\caption{Log $L$}
		\label{fig:l1}
		
	\end{subfigure}
	\quad
	\begin{subfigure}{.4\textwidth}		
		\includegraphics[width=\textwidth]{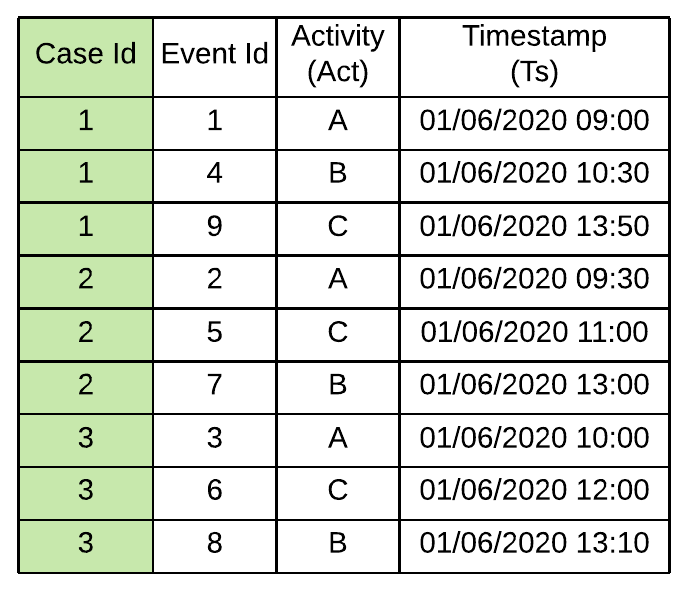}
		\caption{Log $L'$}
		\label{fig:l2}
	\end{subfigure}
	\caption{Event logs $L$ and $L'$ for quality measures example}
	\label{fig:evalQmeasures}
	
\end{figure}
We use the two event logs in \cref{fig:evalQmeasures} to illustrate the different quality measures. We remark that our focus is on logs $L$ and $L'$ that stem from the same uncorrelated log and case id's, but possibly differ for the the way in which events are correlated. Therefore, the cardinality of case sets $S(L)$ and $S(L')$ yet cases in $S(L)$ and $S(L')$ may differ.

\subsection{Log-to-log similarity}\label{sec:log2logsim}



The log-to-log similarity category focuses on measuring the structure similarity between two logs from trace- and case-structure perspectives. This category is comprised of six measures. The values of those measures range from $0$ to $1$, where $1$ indicates the highest similarity. 
In the following, we describe the measures following an order given by a decreasing level of abstraction and aggregation with which the similarity between a pair of logs is established. The higher the level of detail, the more fine-granular differences are considered by the measure.
%


\bigskip 
Inspired by the fitness measure proposed in \cite{AdriansyahDA11}, we define our first similarity measure, \emph{trace-to-trace similarity}. It aims at assessing the extent to which two logs capture the same underlying control-flow through the string-edit distance of their traces. We formally define it as follows.
%
\begin{definition}[Trace-to-trace similarity]\label{def:L2L:T2T}
	Let $L=(\abbrev{UL},I,\ell)$ and $L'=(\abbrev{UL},I,\ell')$ be two event logs.	%
	Let {\SEinsdel} be the string-edit distance based on insertions and deletions~\cite{Navarro}.
	We denote with $T = \lbrace t_{1}, t_{2}, \ldots, t_{|T|} \rbrace $ and $T' = \lbrace t'_{1}, t'_{2}, \ldots, t'_{|T'|} \rbrace $ the set of \emph{distinct} traces that are derived from event logs $L$ and $L'$, respectively, i.e., $T = \bigcup_{\sigma \in S(L)}{\textrm{Act}(\sigma)}$ and
	$T' = \bigcup_{\sigma' \in S(L')}{\textrm{Act}(\sigma')}$. 
	We indicate as the \emph{trace-closest trace} to $t \in T$ a trace in $t'_{\star} \in T'$ derived as follows:
	\begin{equation}
	t'_{\star} = \underset{ t' \in T' }{\operatorname*{arg\,min}} \, \Big\lbrace \SEinsdel\left(t,t'\right)
	\Big\rbrace
	\end{equation}%
	\noindent The \emph{trace-to-trace similarity} {\TTsim} is computed as follows: 
	\begin{equation}
	\TTsim = 1 - \frac{\sum\limits_{t \in T}{\SEinsdel\left(t, t'_\star\right)}}{\sum\limits_{t \in T}{\left(|t|+|t'_\star|\right)}}
	\end{equation}
	
\end{definition}

\begin{table}[tb]
	\caption{Computation of \TTsim for logs $L$ and $L'$ in \cref{fig:evalQmeasures}}
	\label{fig:ex:l2lTrace}
	
	\begin{subtable}[tb]{.40\textwidth}
		\centering
		\caption{Distinct traces in $L$}
		\label{fig:l1:ts}
		\begin{adjustbox}{max width=\textwidth}
			\begin{tabular}{|l|} 
				\hline
				$t_1 = \langle \inst{A},\inst{B},\inst{C} \rangle$  \\ 
				\hline
				$t_2 = \langle \inst{A},\inst{C},\inst{B} \rangle$   \\
				\hline
			\end{tabular}
		\end{adjustbox}	
	\end{subtable}
	\begin{subtable}[tb]{.40\textwidth}
		\centering
		\caption{Distinct traces in $L'$}
		\label{fig:l2:ts}
		\begin{adjustbox}{max width=\textwidth}
			\begin{tabular}{|l|} 
				\hline
				$t'_1 = \langle \inst{A},\inst{C},\inst{B} \rangle$  \\ 
				\hline
				$t'_2 = \langle \inst{A},\inst{B},\inst{C} \rangle$   \\
				\hline
			\end{tabular}
		\end{adjustbox}	
	\end{subtable}
	\\
	\begin{subtable}[tb]{.9\textwidth}
		\centering
		\caption{Matching pair $(t_{\star},t'_{\star})$ for each trace in $L$ and $L'$}
		\label{fig:l2lt}
		\begin{adjustbox}{max width=\textwidth}
			\begin{tabular}[tb]{|l|l|}
				\hline
				${\SEinsdel}(t_{1},t'_{1}) = 1$ & \cellcolor[HTML]{D9E9F9} ${\SEinsdel}(t_{1},t'_{2}) = 0$ \\ \hline
				\cellcolor[HTML]{D9E9F9} ${\SEinsdel}(t_{2},t'_{1}) = 0$ &${\SEinsdel}(t_{2},t'_{2}) = 1$\\ \hline
			\end{tabular}
		\end{adjustbox}
	\end{subtable}
\end{table}

\noindent 

For example, \cref{fig:l1:ts} shows two distinct traces in event log $L$ (depicted in \cref{fig:l1})  and \cref{fig:l2:ts} shows two distinct traces in event log $L'$ (\cref{fig:l2}). The pairs of trace-closest traces $(t,t'_{\star})$ are illustrated in \cref{fig:l2lt}. We select the pairs that minimize the total distance $\Delta_\textrm{total}$ between traces (see the marked cells in \cref{fig:l2lt}). For instance, for $t_1$ we select the trace-closest pair $\left(t_{1}, t'_{2}\right)$ instead of $\left(t_{1}, t'_{1}\right)$ because the distance of the former $\SEinsdel\left(t_{1},t'_{2}\right)=0$ is lower than $\SEinsdel\left(t_{1},t'_{1}\right)=1$.
Finally, we compute {\TTsim} as a fraction having the sum of the string-edit distances between pairs of trace-closest traces ($0+0=0$) as the numerator and the length of the traces in $L$ and their trace-closest traces in $L'$ ($k=|t_1|+|t'_2|+|t_2|+|t'_1|= 12$) as the denominator.

\bigskip

The second measure we introduce is the \emph{trace-to-trace frequency similarity}. It is more fine-granular than trace-to-trace similarity as it also takes into account the frequency with which traces occur. The formal definition follows. 

\begin{definition}[Trace-to-trace frequency similarity]\label{def:L2L:similarity}
	Let $L=(\abbrev{UL},I,\ell)$ and $L'=(\abbrev{UL},I,\ell')$ be two event logs, whose cases are
	$S(L) = \lbrace \sigma_{1}, \sigma_{2}, \ldots, \sigma_{|I|} \rbrace $
	and
	$S(L') = \lbrace \sigma'_{1}, \sigma'_{2}, \ldots, \sigma'_{|I|} \rbrace $
	respectively. 
	Let {\SEinsdel} be the string-edit distance based on insertions and deletions~\cite{Navarro}.
	%
	Let $\mathrm{tcc} : S(L) \to S(L')$ be a bijective function mapping every case in $L$ to exactly one case in $L'$, $\mathrm{TCC}$ the set of all possible such bijective functions definable having $S(L)$ and $S(L')$ as domain and range respectively, and $\mathrm{tcc}_\star \in \mathrm{TCC}$ be such that the total string-edit distance between $\mathrm{tcc}_\star$-mapped pairs of cases is minimal:
	\begin{equation}\label{eq:tcc:star}
	\mathrm{tcc}_\star = \underset{ \mathrm{tcc} \in \mathrm{TCC} }{\operatorname*{arg\,min}} \, \Big\lbrace \sum\limits_{\sigma \in S(L)} \SEinsdel\left(\attr{Act}(\sigma),\attr{Act}(\mathrm{tcc}(\sigma))\right)
	\Big\rbrace
	\end{equation}
	%
	Naming the $\mathrm{tcc}_\star$-mapped cases as trace-closest case pairs, let $\Delta_\mathrm{total}(L, L')$ be the sum of all string-edit distances between trace-closest case pairs:
	\begin{equation}
	\Delta_\mathrm{total} = \sum\limits_{\sigma \in S(L)} \SEinsdel\left(\attr{Act}(\sigma),\attr{Act}(\mathrm{tcc}_\star(\sigma))\right)
	\end{equation}
	The \emph{trace-to-trace frequency similarity}, \LLsim, is the opposite of the average of the total distances between trace-closest case pairs: 
	\begin{equation}
	\TTsim\!\left( L, L' \right) = 1 - \frac{\Delta_\mathrm{total}}{2 \times|E| }
	\end{equation}
\end{definition}
%

\begin{table}[tb]
	\caption{Computation of \LLsim for logs $L$ and $L'$ in \cref{fig:evalQmeasures}}
	\label{fig:ex:l2lfreq}
	
	\begin{subtable}[tb]{.40\textwidth}
		\centering
		\caption{Traces in $L$}
		\label{fig:l1:cT}
		\begin{adjustbox}{max width=\textwidth}
			\begin{tabular}{|l|} 
				\hline
				$\mathrm{Act}(\sigma_1)= \langle \inst{A},\inst{B},\inst{C} \rangle$  \\ 
				\hline
				$\mathrm{Act}(\sigma_2)=\langle \inst{A},\inst{C},\inst{B} \rangle$   \\
				\hline
				$\mathrm{Act}(\sigma_3)=\langle \inst{A},\inst{B},\inst{C} \rangle$   \\
				\hline				
			\end{tabular}
		\end{adjustbox}	
	\end{subtable}
	\begin{subtable}[tb]{.40\textwidth}
		\centering
		\caption{Traces in $L'$}
		\label{fig:l2:cT}
		\begin{adjustbox}{max width=\textwidth}
			\begin{tabular}{|l|} 
				\hline
				$\mathrm{Act}(\sigma'_1)= \langle \inst{A},\inst{B},\inst{C} \rangle$  \\ 
				\hline
				$\mathrm{Act}(\sigma'_2)=\langle \inst{A},\inst{C},\inst{B} \rangle$   \\
				\hline
				$\mathrm{Act}(\sigma'_3)=\langle \inst{A},\inst{C},\inst{B} \rangle$   \\
				\hline
			\end{tabular}
		\end{adjustbox}	
	\end{subtable}
	\\
	\vspace{0.5\baselineskip}
	\begin{subtable}[tb]{.8\textwidth}
		\centering
		\caption{Matching pair $(\sigma_{\star},\sigma'_{\star})$ for each trace in $L$ and $L'$}
		\label{fig:l2lc}
		\begin{adjustbox}{max width=\textwidth}
			\begin{tabular}[tb]{|l|l|l|}
				\hline
				\cellcolor[HTML]{D9E9F9} ${\SEinsdel}(\mathrm{Act}(\sigma_1),\mathrm{Act}(\sigma'_1)) = 0$ & ${\SEinsdel}(\mathrm{Act}(\sigma_2),\mathrm{Act}(\sigma'_1)) = 2$ & ${\SEinsdel}(\mathrm{Act}(\sigma_3),\mathrm{Act}(\sigma'_1)) = 0$ \\ \hline
				${\SEinsdel}(\mathrm{Act}(\sigma_1),\mathrm{Act}(\sigma'_2)) = 2$ & \cellcolor[HTML]{D9E9F9} ${\SEinsdel}(\mathrm{Act}(\sigma_2),\mathrm{Act}(\sigma'_2)) = 0$ & ${\SEinsdel}(\mathrm{Act}(\sigma_3),\mathrm{Act}(\sigma'_2)) = 2$ \\ \hline
				${\SEinsdel}(\mathrm{Act}(\sigma_1),\mathrm{Act}(\sigma'_2)) = 0$ & ${\SEinsdel}(\mathrm{Act}(\sigma_2),\mathrm{Act}(\sigma'_3)) = 0$ & \cellcolor[HTML]{D9E9F9} ${\SEinsdel}(\mathrm{Act}(\sigma_3),\mathrm{Act}(\sigma'_3)) = 2$ \\ \hline				
			\end{tabular}
		\end{adjustbox}
	\end{subtable}
\end{table}

\noindent To compute $\mathrm{tcc}_\star$ as in \cref{eq:tcc:star} and thus find the trace-closest case pairs, we use the Hungarian Algorithm \cite{HungarianAlg}. 
For example, \cref{fig:l1:cT} shows the traces that stem from event log $L$ (depicted in \cref{fig:l1}) and \cref{fig:l2:ts} shows the traces stemming from event log $L'$ (depicted in \cref{fig:l2}). \Cref{fig:l2lc} illustrates the pairs of trace-closest cases that we derive. The selected ones are colored in light blue: 
$\SEinsdel\left(\attr{Act}\left(\sigma_{1}\right),\attr{Act}\left(\sigma'_{1}\right)\right)=0$,
$\SEinsdel\left(\attr{Act}\left(\sigma_{2}\right),\attr{Act}\left(\sigma'_{2}\right)\right)=0$, and
$\SEinsdel\left(\attr{Act}\left(\sigma_{3}\right),\attr{Act}\left(\sigma'_{3}\right)\right)=2$.
These pairs lead to the minimum $\Delta_\mathrm{total} =2$. Finally, we compute $\LLsim = 1-\frac{2}{2 \times 9}  = 0.78$.

\bigskip

The following four measures investigate the similarity between logs at the level of events.
Notice that we compare pairs of cases for which the first event correspond.
%
%
The third measure we describe is the \emph{partial case similarity}, which is based upon the number of events shared by cases that have the same first event. 



\begin{definition}[Partial case similarity]\label{def:L2L:E2F}
	Let $L=(\abbrev{UL},I,\ell)$ and $L'=(\abbrev{UL},I,\ell')$ be two event logs, whose case sets are
	$S(L)$ and $S(L')$
	respectively. 
	%
	We indicate with $\mathrm{intersect}: E^* \times E^* \to \mathbb{N} \cup {0} $ a function that takes two cases $\sigma$ and $\sigma'$ as input and returns the number of events that $\sigma$ and $\sigma'$ have in common. 
	\begin{equation}\label{eq:intersect}
	\mathrm{intersect}\!\left(\sigma,\sigma'\right) = \left| \left\{ e \in \sigma : e \in \sigma' \right\} \right|
	\end{equation}
	
	\noindent The partial case similarity distance {\FEsim} averages the number of events in common (except the first one) that cases in $L$ and $L'$ have when they share the same first event over the number of events in $L$ (except the first ones of the cases), as follows:
	\begin{equation}\label{eq:first}
	\FEsim\!\left(L,L'\right)= 
	\frac%
	{\underset{\substack{ \sigma \in S(L), \\ \sigma' \in S(L'): \\ \sigma(1) = \sigma'(1) }}{\sum}%
		{ \mathrm{intersect}\!\left(\sigma\left[2,|\sigma|\right],\sigma'\!\left[2,\left|\sigma'\right|\right]\right)}}{ |E| - |I| } 
	\end{equation}
	
\end{definition}

\begin{table}
	\centering
	\caption{Intersect between cases with the same start event in $L$ and $L'$ }
	\label{fig:ex:l2lFirst}
	\begin{tabular}{|l|l|l|} 
		\hline
		$\mathrm{intersect}(\sigma_1,\sigma'_1)=1$ &	$\mathrm{intersect}(\sigma_2,\sigma'_2)=1$ &
		$\mathrm{intersect}(\sigma_3,\sigma'_3)=1$   \\
		\hline
	\end{tabular}
\end{table}
\noindent For instance, given the event logs $L$ and $L'$ in \cref{fig:l1,fig:l2}, we compute the elements of the sum in the numerator of $\FEsim\left(L,L'\right)$ as depicted in \cref{fig:ex:l2lFirst}. In the example, $\sigma_{1}$ and $\sigma'_{1}$ have the same start event $\inst{A_1}$, thus, we check the occurrence of events in $\sigma_{1}[2,3]$ in $\sigma'_{1}[2,3]$ and find that only $\inst{B_4}$ occur in both cases, so $\mathrm{intersect}\left(\sigma_1[2,3],\sigma'_1[2,3]\right) =1$. Finally, \FEsim is computed by averaging the non-start events in common over the total number of the non-start events: $\FEsim = \frac{1+1+1}{9-3} = 0.5$.

\bigskip

The fourth measure we describe is the \emph{bigram similarity}. Inspired by \cite{cMiner}, it is based on the number of sequences of two events (henceforth, bigrams, i.e., n-grams of length $2$) that occur in both logs. 
We formally define it as follows.  

\begin{definition}[Bigram similarity]\label{def:L2L:E2p}
	Let $L=(\abbrev{UL},I,\ell)$ and $L'=(\abbrev{UL},I,\ell')$ be two event logs, whose case sets are
	$S(L)$ and $S(L')$
	respectively. 
	We denote as $\mathrm{occurs2}(\langle e,e'\rangle,L)$ the indicator function that returns $1$ if there exists a case $\sigma \in S(L)$ such that $\left\langle e,e' \right\rangle$ is a segment of it:
	\begin{equation}
	\resizebox{.98\columnwidth}{!}{$%
		\mathrm{occurs2}\!\left(\left\langle e,e'\right\rangle,L\right)  = 
		\begin{cases}
		1 & \text{if there exists } \sigma \in S(L) \text{ s.t. } \left\langle e,e'\right\rangle \subseteq \sigma \\
		0 & \text{otherwise}
		\end{cases}
		$}
	\end{equation}
	The \emph{bigram similarity} $\PEsim$ is computed dividing by the cardinality of $S(L)$ the average of bigrams in the cases of $L$ that also occur in $L'$ as follows: 
	%
	%
	%
	%
	%
	
	\begin{equation}\label{eq:pair}%
	\resizebox{.98\columnwidth}{!}{$%
		\PEsim\!\left(L,L'\right)= 
		\frac{1}{|I|} 
		{\sum\limits_{\sigma \in S(L)}{ 
				\frac{1}{|\sigma| - 1} \left(\sum\limits^{|\sigma|-1}_{i=1}%
				{\mathrm{occurs2}\left(\langle \sigma(i),\sigma(i+1)\rangle,L'\right)}
				\right)}}
		$}
	\end{equation}

\end{definition}

\begin{table}
	\centering
	\caption{Computation of \PEsim for logs $L$ and $L'$ in \cref{fig:evalQmeasures}}
	\label{fig:ex:l2lpair}
	\begin{tabular}{|l|l|} 
		\hline
		$\mathrm{occurs2}(\langle\inst{A_1},\inst{B_4}\rangle.L')=1$ &$\mathrm{occurs2}(\langle\inst{B_4},\inst{C_5}\rangle.L')=0$\\
		\hline
		$\mathrm{occurs2}(\langle\inst{A_2},\inst{C_6}\rangle.L')=0$ &$\mathrm{occurs2}(\langle\inst{C_6},\inst{B_7}\rangle.L')=0$\\
		\hline
		$\mathrm{occurs2}(\langle\inst{A_3},\inst{B_8}\rangle.L')=0$ &$\mathrm{occurs2}(\langle\inst{B_8},\inst{C_9}\rangle.L')=0$\\
		\hline
	\end{tabular}
\end{table}
\noindent For example, for every pair ¨ in $L$ (depicted in \cref{fig:l1}), we check if $\left\langle e,e' \right\rangle$ occurs in $L'$ (depicted in \cref{fig:l2}) as shown in \cref{fig:ex:l2lpair}. Notice that $L$ and $L'$ 
have only one bigram in common, that is $\langle \inst{A_1},\inst{B_4} \rangle$.
Therefore,
$\PEsim$ = $\frac{1}{3} \left(\frac{1+0}{3-1} + \frac{0+0}{3-1} + \frac{0+0}{3-1} \right) = 0.167$.

\bigskip

The fifth measure we describe is the \emph{trigram similarity}, which extends the bigram similarity by considering n-grams of length $3$ (trigrams) in place of bigrams. We formally define it as follows. 

\begin{definition}[Trigram similarity]\label{def:L2L:E3}
	
	Let $L=(\abbrev{UL},I,\ell)$ and $L'=(\abbrev{UL},I,\ell')$ be two event logs, whose case sets are
	$S(L)$ and $S(L')$
	respectively. 
	We denote as $\mathrm{occurs3}(\langle e,e',e''\rangle,L)$ the indicator function that returns $1$ if there exists a case $\sigma \in S(L)$ such that $\left\langle e,e',e'' \right\rangle$ is a segment of it.
	\begin{equation}
	\resizebox{.98\columnwidth}{!}{$%
		\mathrm{occurs3}\!\left(\left\langle e,e',e'' \right\rangle,L\right)  = 
		\begin{cases}
		1 & \text{if there exists } \sigma \in S(L) \text{ s.t. } \left\langle e,e',e'' \right\rangle \subseteq \sigma \\
		0 & \text{otherwise}
		\end{cases}
		$}
	\end{equation}
	The \emph{trigram similarity} $\TEsim$ is computed dividing by the cardinality of $S(L)$ the average of trigrams in the cases of $L$ that also occur in $L'$ as follows: 
	%
	%
	\begin{equation}
	\resizebox{.99\columnwidth}{!}{$%
		\TEsim\!\left(L,L'\right) =
		\frac%
		{\sum\limits_{\sigma \in S(L)}{ \left(
				\frac{1}{|\sigma| - 2} \sum\limits^{|\sigma|-1}_{i=2}{	\mathrm{occurs3}\left(\langle\sigma(i-1),\sigma(i),\sigma(i+1)\rangle,L'\right)}
				\right)}}
		{|I|}
		$}
	\end{equation}

\end{definition}

\begin{table}
	\centering
	\caption{Computation of \TEsim for logs $L$ and $L'$ in \cref{fig:evalQmeasures}}
	\label{fig:ex:l2ltriple}
	\begin{tabular}{|l|} 
		\hline
		$\mathrm{occurs3}(\langle\inst{A_1},\inst{B_4},\inst{C_5}\rangle.L')=0$\\
		\hline
		$\mathrm{occurs3}(\langle\inst{A_2},\inst{C_6},\inst{B_7}\rangle.L')=0$\\
		\hline
		$\mathrm{occurs3}(\langle\inst{A_3},\inst{B_8},,\inst{C_9}\rangle.L')=0$\\
		\hline
	\end{tabular}
\end{table}

\noindent For example, for every trigram $\langle e,e',e''\rangle$ in $L$ (depicted in \cref{fig:l1}), we check if it occurs in $L'$ (depicted in \cref{fig:l2}). As shown in \cref{fig:ex:l2ltriple}, $L$ and $L'$ do not have trigrams in common. Therefore,  $\TEsim = 0$.
If the case assignments of $e_5$ and $e_6$ were swapped (i.e., $e_5$ and $e_6$ had been assigned with $\sigma'_{3}$ and $\sigma'_{2}$, respectively), then the value of $\TEsim$ would be $ \frac{1}{3} $.

\bigskip

The last measure we describe is the \emph{case similarity}, which checks the extent to which a pair of logs identically correlate cases as a whole. We formally define it as follows. 

\begin{definition}[Case similarity, \Csim]\label{def:L2L:case}
	Let $L=(\abbrev{UL},I,\ell)$ and $L'=(\abbrev{UL},I,\ell')$ be two event logs, whose case sets are
	$S(L)$ and $S(L')$
	respectively. 
	%
	%
	%
	%
	%
	%
	The \emph{case similarity}, $\Csim$, amounts to the number of cases that are equal in $L$ and $L'$ divided by the total number of cases: 
	\begin{equation}
	\Csim\!\left(L,L'\right) = \frac %
	{\left| S(L) \cap S(L') \right|}
	{|I|}
	\end{equation}
\end{definition}

%
%

\noindent Notice that we indicate with $S(L) \cap S(L')$ the cases in $L$ that have an equal one in $L'$. As the number of cases is the same in $L$ and $L'$, $\Csim$ can be considered as a S{\o}rensen-Dice coefficient for $S(L)$ and $S(L')$~\cite{DezaDeza/2006:DictionaryOfDistances}. In the example, given $L$ (depicted in \cref{fig:l1}) and $L'$ (depicted in \cref{fig:l2}), there are no equal cases occurring in $L$ and $L'$. 
Therefore, $\Csim = 0$.
If the case assignments of $e_5$ and $e_6$ were swapped (i.e., $\ell(e_5) = \sigma'_{3}$ and $\ell(e_6) = \sigma'_{2}$), then the value of $\Csim$ would be $ \frac{1}{3} $.

\bigskip

Up to this point, we have presented the measures following an order given by the increasing amount of details they consider in the comparison.
We shall refer to the ones at a higher level of abstraction as more \emph{relaxed} (e.g., \TTsim), as opposed to the \emph{strict}er ones (e.g., \Csim).
In the following, we present measures that deal with time deviations.

\subsection{Log-to-log time deviation}
%
The log-to-log time deviation category consists of two measures which use the symmetric mean absolute percentage error (SMAPE) to compute the time deviation between a pair of logs. Their values range from $0$ to $1$, where $0$ indicates the highest similarity between the two logs. 

\bigskip

The first measure we describe here is the \emph{event time deviation}. It investigates the extent to which the two logs deviate in terms of the elapsed time of events. We formally define it as follows.


%

\begin{definition}[Event time deviation]\label{def:L2L:SMAPE} 
	Let $L=(\abbrev{UL},I,\ell)$ and $L'=(\abbrev{UL},I,\ell')$ be two event logs defined over a common universe of events $E$. 
	Let $\mathrm{ET}(\sigma,\Evt)$ be the elapsed time (ET) of event $\Evt$ in case $\sigma$ as per \cref{eq:elapsedTime}.
	The \emph{event time deviation}, {\LLsmape}, is the Symmetric Mean Absolute Percentage Error (SMAPE) of the elapsed time of events between $L$ and $L'$, computed as follows:
	%
	\\
	\resizebox{\hsize}{!}{\begin{minipage}{\linewidth}
			\begin{flalign}\label{eq:SMAPE}%
			\LLsmape(L,L')=&
			\frac
			{\sum_{\Evt \in E}{\frac{\left| \mathrm{ET}(\sigma,\Evt) - \mathrm{ET}(\sigma',\Evt)\right|}
					{|\mathrm{ET}(\sigma,\Evt)|+|\mathrm{ET}(\sigma',\Evt)|}}}{|E|- |I|} \nonumber\\
			& \textrm{with} \: \sigma = L(\ell(\Evt)), \; \sigma' = L'(\ell(\Evt))
			\end{flalign}
	\end{minipage}}
\end{definition}

\begin{table}[tb]
	\caption{Computation of \LLsmape for logs $L$ and $L'$ in \cref{fig:evalQmeasures}}
	\label{fig:ex:l2l:smape}
	
	\begin{subtable}[tb]{.45\textwidth}
		\centering
		\caption{Elapsed time (in min) of events in $L$}
		\label{fig:l1:ET}
		\begin{adjustbox}{max width=\textwidth}
			\begin{tabular}{|l|l|} 
				\hline
				$\mathrm{ET}(\sigma_1,\inst{B_4}) = 90 $&  $\mathrm{ET}(\sigma_1,\inst{C_5}) = 30 $\\ 
				\hline
				$\mathrm{ET}(\sigma_1,\inst{C_6}) = 150 $&  $\mathrm{ET}(\sigma_1,\inst{B_7}) = 60 $\\ 
				\hline
				$\mathrm{ET}(\sigma_1,\inst{B_8}) = 190 $&  $\mathrm{ET}(\sigma_1,\inst{C_9}) = 40 $\\ 
				\hline		
			\end{tabular}
		\end{adjustbox}	
	\end{subtable}
	\hfill
	\begin{subtable}[tb]{.45\textwidth}
		\centering
		\caption{Elapsed time (in min) of events in $L'$}
		\label{fig:l2:ET}
		\begin{adjustbox}{max width=\textwidth}
			\begin{tabular}{|l|l|} 
				\hline
				$\mathrm{ET}(\sigma_1',\inst{B_4}) = 90 $&  $\mathrm{ET}(\sigma_1',\inst{C_5}) = 90 $\\ 
				\hline
				$\mathrm{ET}(\sigma_1',\inst{C_6}) = 120 $&  $\mathrm{ET}(\sigma_1',\inst{B_7}) = 120 $\\ 
				\hline
				$\mathrm{ET}(\sigma_1',\inst{B_8}) = 70 $&  $\mathrm{ET}(\sigma_1',\inst{C_9}) = 200 $\\ 
				\hline	
			\end{tabular}
		\end{adjustbox}	
	\end{subtable}
	
\end{table}

\noindent
For example, the elapsed time of  $\inst{C_5}$ in $L$, given that $L(\ell(\mathsf{C_5})) = \sigma_{1}$, is $\mathsf{C_5}.\mathrm{Ts} - \mathsf{B_4}.\mathrm{Ts} = \SI{30}{\minute}$,
whereas the elapsed time of $\inst{C_5}$ in $L'$, given that $L'(\ell(\mathsf{C_5})) = \sigma'_{2}$, is $\mathsf{C_5}.\mathrm{Ts} - \mathsf{A_2}.\mathrm{Ts} = \SI{90}{\minute}$.
We compute \LLsmape using the elapsed time of the events in $L$ (depicted in \cref{fig:l1}), and the elapsed time of the events in $L'$ (depicted in \cref{fig:l2}) as shown in \cref{fig:ex:l2l:smape}.
As a result, $\LLsmape = 0.35$.


\bigskip 
The second measure we describe is the \emph{case cycle time deviation}, assessing the extent to which two logs differ in terms of the cases' cycle time. To compare pairs of cases, we consider those that begin with the same start event, as seen in \cref{sec:log2logsim} with {\FEsim}. We formally define the measure as follows.

\begin{definition}[Case cycle time deviation]\label{def:L2L:CTsmape} 
	Let $L=(\abbrev{UL},I,\ell)$ and $L'=(\abbrev{UL},I,\ell')$ be two event logs defined over a common universe of events $E$. 
	Let $\mathrm{CT}(\sigma)$ be the cycle time of case $\sigma$, computed as follows~\cite{VanLooy2016,DBLP:books/sp/DumasRMR18}:
	\begin{equation}
	\mathrm{CT}(\sigma)= \sigma(|\sigma|).Ts - \sigma(1).Ts
	\end{equation}
	

	\noindent The \emph{case cycle time deviation} {\LLCTsmape} is the symmetric mean absolute percentage error of the cycle time between cases in $L$ and $L'$:
	\begin{equation}\label{eq:CTSMAPE}
	\LLCTsmape(L,L')=
	\frac{1}{|I|} \times {\sum_{\substack{\sigma \in L, \\
				\sigma' in L': \\ \sigma(1) = \sigma'(1)}}
		{\frac{\left| \mathrm{CT}(\sigma) - \mathrm{CT}(\sigma')\right|}
			{|\mathrm{CT}(\sigma)|+|\mathrm{CT}(\sigma')|}}} 
	\end{equation}
\end{definition}

\begin{table}[tb]
	\caption{Computation of \LLCTsmape for logs $L$ and $L'$ in \cref{fig:evalQmeasures}}
	\label{fig:ex:l2l:ctsmape}
	
	\begin{subtable}[tb]{.45\textwidth}
		\centering
		\caption{Cycle time (in hours) of events in $L$}
		\label{fig:l1:CT}
		\begin{adjustbox}{max width=\textwidth}
			\begin{tabular}{|l|} 
				\hline
				$ \mathrm{CT}(\sigma_1) = \inst{C_5}.\attr{Ts} - \inst{A_1}.\attr{Ts} = 1$\\ 
				\hline
				$ \mathrm{CT}(\sigma_2) = \inst{B_7}.\attr{Ts} - \inst{A_2}.\attr{Ts} = 3.5$\\ 
				\hline
				$ \mathrm{CT}(\sigma_3) = \inst{C_9}.\attr{Ts} - \inst{A_3}.\attr{Ts} =3.8 $\\ 
				\hline		
			\end{tabular}
		\end{adjustbox}	
	\end{subtable}
	\hfill
	\begin{subtable}[tb]{.45\textwidth}
		\centering
		\caption{Cycle time (in hours) of events in $L'$}
		\label{fig:l2:CT}
		\begin{adjustbox}{max width=\textwidth}
			\begin{tabular}{|l|} 
				\hline
				$ \mathrm{CT}(\sigma'_1) = \inst{C_9}.\attr{Ts} - \inst{A_1}.\attr{Ts} = 4.8$\\ 
				\hline
				$ \mathrm{CT}(\sigma'_2) = \inst{B_7}.\attr{Ts} - \inst{A_2}.\attr{Ts} = 3.5$\\ 
				\hline
				$ \mathrm{CT}(\sigma'_3) = \inst{B_8}.\attr{Ts} - \inst{A_3}.\attr{Ts} =3.17$\\ 
				\hline		
			\end{tabular}
		\end{adjustbox}	
	\end{subtable}
	
\end{table}

\noindent
For example, we compute the cycle time of case $\sigma_1$ in $L$ based on the first and the last events in the case: $\mathrm{CT}(\sigma_1)=\mathsf{C_5}.\mathrm{Ts} - \mathsf{A_1}.\mathrm{Ts} =\SI{1}{\hour}$. The cycle time of case $\sigma'_1$ in $L'$ is  $\mathrm{CT}(\sigma'_1)$ = $\mathsf{C_9}.\mathrm{Ts} - \mathsf{A_1}.\mathrm{Ts}  =\SI{4.8}{\hour}$. We compute {\LLCTsmape} comparing the cycle time of the cases in event log $L$ (depicted in \cref{fig:l1}) and the ones in event log $L'$ (depicted in \cref{fig:l2}) having the same start event, as shown in \cref{fig:ex:l2l:ctsmape}. As a result, $\LLCTsmape = 0.25$.

\bigskip
Thus far, we have presented the measures we use to assess the outcome of our approach.
Next, we illustrate our experimental results and evaluate them by means of the aforementioned measures.

\section{Evaluation}\label{evaluation}
%

We implemented a prototype tool for \ECSAnew.%
\footnote{\url{https://github.com/DinaBayomie/EC-SA/releases/tag/untagged-b9067967b502ea98491a}}
Using this tool, we conducted three experiments to evaluate the accuracy and time performance of our approach, and compared the results with \ECSA~\cite{ECSA-ER} used as a baseline.

\subsection{Design}\label{subSec:Eval:setup}
\begin{figure}[pbt!]
	\centering
	\includegraphics[width=\textwidth]{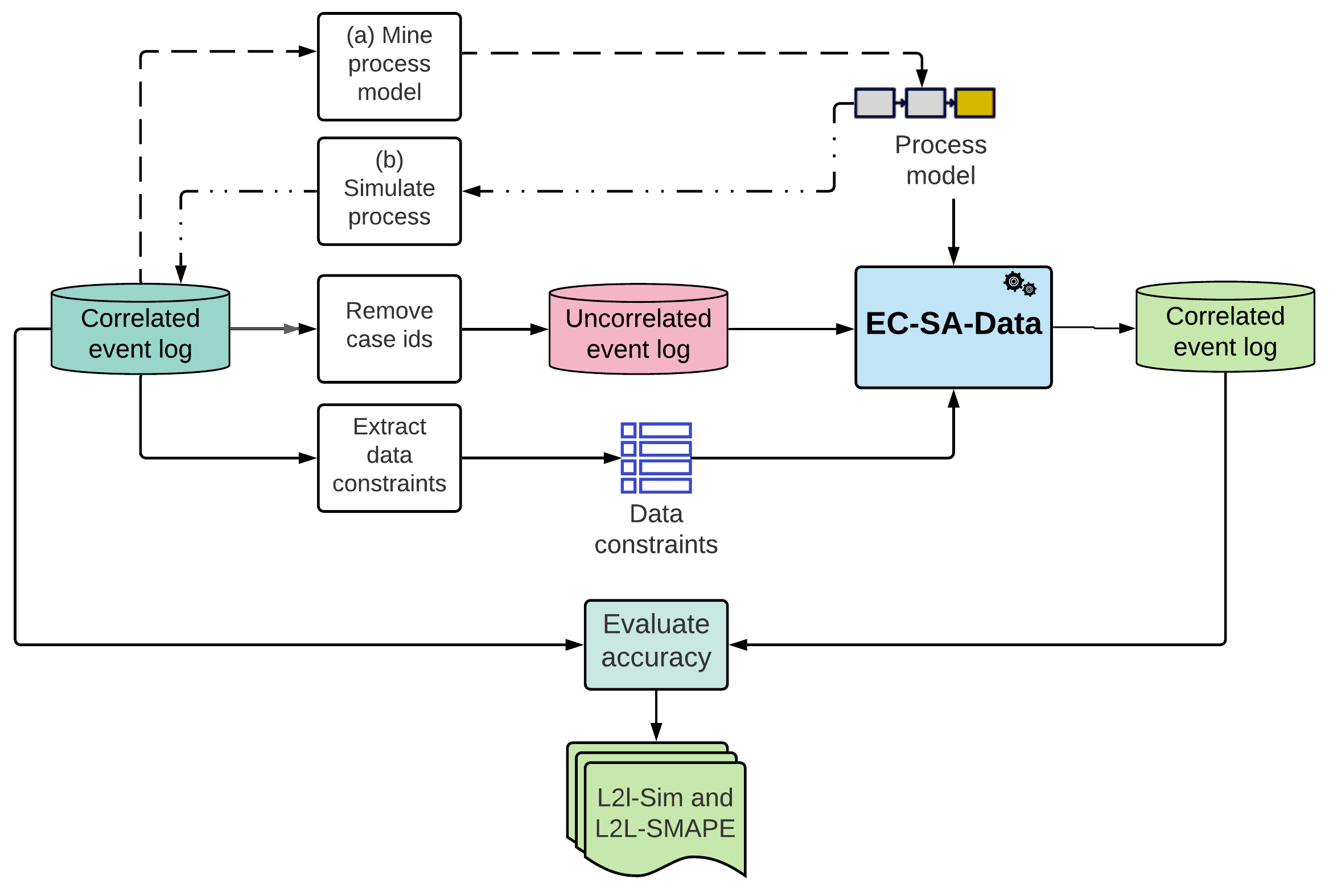}
	\caption{Evaluation steps}
	\label{fig:eval-steps}
\end{figure}

\Cref{fig:eval-steps} illustrates our evaluation process.
The primary input is a correlated event log with defined cases;
we refer to it as the \emph{original log}.
We remove the case identifiers from it and thereby create an uncorrelated event log. 
Thereupon, we run our implemented technique and measure its accuracy using the two categories of measures defined in \cref{qualitymeasures}. The first category is the \emph{log-to-log similarity}, which assesses the extent to which \ECSAnew generates a correlated log (we refer to it as $L'$ in \Cref{def:L2L:T2T,def:L2L:similarity,def:L2L:E2F,def:L2L:E2p,def:L2L:E3,def:L2L:case,def:L2L:SMAPE,def:L2L:CTsmape}) that is consistent with the original log (we refer to it as $L$ in \Cref{def:L2L:T2T,def:L2L:similarity,def:L2L:E2F,def:L2L:E2p,def:L2L:E3,def:L2L:case,def:L2L:SMAPE,def:L2L:CTsmape}) in terms of  trace-based and case-based distances. The second category is \emph{log-to-log time deviation}, which considers the temporal distance of events' elapsed times and cases' cycle times. 

The three experiments differ in terms of their input and research objectives. Overall, we aim to assess the effectiveness of our technique, taking into account evaluation principles as described in~\cite{DBLP:journals/corr/abs-2107-10675}.

The first experiment performs a sensitivity analysis on the impact on the accuracy of the output log that increasing the number of cases overall and of the average work-in-progress cases (WIP, i.e., the density of the overlapping cases at a point in time) entails. We tuned the inter-arrival time between starting cases to setup the WIP. 
To generate the input logs, we simulated a process that includes basic process behavior patterns, i.e., sequence, concurrency, exclusiveness, and cyclic runs.
%
%
The produced logs present the following characteristics:
\begin{inparaenum}[(a)]
	\item a varying number of cases, ranging between \num{100} and \num{1000} at steps of \num{100};
	\item different inter-arrival time based on the cycle time (CT) of the process (i.e., the time spent by a case from start to end), ranging between $\sfrac{\textrm{CT}}{32}$ and CT, at multiplicative steps of $\num{2}$.
\end{inparaenum}

\begin{table}
	\begin{tabular}{l S[table-format=6] S[table-format=2.1] S[table-format=6] S[table-format=2] S[table-format=2] S[table-format=2.1] S[table-format=3]}
		\toprule
		& \multicolumn{2}{c}{\textbf{Traces}} & \multicolumn{2}{c}{\textbf{Events}} & \multicolumn{3}{c}{\textbf{Trace length}}  \\
		\cmidrule(r){2-3} \cmidrule(r){4-5} \cmidrule{6-8}
		\textbf{Event log} & \textbf{Total} & \textbf{Dst.\%}    & \textbf{Total} & \textbf{Dst.\%}    & \textbf{Min} & \textbf{Avg} & \textbf{Max} \\ \midrule
		BPIC13\textsubscript{cp} \cite{BPIC13cp}                              & 1487           & 12.3               & 6660           & 7                  & 1            & 4            & 35           \\
		BPIC13\textsubscript{inc} \cite{BPIC13inc}                            & 7554           & 20.0               & 65533          & 13                 & 1            & 9            & 123          \\
		BPIC15\textsubscript{1f} \cite{BPIC151f}                              & 902            & 32.7               & 21656          & 70                 & 5            & 24           & 50           \\
		BPIC17\textsubscript{f} \cite{BPIC17}                                 & 21861          & 40.1               & 714198         & 41                 & 11           & 33           & 113          \\ \bottomrule
	\end{tabular}
	\caption{Descriptive statistics of real logs}
	\label{tbl:realL}
\end{table}
\begin{table}
	\begin{tabular}{l S[table-format=6] S[table-format=2.1] S[table-format=6]  }
		\toprule
		\textbf{Event\ log}                        & \textbf{Equality constraints} & {\textbf{{\ifthencon} constraints}} &  \\ \midrule
		BPIC13\textsubscript{cp} \cite{BPIC13cp}   & 3                             & 3                                  &  \\
		BPIC13\textsubscript{inc} \cite{BPIC13inc} & 3                             & 2                                  &  \\
		BPIC15\textsubscript{1f} \cite{BPIC151f}   & 4                             & 0                                  &  \\
		BPIC17\textsubscript{f} \cite{BPIC17}      & 3                             & 7                                  &  \\ \bottomrule
	\end{tabular}
	\caption{Data constraints per event log}
	\label{tbl:const}
\end{table}
The second experiment assesses the accuracy improvement our approach yields when data constraints are provided together with an input process model. 
For this experiment, we used four real-world datasets from the benchmark of Augusto et al.~\cite{benchmark} based on the publicly available event logs in the BPIC repository. \Cref{tbl:realL} shows some descriptive statistics about them. %
We mined the process models from the original logs using a state-of-the-art discovery technique, namely Split Miner~\cite{augusto2018split}.
We extracted the data constraints by visual inspection and analysis of those event logs -- \cref{tbl:const} summarizes our findings. 
Thereupon, we compared the results attained with our approach with those of \ECSA, as the latter does not provide the capability of including data constraints to steer the assignment of case identifiers to the events.

The third experiment performs a sensitivity analysis that investigates the effect of the constraints on the accuracy of the log generated by \ECSAnew. We used the BPIC17 event log~\cite{BPIC17} as filtered by Augusto et al.~\cite{benchmark} (hence the `f' subscript in ``BPIC17\textsubscript{f}'' in the tables and figures),
as we observed that a relatively high number of data constraints define its behavior, compared to other real-world event logs as shown in \cref{tbl:const}. In particular, we inferred ten rules that regulate the behavior of the process behind the BPIC17 log (see \cref{rules}). There are three data-attribute equality rules and seven {\ifthencon} constraints. Six {\ifthencon} rules are based on the matching of the operating resources over some activities within a case. The seventh {\ifthencon} rule is a correlation rule based on the equality of two different data attributes over some activities within a case. 

This experiment investigates two aspects. 
%
The first aspect pertains to the effect of an increase in the number of used constraints on the accuracy of the generated log. We gradually increase the number of used constraints from zero to ten.  Notice that the order of the constraints does not affect the accuracy of the output. Thereby, we investigate the impact of increasing the knowledge on the accuracy of the generated logs.

The second aspect concerns the impact of the knowledge quality on the accuracy of the generated logs. We impersonate the business analysts in their iterative endeavor. While inspecting the data, some rules occur as evident. Other ones are less certain or harder to confirm. Nevertheless, they could use all of them in an attempt to drive the automated correlation, although some could possibly misrepresent the data, thereby misleading the technique.
To mimic this situation, we run successive tests adding
\begin{inparaenum}[(a)]
	\item four correct rules (three of which are data-attribute equality constraints and one is an {\ifthencon} constraint), and then
	\item three inexact rules. 
\end{inparaenum}
Notice that we omit three correct rules out of the ten aforementioned ones to mimic the missing knowledge. 

Next, we describe in details the measures we used to evaluate accuracy and the results of our experiments.

\subsection{Results}
\begin{figure*}[ptb!]
	\begin{subfigure}{.4\textwidth}
		\includegraphics[width=\textwidth]{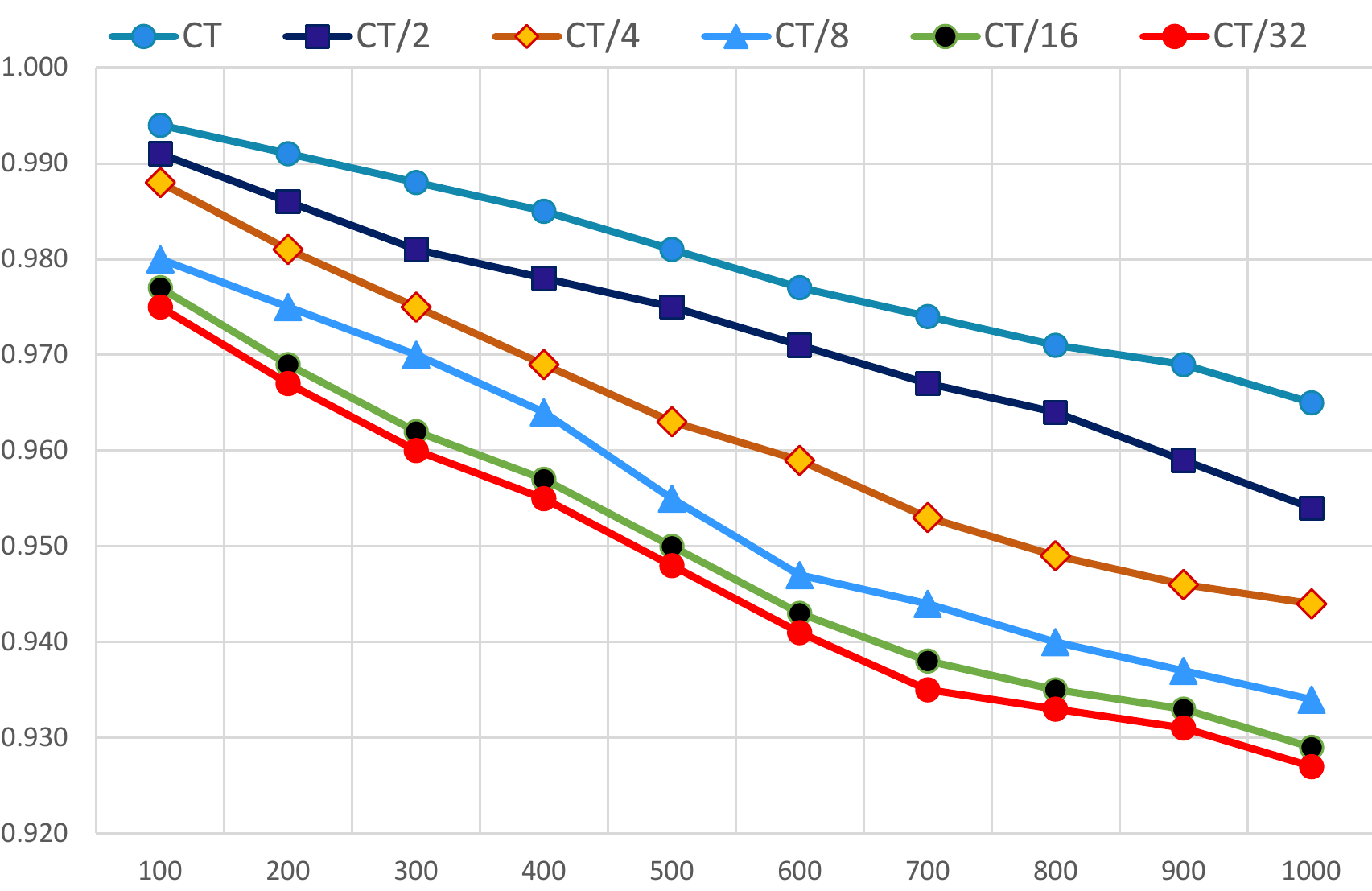}
		\caption{\ECSAnew - \TTsim}
		\label{fig:exp1:traces}
	\end{subfigure}
	\begin{subfigure}{.4\textwidth}
		\includegraphics[width=\textwidth]{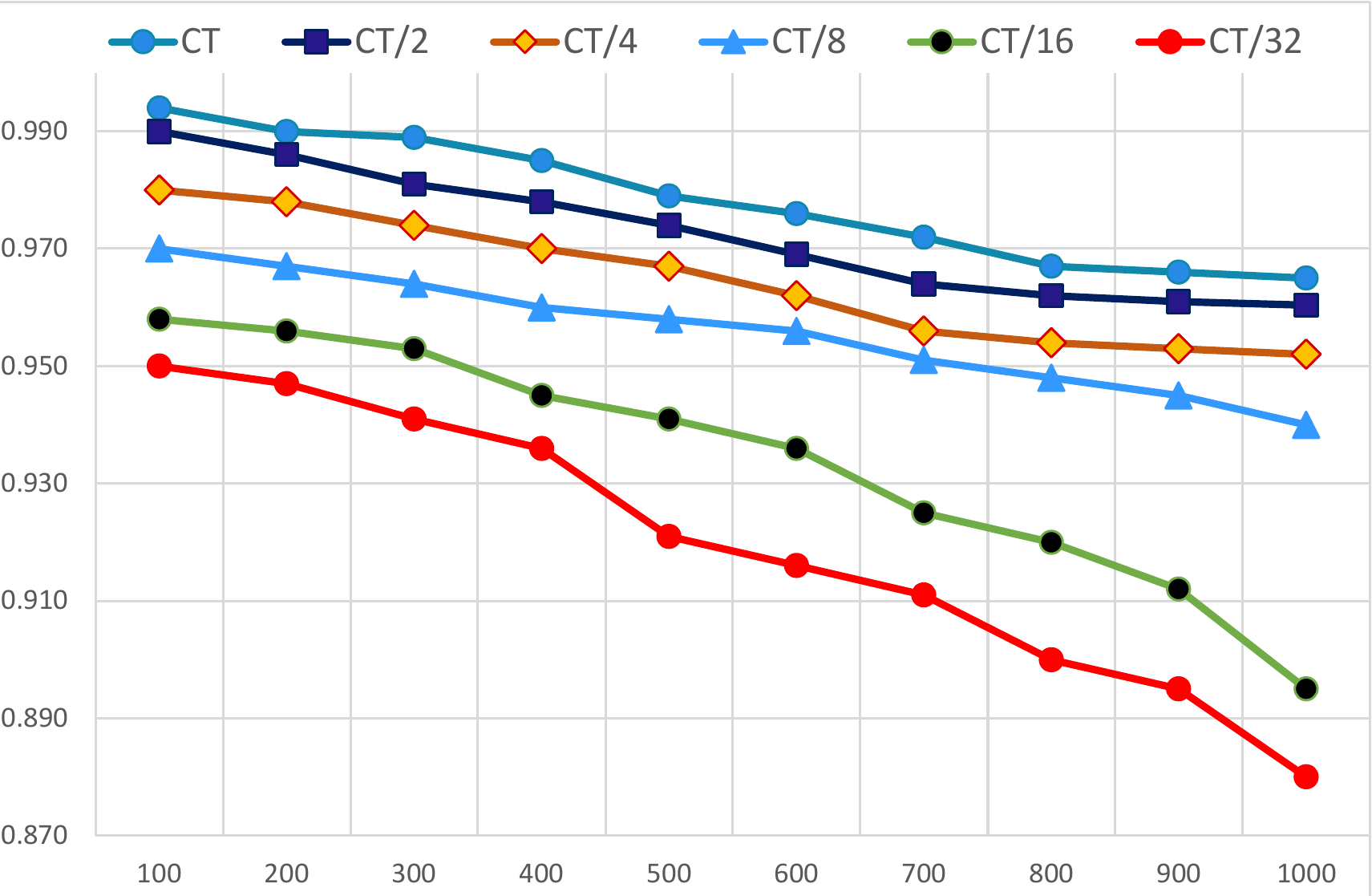}
		\caption{\LLsim}
		\label{fig:exp1:freq}
	\end{subfigure}
	\quad
	\begin{subfigure}{.4\textwidth}
		\includegraphics[width=\textwidth]{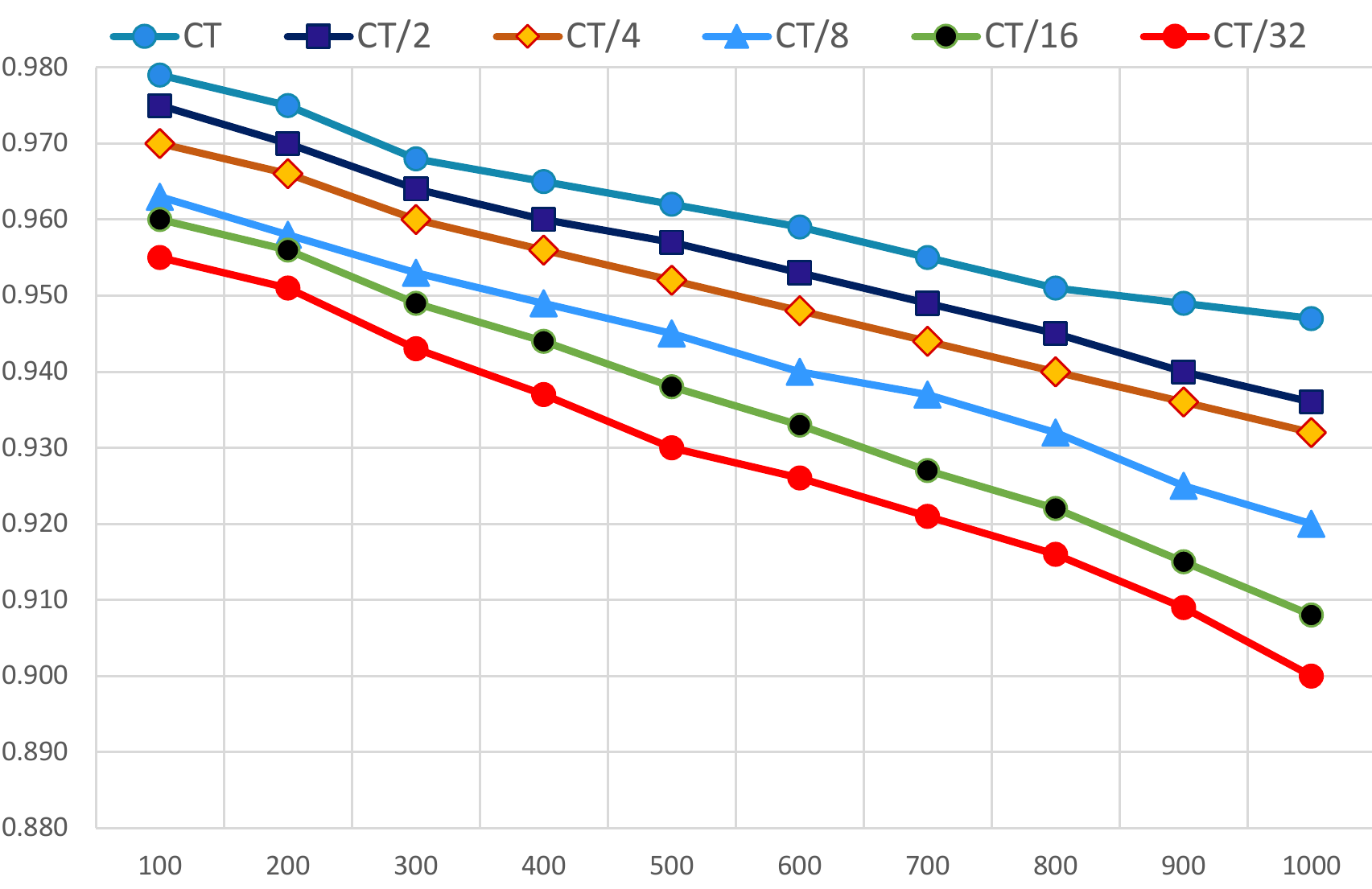}
		\caption{\FEsim}
		\label{fig:exp1:first}
	\end{subfigure}
	\begin{subfigure}{.4\textwidth}
		\includegraphics[width=\textwidth]{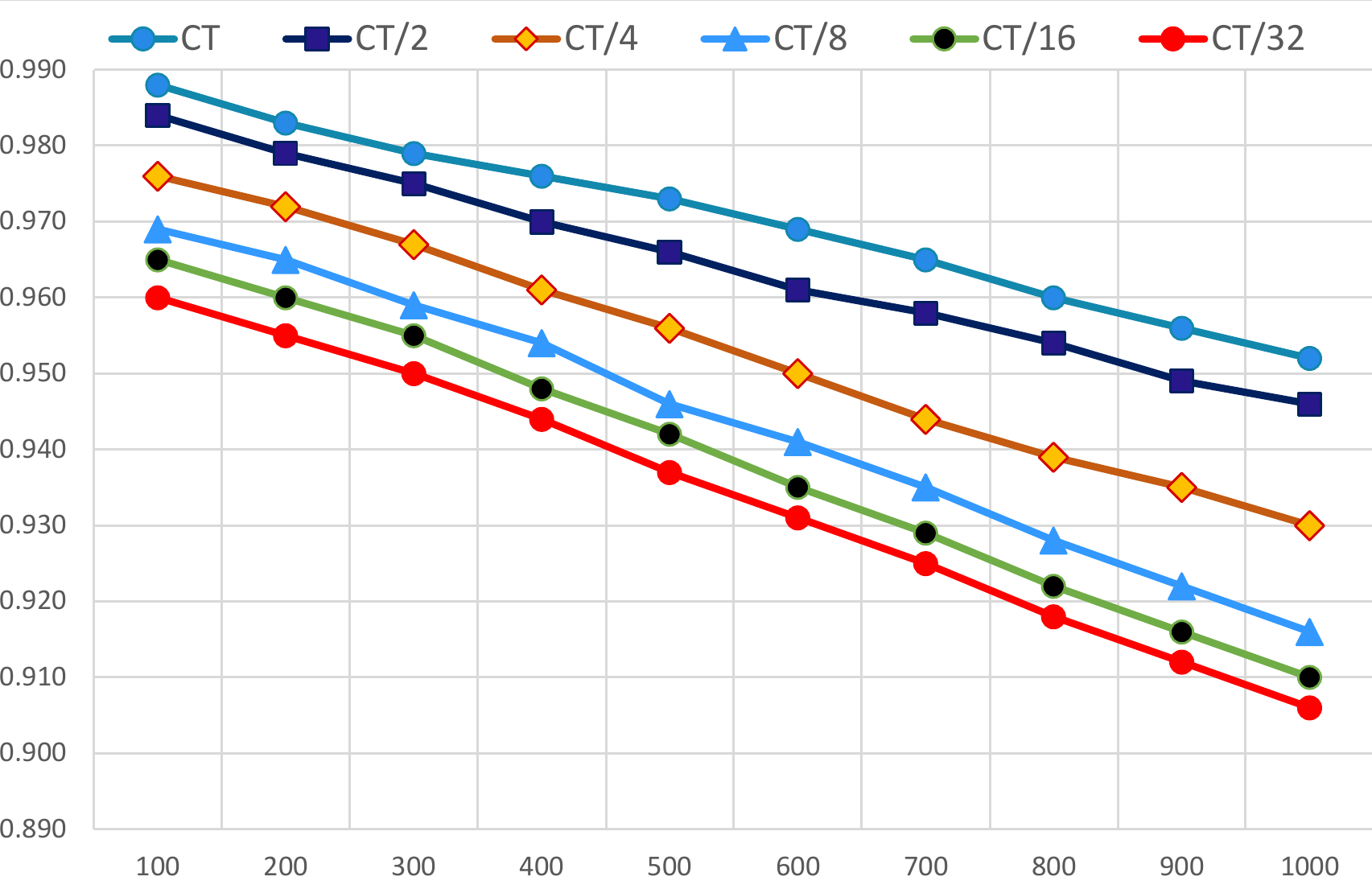}
		\caption{\PEsim}
		\label{fig:exp1:pair}
	\end{subfigure}
	\quad
	\begin{subfigure}{.4\textwidth}
		\includegraphics[width=\textwidth]{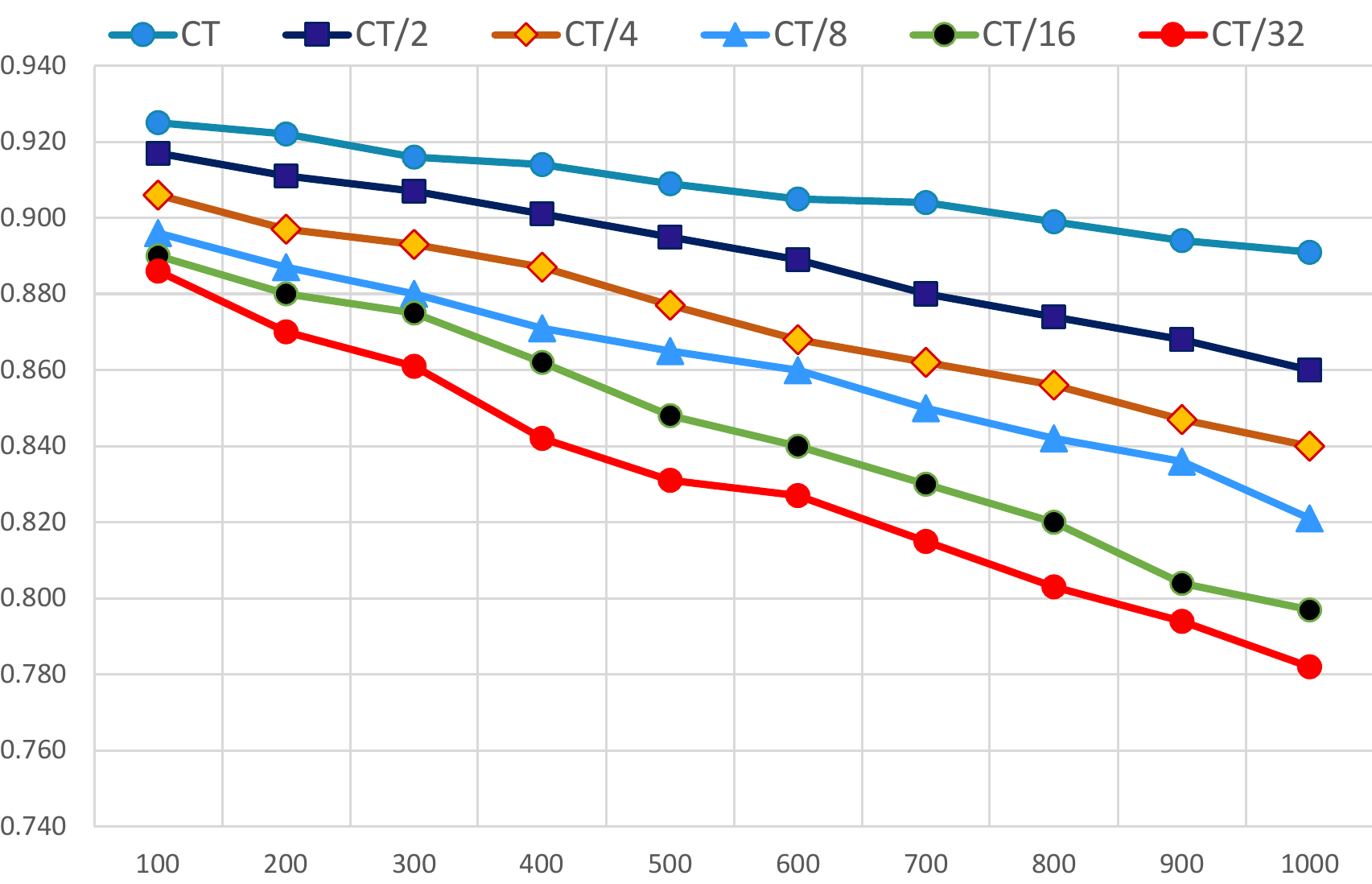}
		\caption{\TEsim}
		\label{fig:exp1:triple}
	\end{subfigure}
	\begin{subfigure}{.4\textwidth}
		\includegraphics[width=\textwidth]{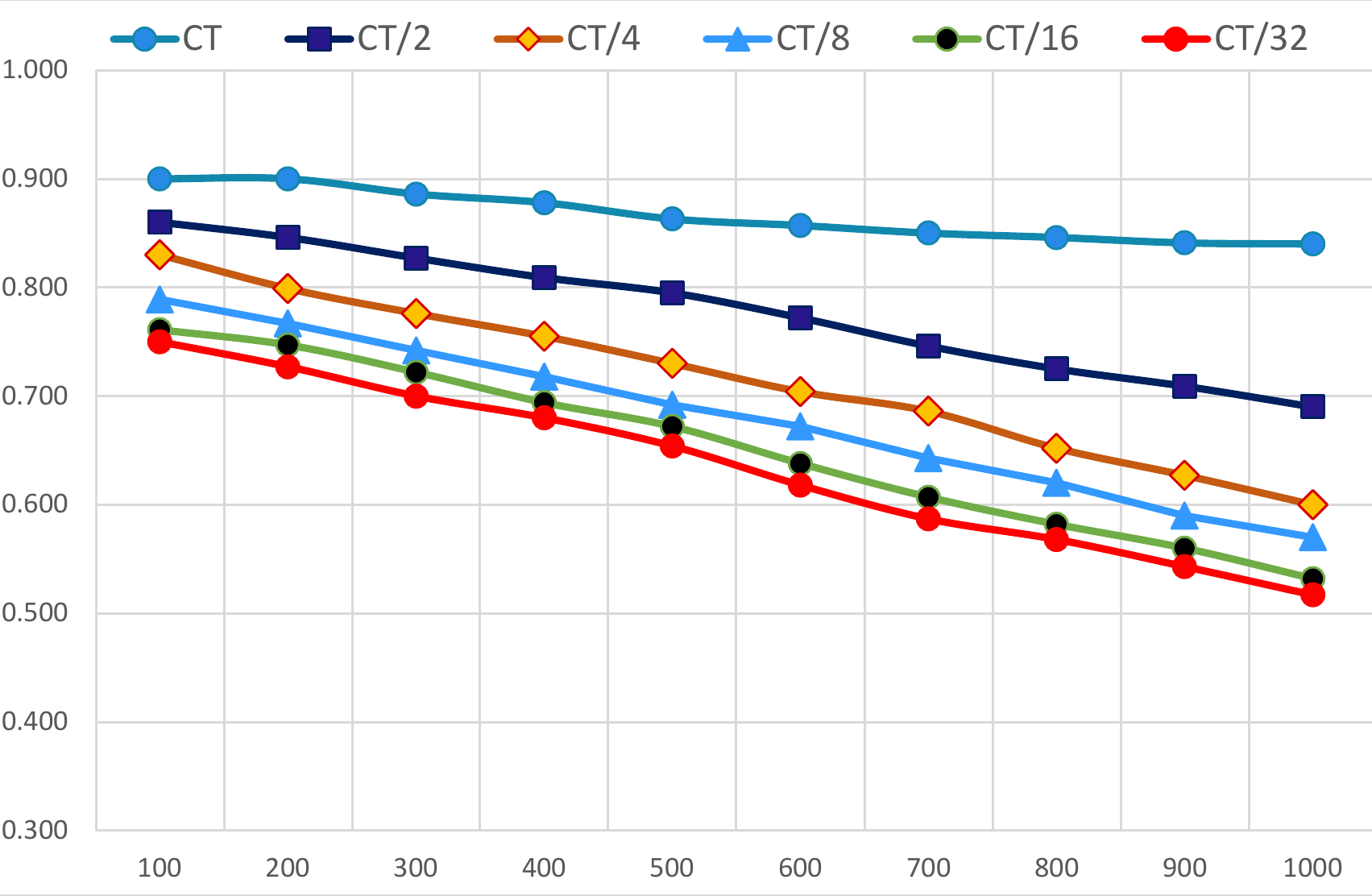}
		\caption{\Csim}
		\label{fig:exp1:case}
	\end{subfigure}
	\caption{The impact of increasing log size (on the horizontal axis) and WIP (differentiated by the color and shape of the poly-lines) on log-to-log similarity measures (on the vertical axis)}
	\label{fig:exp1:l2l}
\end{figure*}

\begin{figure*} [htb!]
	\begin{subfigure}{.4\textwidth}
		\includegraphics[width=\textwidth]{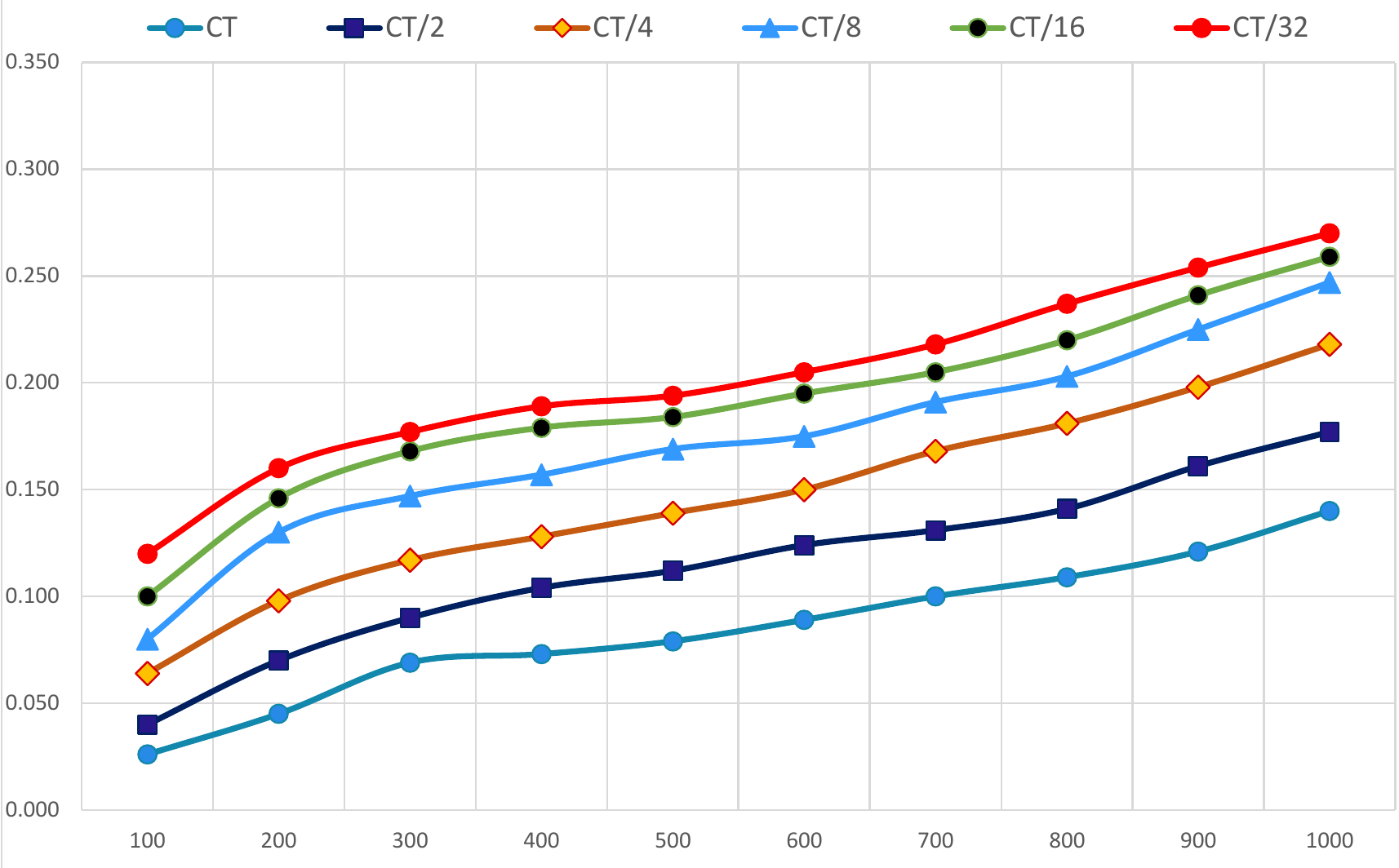}
		\caption{\LLsmape}
		\label{fig:exp1:etime}
	\end{subfigure}
	\begin{subfigure}{.4\textwidth}
		\includegraphics[width=\textwidth]{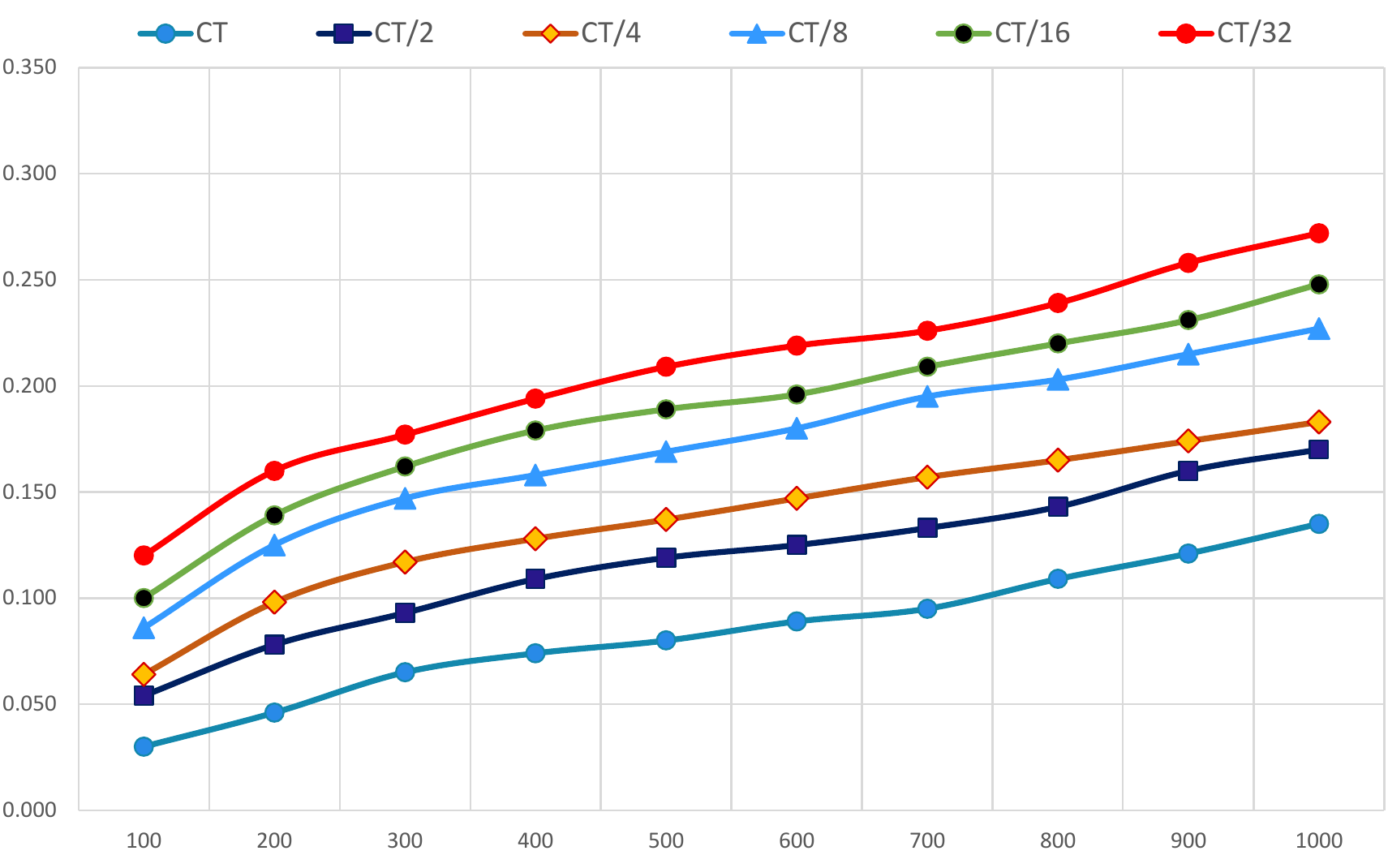}
		\caption{\LLCTsmape}
		\label{fig:exp1:ct}
	\end{subfigure}
	
	\caption{The impact of increasing log size (on the horizontal axis) and WIP (differentiated by the color and shape of the poly-lines) on log-to-log time deviation measures (on the vertical axis)}
	\label{fig:exp1:time}
\end{figure*}

\Cref{fig:exp1:l2l,fig:exp1:time} plot the results of the first experiment, which studies the impact of increasing log size and WIP on the accuracy of \ECSAnew. As we detail in the following, we observe that the stricter the measure is, the steeper the decline gets as those parameters increase.

\Cref{fig:exp1:l2l} shows how log size and WIP affect log-to-log similarity measures.
Markedly, a relaxed similarity measure such as {\TTsim} drops by around \SI{4}{\percent} from the situation in which logs consist of \num{100} cases to when logs contain \num{1000} cases, as shown in \cref{fig:exp1:traces}.
Also, \TTsim\ drops by around \SI{2}{\percent} when the inter-arrival rate goes from a value that equates the process cycle time (\num{1}~CT) to one thirty-second thereof ($\sfrac{1}{32}$~CT).
The {\PEsim} similarity measure decreases by around \SI{5}{\percent} having logs whose cardinality increases from \num{100} cases to \num{1000} cases. Also, its value decreases by around \SI{3}{\percent} as the inter-arrival rate goes from \num{1}~CT to $\sfrac{1}{32}$~CT, as shown in \cref{fig:exp1:pair}. 
The strictest similarity measure, {\Csim}, falls by around \SI{19}{\percent} when cases increase from \num{100} to \num{1000}, as shown in \cref{fig:exp1:case} and by around \SI{18}{\percent} 
as the inter-arrival rate reaches a thirty-second of the process cycle time from the initial value of a whole cycle time. 

\Cref{fig:exp1:time} illustrates how log size and WIP affect log-to-log time deviation measures. As \cref{fig:exp1:etime} depicts, {\LLsmape} and {\LLCTsmape} drop by around \SI{14}{\percent} and \SI{13}{\percent}, respectively, when the number of cases in the logs decreases from \num{100} to \num{1000}.
{\LLsmape} and {\LLCTsmape} drops by around \SI{8}{\percent} and \SI{15}{\percent}, respectively, as the inter-arrival goes from \num{1}~CT to $\sfrac{1}{32}$~CT, as shown in \cref{fig:exp1:ct}.

These drops occur because bigger logs bring more options to assign events with, and the inter-arrival rate has an influence on the number of overlapping cases.
The combination of higher volumes of cases and their density 
increases the uncertainty of the correlation decision step, thereby affecting the accuracy of the technique.

\medskip

\begin{figure*}[ptb!]
	\begin{subfigure}{.45\textwidth}
		\includegraphics[width=\textwidth]{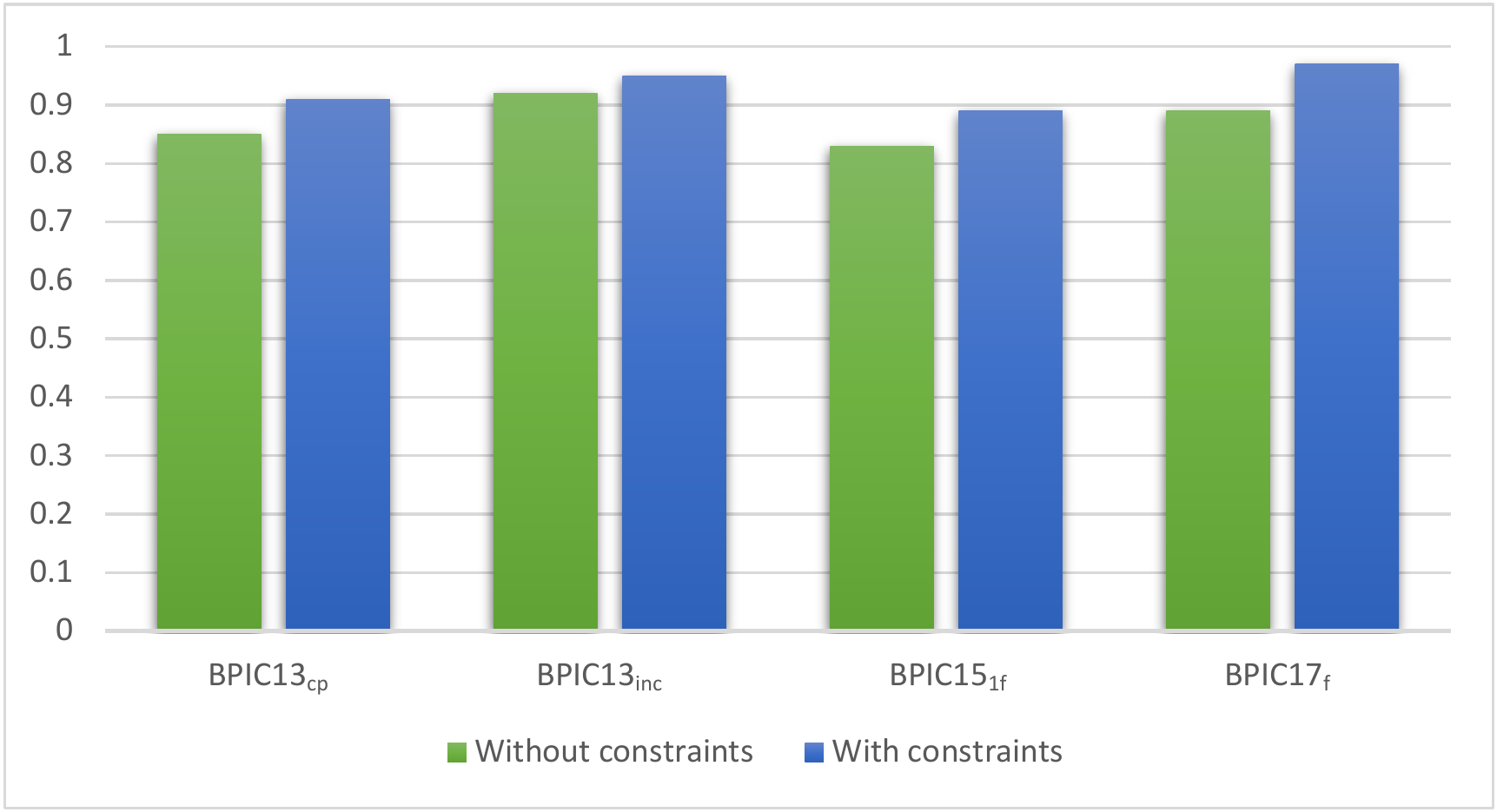}
		\caption{\TTsim}
		\label{fig:exp2:trace}
	\end{subfigure}
	\begin{subfigure}{.45\textwidth}
		\includegraphics[width=\textwidth]{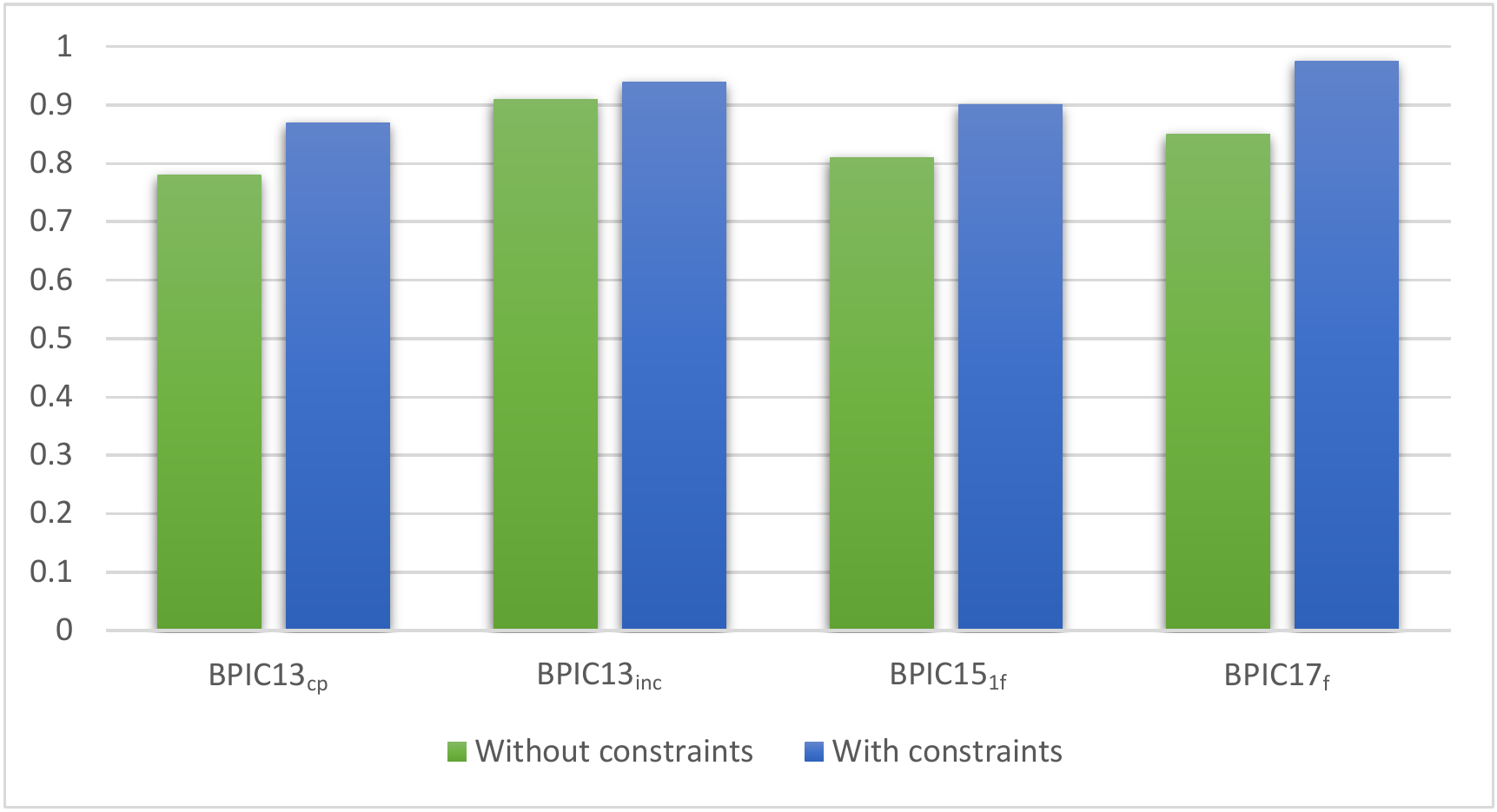}
		\caption{\LLsim}
		\label{fig:exp2:freq}
	\end{subfigure}
	\quad
	\begin{subfigure}{.45\textwidth}
		\includegraphics[width=\textwidth]{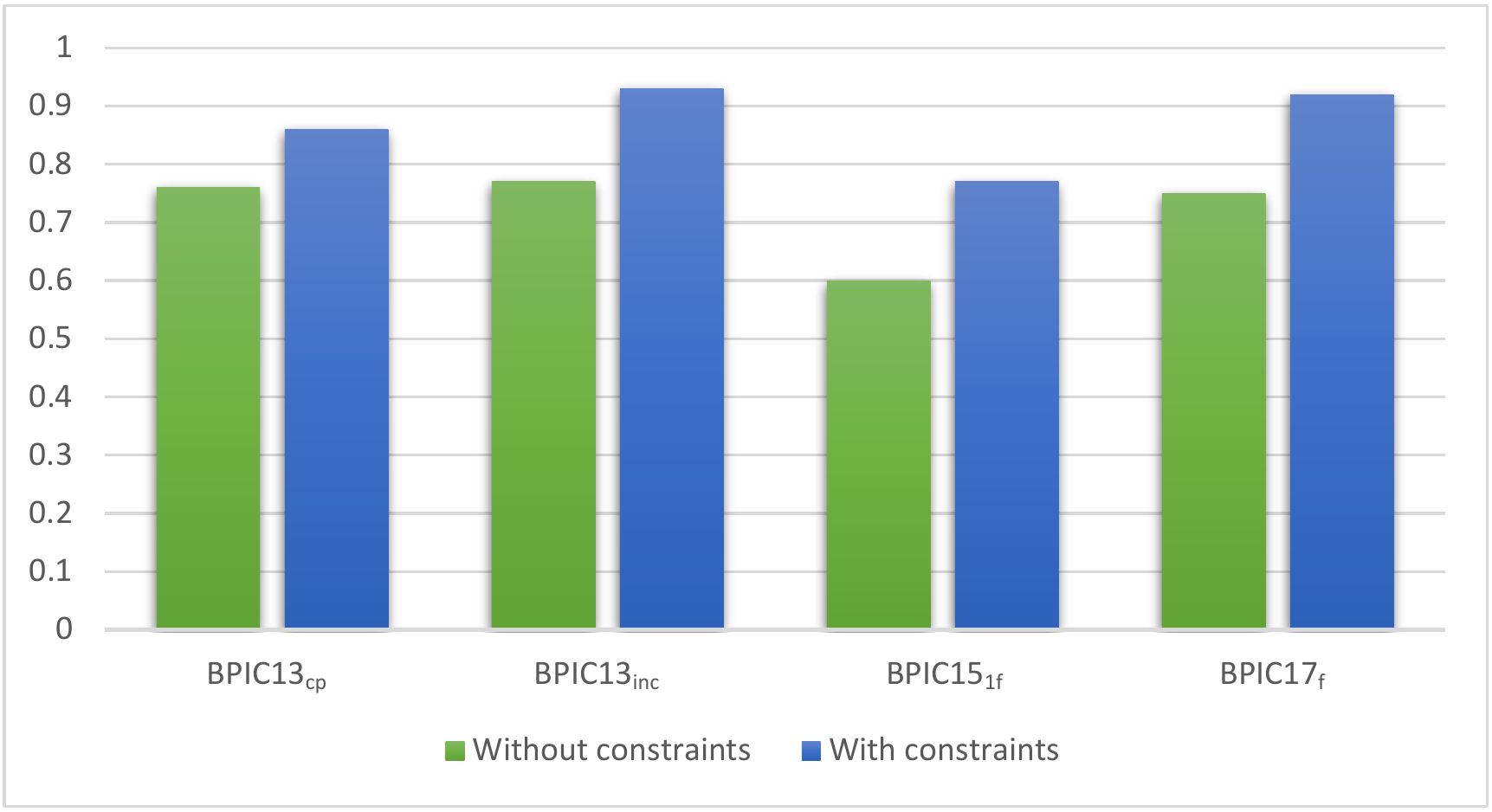}
		\caption{\FEsim}
		\label{fig:exp2:first}
	\end{subfigure}
	\begin{subfigure}{.45\textwidth}
		\includegraphics[width=\textwidth]{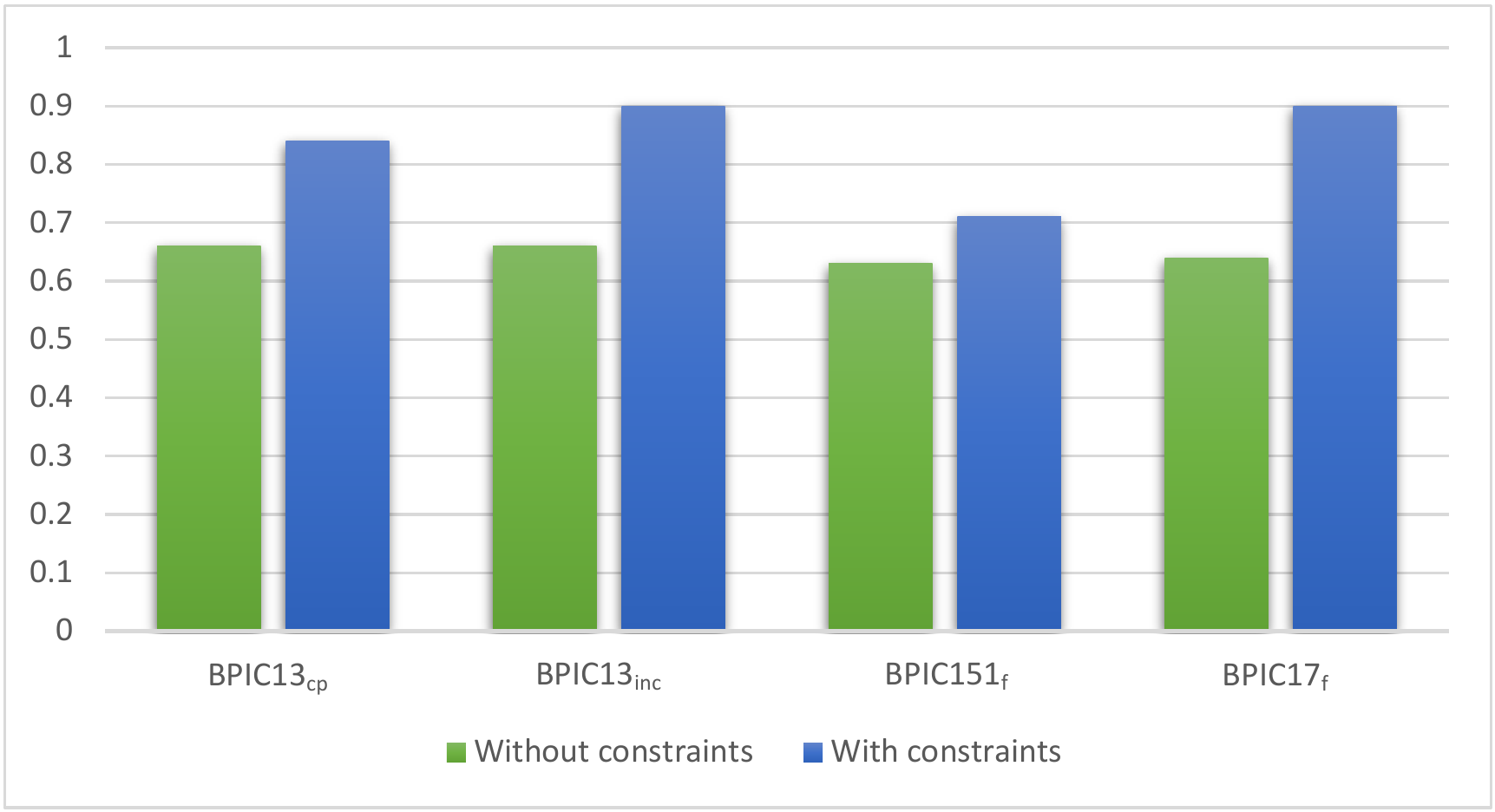}
		\caption{\PEsim}
		\label{fig:exp2:pair}
	\end{subfigure}
	\quad
	\begin{subfigure}{.45\textwidth}
		\includegraphics[width=\textwidth]{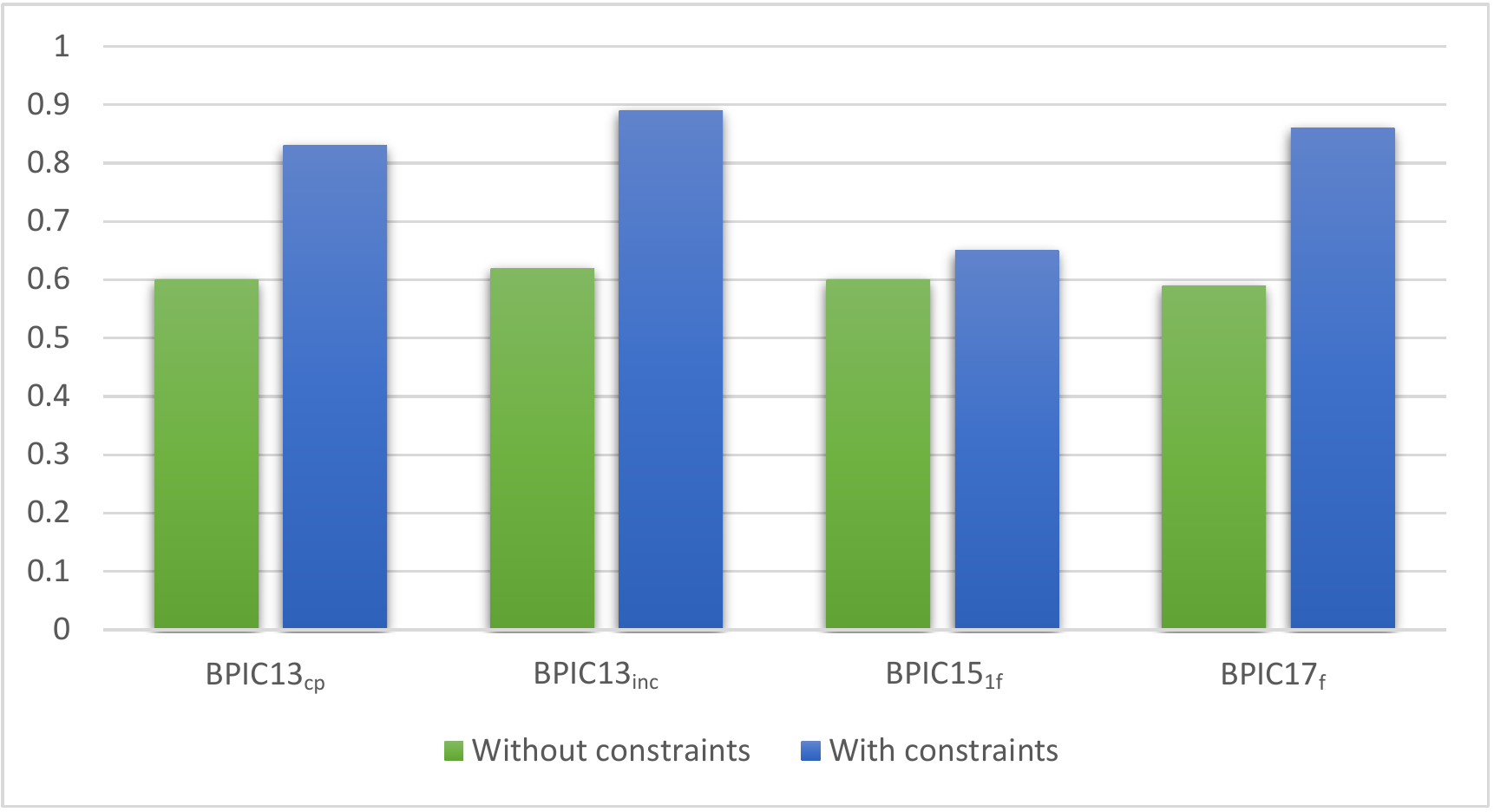}
		\caption{\TEsim}
		\label{fig:exp2:triple}
	\end{subfigure}
	\begin{subfigure}{.45\textwidth}
		\includegraphics[width=\textwidth]{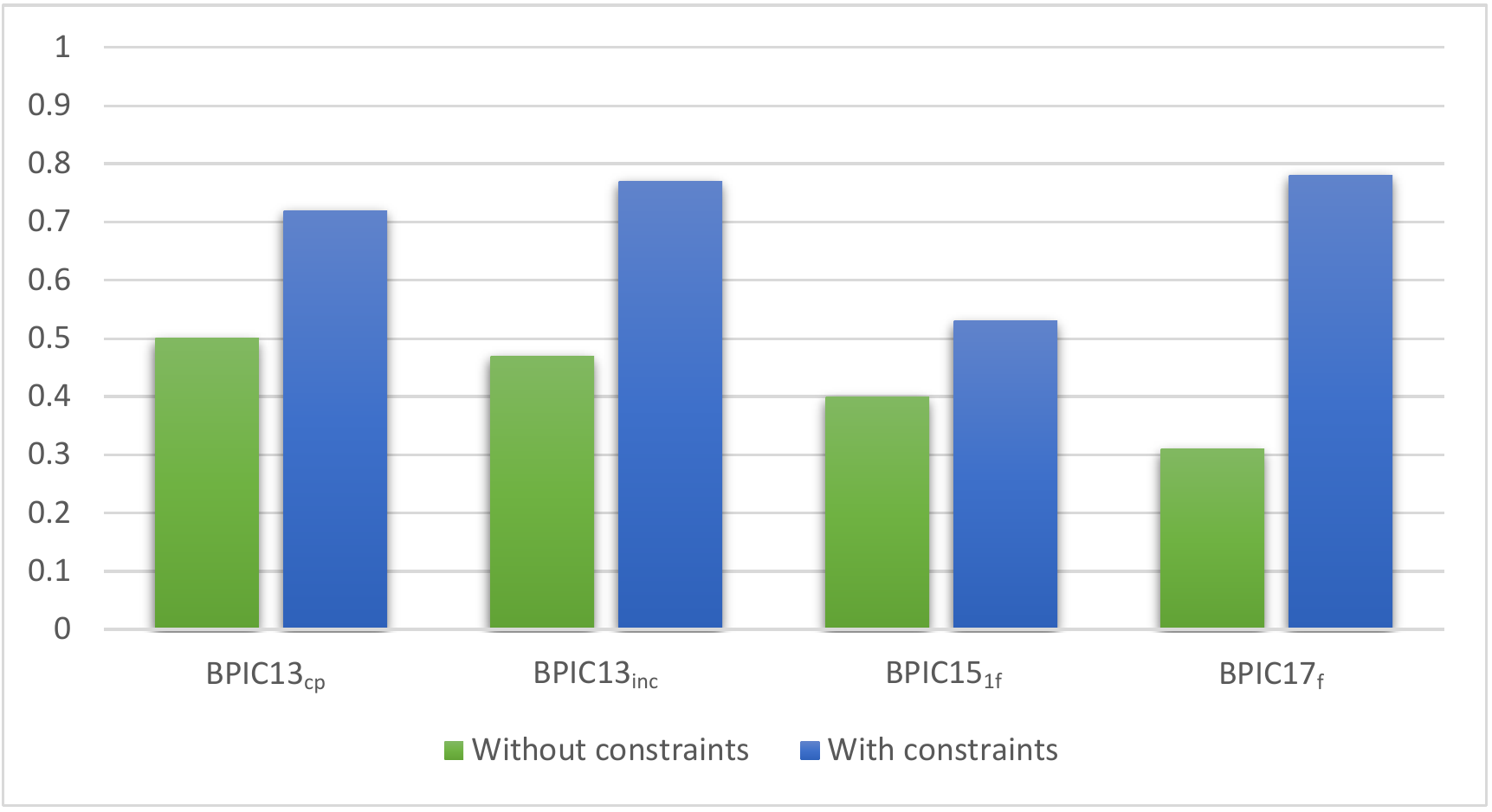}
		\caption{\Csim}
		\label{fig:exp2:case}
	\end{subfigure}
	\quad
	
	\caption{Impact of using data constraints with real-world logs on log-to-log similarity measures}
	\label{fig:exp2}
\end{figure*}
\begin{figure*} [ptb!]
	\centering
	\begin{subfigure}{.45\textwidth}
		\includegraphics[width=\textwidth]{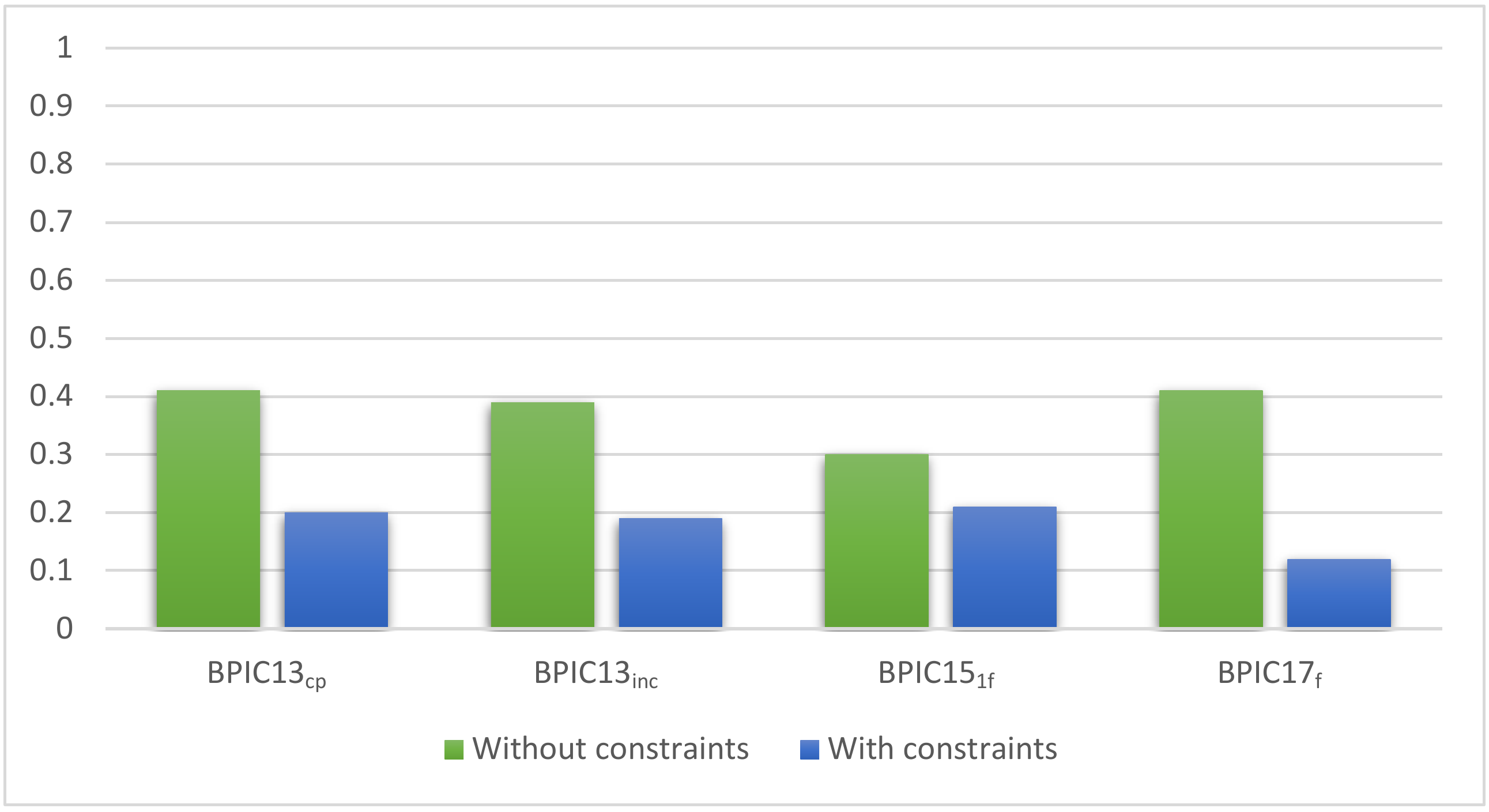}
		\caption{\LLsmape}
		\label{fig:exp2:etime}
	\end{subfigure}
	\begin{subfigure}{.45\textwidth}
		\includegraphics[width=\textwidth]{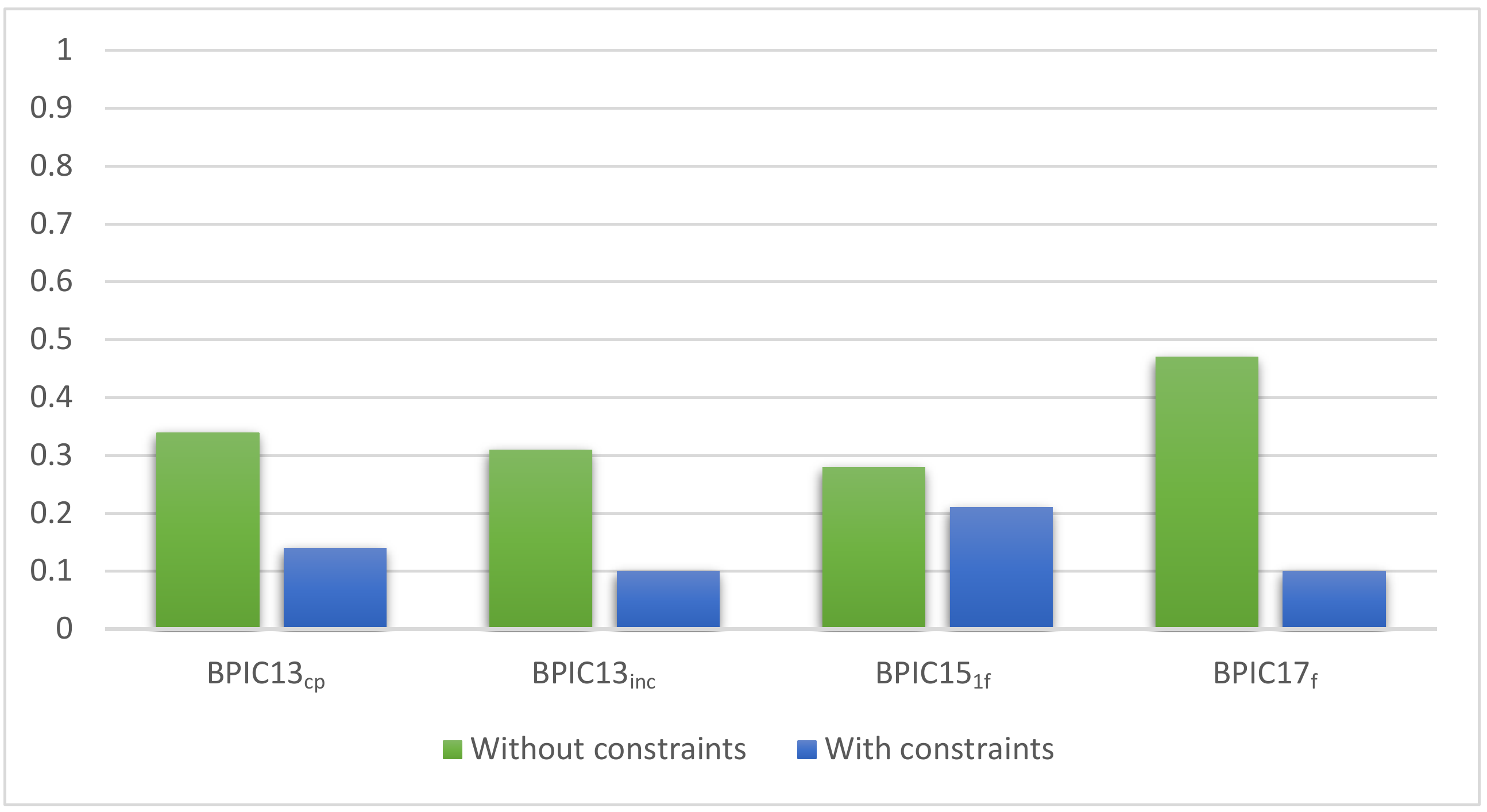}
		\caption{\LLCTsmape}
		\label{fig:exp2:ct}
	\end{subfigure}
	
	\caption{Impact of using data constraints with real-world logs on time deviation measures}
	\label{fig:exp2:dtime}
\end{figure*}

\Cref{fig:exp2,fig:exp2:dtime} show the results of the second experiment, which studies the impact that adding data constraints has on the correlation accuracy and compares the results with \ECSA (which, in contrast, did not allow for the inclusion of constraints).  We can see that doing so improves the accuracy and, indeed, \ECSAnew outperforms \ECSA. 
\Cref{fig:exp2} shows that {\TTsim}, {\PEsim} and {\Csim} increase by around \SIlist{6;15;28}{\percent} on average when constraints are in use, respectively. Notably, using data constraints with {\ECSAnew} on the BPIC17 log dramatically improves the event correlation quality 
as it can be observed in \cref{fig:exp2:case} -- notice that {\Csim} increases by \SI{46}{\percent}.
\Cref{fig:exp2:dtime} evidences that also the time deviation decreases when constraints are in use, as {\LLsmape} and {\LLCTsmape} rise by \SIlist{19;21}{\percent}, respectively. 

Using constraints enhances the correlation process as they 
prune out the violating options for case assignment. Consequently, the uncertainty of the correlation decision step decreases and this positively affects the quality of the generated log. 
However, the usage of data constraints affects the performance of {\ECSAnew} particularly in terms of execution time. 
The reason is, constraints must be verified at every assignment step. Therefore, the more the constraints to check, the higher the overall computation time. For instance, {\ECSAnew} ran for \SI{13}{\hour} to complete the execution with the BPIC17 event log using \num{10} constraints, in contrast with the \SI{8.6}{\hour} needed in absence of constraints. The processing of the BPIC15\textsubscript{1f} log required \SI{6.2}{\hour} with \num{4} constraints and \SI{5.5}{\hour} without constraints.

\medskip

\begin{figure*}[ptb!] 
	\begin{subfigure}{.45\textwidth}
		\includegraphics[width=\textwidth]{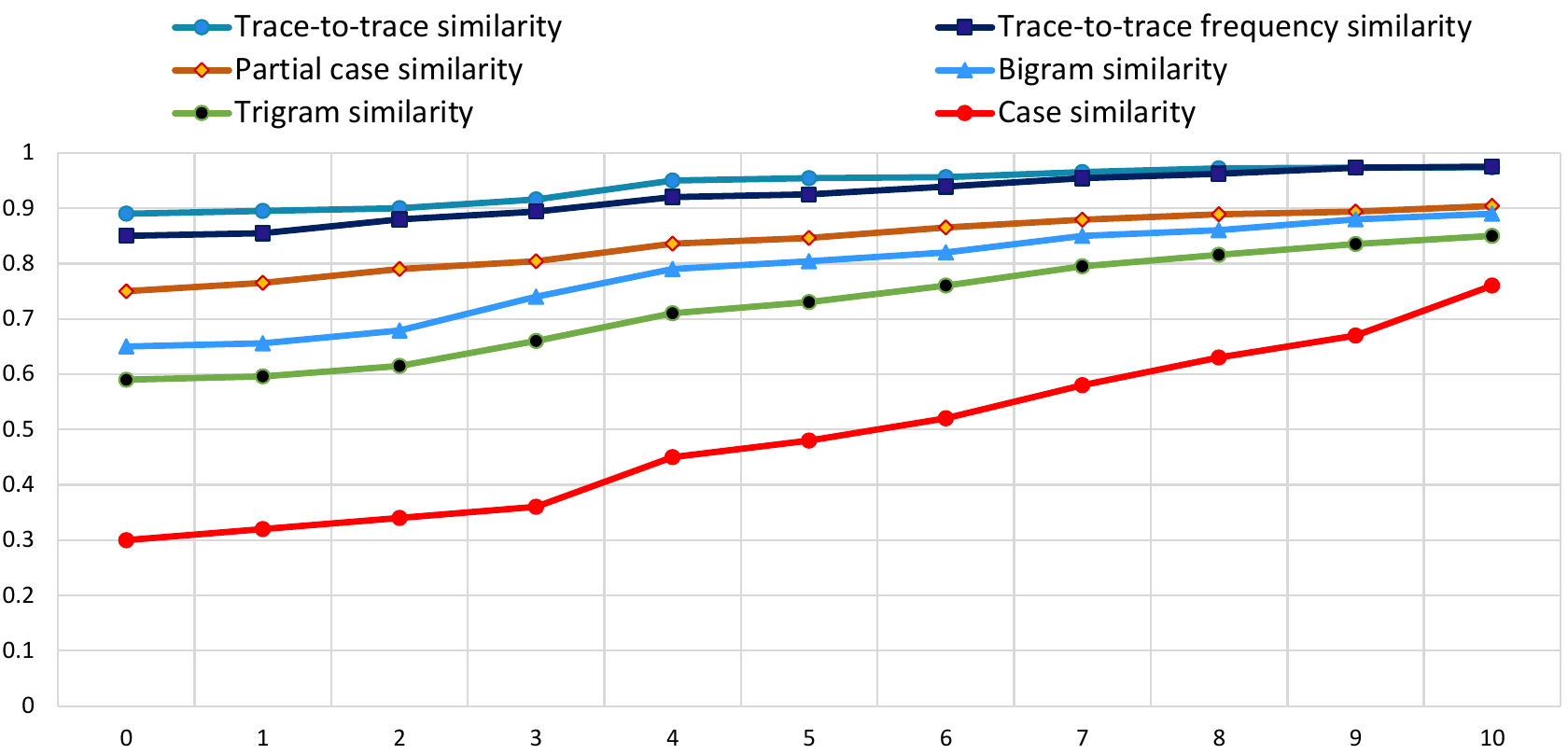}
		\caption{Log similarity measures}
		\label{fig:exp3:l2l}
	\end{subfigure}
	\begin{subfigure}{.45\textwidth}
		\includegraphics[width=\textwidth]{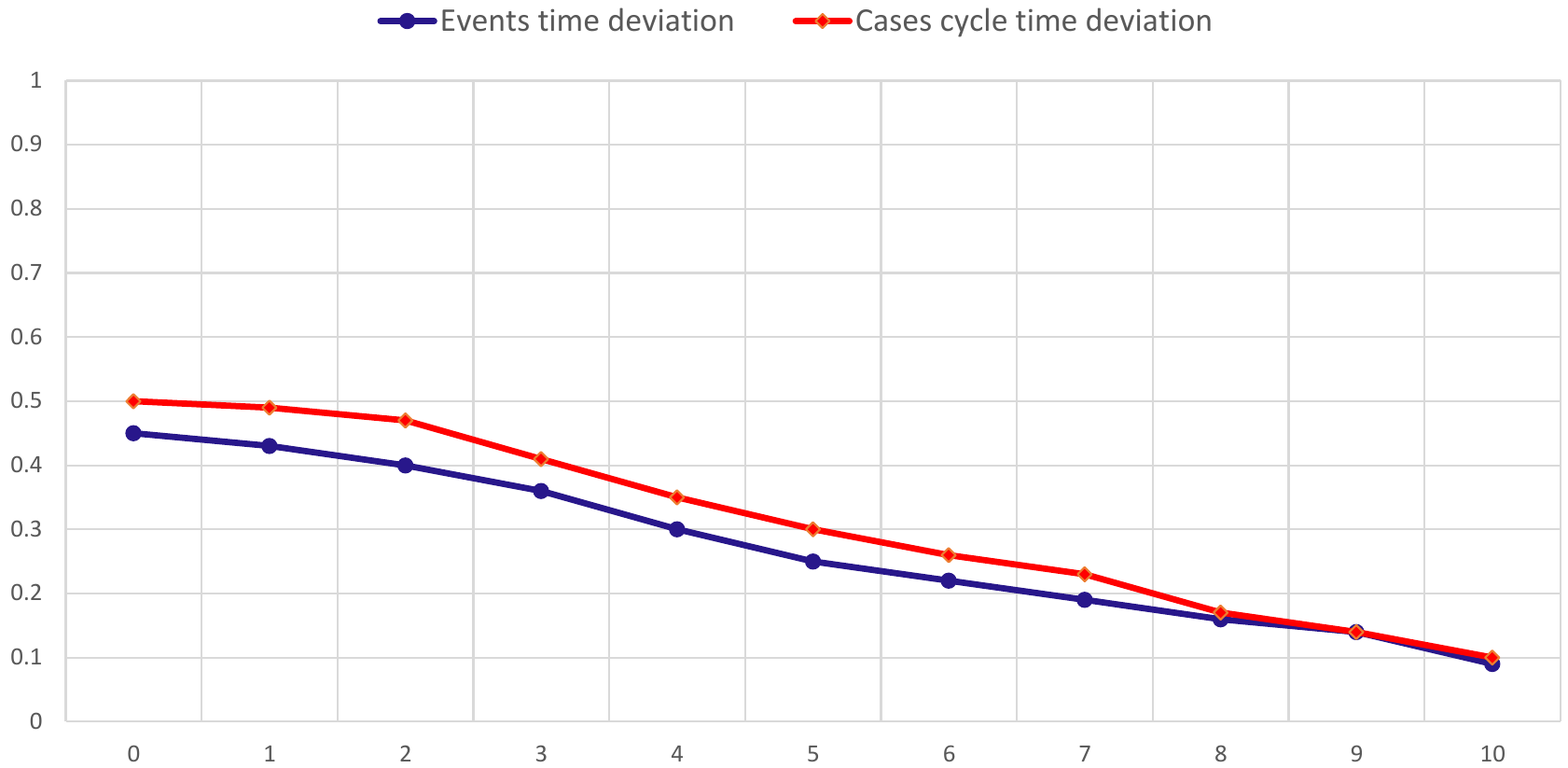}
		\caption{Time variance measures}
		\label{fig:exp3:smap}
	\end{subfigure}
	\quad
	\begin{subfigure}{.45\textwidth}
		\includegraphics[width=\textwidth]{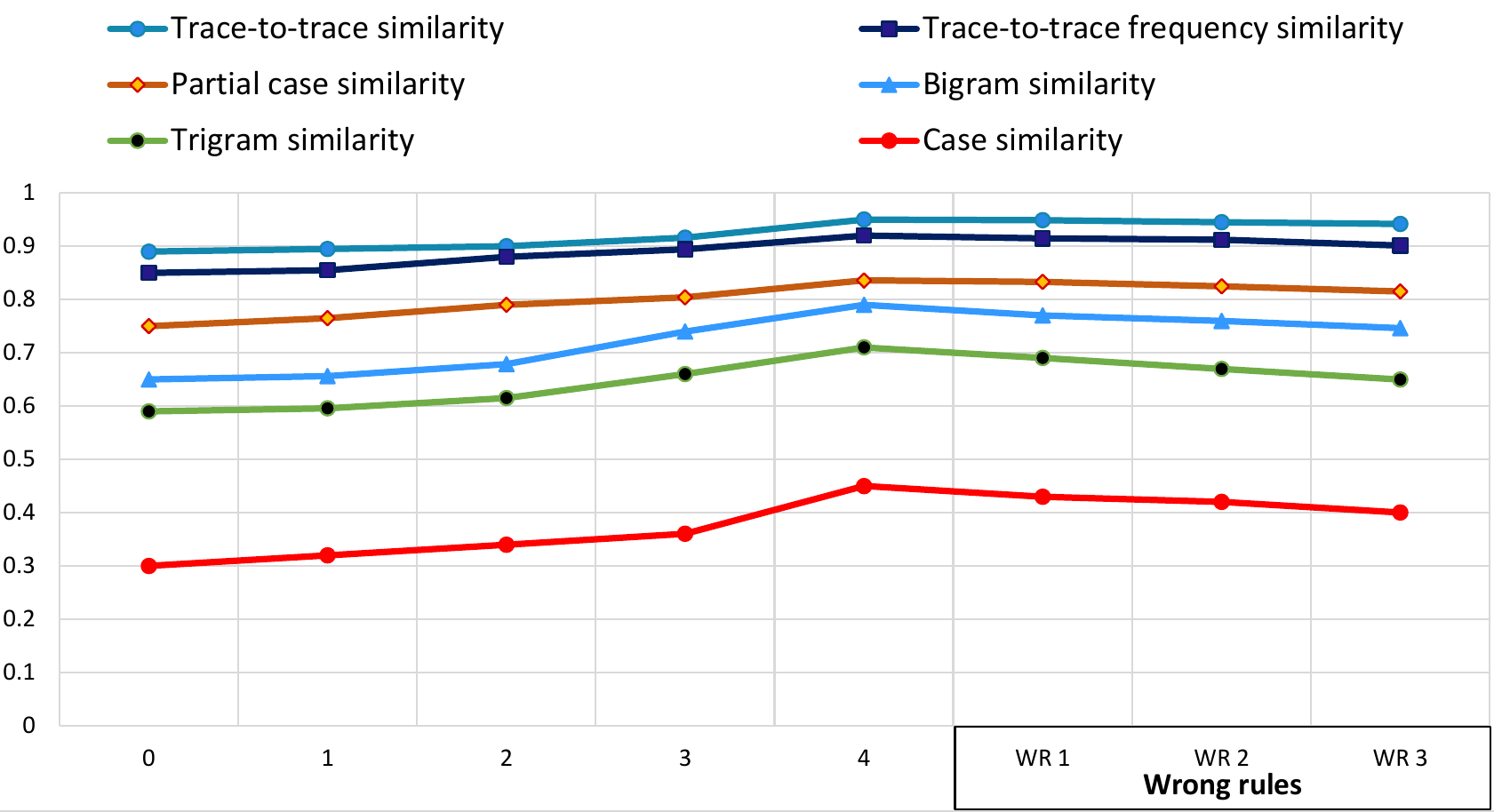}
		\caption{Log similarity measures -- using wrong rules}
		\label{fig:exp3:fl2l}
	\end{subfigure}
	\begin{subfigure}{.45\textwidth}
		\includegraphics[width=\textwidth]{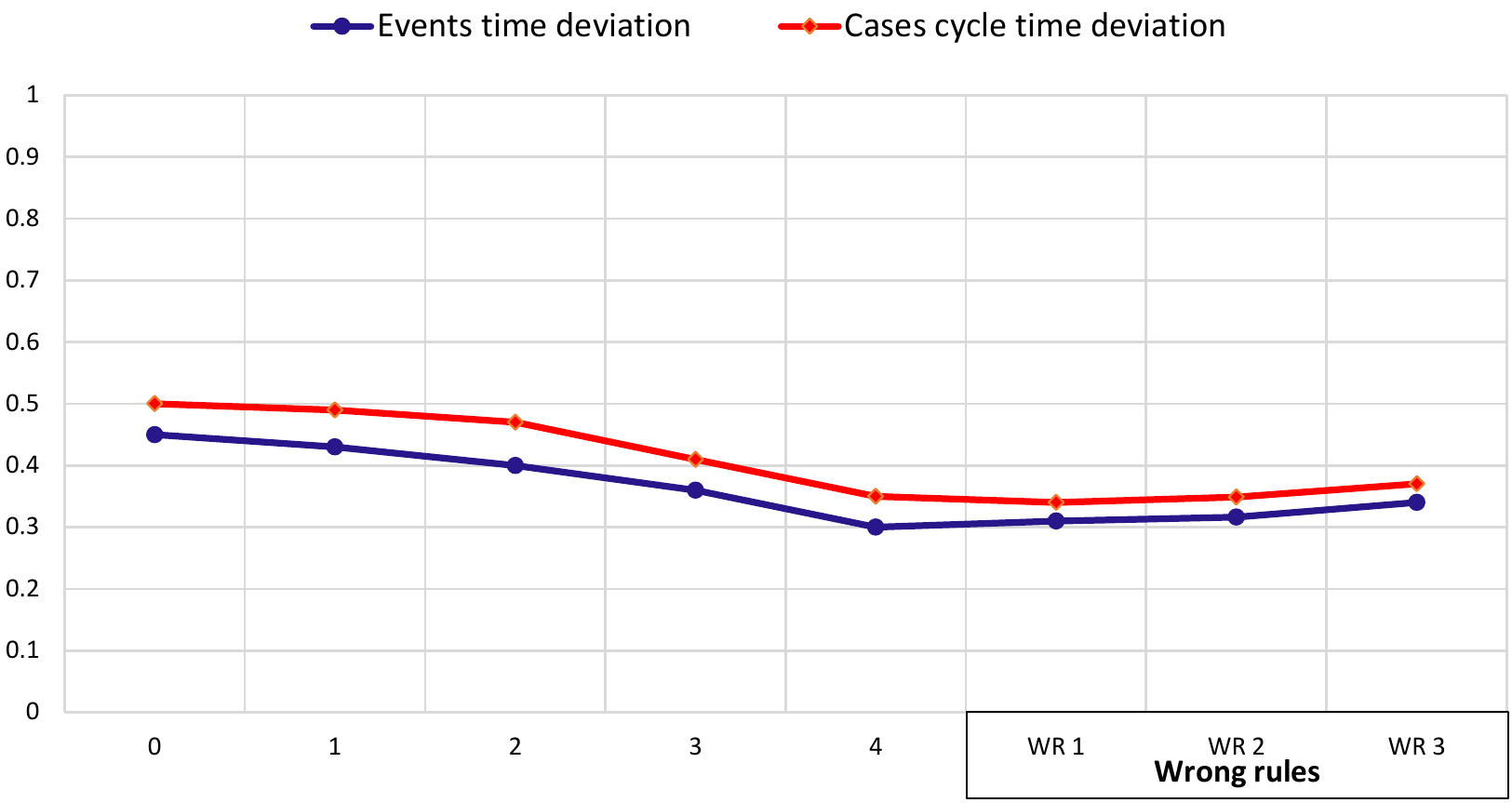}
		\caption{Time variance measures -- using wrong rules}
		\label{fig:exp3:fsmape}
	\end{subfigure}
	\caption{Sensitivity analysis on the usage of constraints}
	\label{fig:exp3}
\end{figure*}

\Cref{fig:exp3} shows the results of the third experiment, which studies the effect of the number and correctness of constraints on the accuracy of the correlation process of \ECSAnew. \Cref{fig:exp3:l2l} shows that using more constraints improves the log-to-log similarity accuracy measures of the generated logs. For instance, {\TTsim}, {\PEsim} and {\Csim} increase by around \SIlist{8;24;46}{\percent} when the constraints reach the peak of \num{10} correct ones. \Cref{fig:exp3:smap} evidences that the log-to-log time deviation accuracy measures decrease too: in particular, {\LLsmape} and {\LLCTsmape} drop by up to \SIlist{36;40}{\percent}, respectively.
These improvements materialize as more knowledge in terms of data constraints helps to discard event-case assignment possibilities and, therefore, decrease the uncertainty of the correlation decision.

\Cref{fig:exp3:fl2l,fig:exp3:fsmape} illustrate the effect of using incorrect constraints. These constraints are intentionally inconsistent with data in order to study the impact of the knowledge quality on the
accuracy of the technique.
We observe that using \num{3} wrong constraints makes {\TTsim} reduce by \SI{1}{\percent}, {\PEsim} by \SI{4}{\percent} and {\Csim} by \SI{5}{\percent}. Also, it affects the time deviation measures as {\LLsmape} and {\LLCTsmape} increase by \SIlist{4;2}{\percent}, respectively.
We observe that invalid constraints affect the overall accuracy though not severely thanks to the positive impact of other valid ones. This scenario is meant to resemble a realistic scenario in which constraints are available but without the certainty that all of those are consistent with the event data at hand. We can see that {\ECSAnew} still provides a valid correlated log with sufficient accuracy.

\subsection{Discussion}
Based on the different sensitivity analyses we conducted, we found that using constraints improves correlation accuracy due to a reduction of possible assignments per event. Also, we see that by using constraints, we reduce the arbitrary decisions taken based on lowly accurate models. Our approach is mildly sensitive to the accuracy of the given data. The usage of various constraints with different support and correctness influence the event correlation decision. However, based on the third experiment, the correctness of some rules can balance the negative impact of the incorrect ones so as not to dramatically affect the correlation process. The accuracy of our approach tends to be partially prone to a worsening when the cases' density and log size increase, as these parameters increase the number of options available thereby reducing certainty in the event assigning decision. 

We remark that {\ECSAnew} can handle cyclic and parallel behavior, largely present in the real-world event logs we analyzed, as opposed to other proposed techniques~\cite{cMiner,eMax,seq}..
Also, {\ECSAnew} allows users to provide rules that constrain the process behavior; however, it still consider those rules as a complement to the control-flow knowledge in order not to be entirely dependent on the accuracy of the event log data, as discussed in \cref{related}. Thus, combining the control-flow and data-domain knowledge helps to balance the inaccuracy of the input data and supports the correlation process in generating a sufficiently reliable output log. The investigation of possible solutions solely resorting to constraints paves the path for future work, as we will discuss in the next section.
\section{Conclusion}\label{conclusion}
The research presented in this paper addresses the event correlation problem.
To automatically correlate the events to their proper cases, our approach (\ECSAnew) resorts to data constraints to model domain knowledge, in addition to process models that define the control-flow of the original process. Our approach uses multi-level objective simulated annealing to map every event to a case. 
We use trace alignment cost, support of data constraints, and activity execution time variance for optimization. Our evaluation on real-world event logs demonstrates that using data constraints as input in addition to the process model improves the generated log quality positively.

There are several directions for future research. A first option is to support a broader spectrum of data knowledge constraints, such as the inter-case constraints~\cite{DBLP:journals/is/WinterSR20}. 
Also, extending the set of quality measures for correlated event logs is an objective that can be pursued in following endeavors.
Finally, an interesting avenue for future work is to explore the possibility of correlating the events solely based on constraints, without prior knowledge of the whole process model or by means of 
declarative process rules~\cite{BaierDMW18}.
%
\subsubsection*{Acknowledgements} \label{sec:ack}
The work of Claudio Di Ciccio was partly supported by the Italian Ministry of University and Research (MUR) under grant ``Dipartimenti di eccellenza 2018-2022'' of the Department of Computer Science at Sapienza University of Rome and by the Sapienza SPECTRA research project. The research by Jan Mendling was supported by the Einstein Foundation Berlin.
%
%
%
%
%
%
\appendix
\section{BPIC-2017 rules} \label{rules}
\Cref{tbl:2017Rules} reports the data constraints we retrieved from visual inspection of the BPIC17\textsubscript{f} event log to run our experiments.
\begin{table*}[htb]
	\resizebox{0.9\textwidth}{!}{%
		\begin{tabular}{|l|} 
			\hline
			$C_1:\sigma(i-1).\mathrm{ApplicationType} = \sigma(i).\mathrm{ApplicationType}$                                                                                                                                                \\ 
			\hline
			$C_2:\sigma(i-1).\mathrm{LoanGoal} = \sigma(i).\mathrm{LoanGoal}$                                                                                                                                                              \\ 
			\hline
			$C_3:\sigma(i-1).\mathrm{RequestedAmount} = \sigma(i).\mathrm{RequestedAmount}$                                                                                                                                                \\ 
			\hline
			$C_4:\ifcon~ \sigma(i).\attr{EventOrigin} = \textrm{``}\textsf{Offer}\textrm{''} \wedge \sigma(i).\attr{Act}  \neq  \textrm{``}\textsf{O\_Create~offer}\textrm{''} \wedge \sigma(j).\attr{Act} =  \textrm{``}\textsf{O\_Create~offer}\textrm{''} ~\thencon~\sigma(i).\attr{OfferID} = \sigma(j).\attr{EventId}$  \\ 
			\hline
			$C_5:\ifcon~ \sigma(i).\mathrm{Act} = \textrm{``}\textsf{A\_Complete}\textrm{''} \wedge \sigma(j).\mathrm{Act} =  \textrm{``}\textsf{W\_Call~after~offers}\textrm{''}  ~\thencon~ \sigma(i).\attr{Resource} = \sigma(j).\attr{Resource} $                                                     \\
			\hline
			
			$C_6:\ifcon~ \sigma(i).\mathrm{Act} = \textrm{``}\textsf{W\_Call~after~offers}\textrm{''} \wedge \sigma(i).\mathrm{Act} =  \textrm{``}\textsf{O\_Sent~(mail~and~online)}\textrm{''}   ~\thencon~\sigma(i).\attr{Resource} = \sigma(j).\attr{Resource} $                                        \\ 
			\hline
			$C_7:\ifcon~ \sigma(i).\mathrm{Act} = \textrm{``}\textsf{O\_Returned}\textrm{''}  \wedge  \sigma(i).\mathrm{Act} =  \textrm{``}\textsf{A\_Validating}\textrm{''}   ~\thencon~\sigma(i).\attr{Resource} = \sigma(j).\attr{Resource} $                                                          \\ 
			\hline
			$C_8: \ifcon~ \sigma(i).\mathrm{Act} = \textrm{``}\textsf{A\_Pending}\textrm{''}  \wedge  \sigma(i).\mathrm{Act} =  \textrm{``}\textsf{O\_Accepted~(mail~and~online)}\textrm{''}  ~\thencon~\sigma(i).\attr{Resource} = \sigma(j).\attr{Resource} $                                           \\ 
			\hline
			$C_9: \ifcon~ \sigma(i).\mathrm{Act} = \textrm{``}\textsf{O\_Refused}\textrm{''}  \wedge  \sigma(i).\mathrm{Act} =  \textrm{``}\textsf{A\_Denied}\textrm{''}   ~\thencon~\sigma(i).\attr{Resource} = \sigma(j).\attr{Resource} $                                                              \\ 
			\hline
			$C_{10}:   \sigma(i).\mathrm{Act} = \textrm{``}\textsf{O\_Create~offer}\textrm{''}  \wedge  \sigma(i).\mathrm{Act} =  \textrm{``}\textsf{A\_Accepted}\textrm{''}   ~\thencon~\sigma(i).\attr{Resource} = \sigma(j).\attr{Resource} $                                                      \\
			\hline
		\end{tabular}
	}
	\caption{Observed rules in the BPIC17\textsubscript{f} event log}
	\label{tbl:2017Rules}
\end{table*}
%
%
%
%
%
%
\bibliographystyle{model1-num-names}
\bibliography{bib}

\end{document}